# Chapter 1

# Crowded Charges in Ion Channels

***Bob Eisenberg***

Department of Molecular Biophysics and Physiology
Rush University
Chicago IL 60305
USA

Mathematics and Computer Science Division
Argonne National Laboratory
9700 South Cass Avenue
Argonne, IL 60439
USA

**File Name: IonChannels.docx**

**Physical Chemistry and Life**. Ions in water are the liquid of life. Life occurs almost entirely in 'salt water'. Life began in salty oceans. Animals kept that salt water within them when they moved out of the ocean to drier surroundings. The plasma and blood that surrounds all cells are electrolytes more or less resembling sea water. The plasma inside cells is an electrolyte solution that more or less resembles the sea water in which life began. Water itself (without ions) is lethal to animal cells and damaging for most proteins. Water must contain the right ions in the right amounts if it is to sustain life.

Physical chemistry is the language of electrolyte solutions and so physical chemistry, and biology, particularly physiology, have been intertwined since physical chemistry was developed some one hundred fifty years ago. Physiology, of course, was studied by the Greeks some millennia earlier, but the biological role of electrolyte solutions could not be understood until ions were discovered by chemists some 2,000 years later.

Physical chemists and biologists come from different traditions that separated for several decades as biologists identified and described the molecules of life. Communication is not easy between a fundamentally descriptive tradition and a fundamentally analytical one. Biologists have now learned to study their well defined systems with physical techniques, of considerable interest to physical chemists. Physical chemists are increasingly interested in spatially inhomogeneous systems with structures on the atomic scale so common in biology. Physical chemists will find it productive to work on well defined systems built by evolution to be reasonably robust, with input output relations insensitive to environmental insults. The overlap in science is clear. The human overlap is harder because the fields have grown independently for some time, and the knowledge base, assumptions, and jargon of the fields do not coincide. Indeed, they sometimes seem disjoint, without overlap.

This article deals with properties of ion channels that in my view can be dealt with by 'physics as usual', with much the same tools that physical chemists apply to other systems. Indeed, I introduce and use a tool of physicists—a field theory (and boundary conditions) based on an energy variational approach developed by Chun Liu[141,575,766,806,944]—not too widely used among physical chemists. My goal is to provide the knowledge base, and identify the assumptions, that biologists use in studying ion channels, avoiding jargon. Although we do not know enough to write atomic, detailed physical models of the process by which ions move through channels, rather simple models of selectivity and permeation work quite well in important cases. Those physical models and cases are the main focus of this review because they demonstrate the strong essential link between the traditional treatments of ions in chemical physics, and the biological function of ion channels.

At first, ion channels may seem to be an extreme system. They are as small as they can be, given the particulate nature of matter. Ion channels are atomic valves, that allow a handful of atoms to control macroscopic flows of current, and thus macroscopic properties of cells, tissues, animals and life. They do this by working at the extremes of forces as well as sizes. They have enormous densities of ions crowded into tiny spaces with huge electric and chemical fields and forces of excluded volume. These enormous densities are as far as one can imagine from the vanishing densities of simple or ideal fluids. So ion channels may seem a special case not of general interest to physical chemists.

I hope to show that the special case of ion channels gives general insight. "If you look closely enough at a keyhole, you can look through it, and sometimes even glimpse an horizon or even stars" (John Edsall, personal communication). Whenever a physical system is controlled by a small space, whenever a chemical engineer uses a tiny valve, whenever a boundary layer near an electrode is a determinant of electrochemical function, one can expect crowded charges in tiny spaces. In those physical systems, crowded charges are likely to involve the same physics as crowded charges in biological channels. A general theme can be viewed through the biological channel.

The general theme that emerges is that everything interacts with everything else in any crowded environment, including ions in channels. I will argue that crowded conditions require a mathematics that deals naturally with interactions. I will argue that the law of mass action (with constant constants) does not apply in crowded cases. I will argue that crowded systems are complex, not simple fluids. Interactions in complex fluids have been analyzed with Chun Liu's variational method *EnVarA* that naturally deals with interactions[575,766,806,944] and we are now applying that approach to ionic solutions.[280,467]

A number of topics are discussed several times from different perspectives in this review more in the tradition of an essay than a scientific paper. The motivation is to provide physical, chemical, and biological views of the key topics. I hope those already familiar with these ideas will have patience with this approach.

**<u>Ion channels</u>.** Ion channels are the recurrent theme in the passacaglia of this essay, because they provide a natural link between physical chemistry and molecular biology, as we shall see. Ion channels have been studied in astonishing detail[29,415] despite their staggering diversity.[4,174,443,730] Ion channels have enormous biological and medical importance[29]. Thousands of diseases are produced by genetic defects in channels, including many diseases of profound importance, like cystic fibrosis, epilepsy, atrial and ventricular fibrillation, and so on, as documented in many papers[6,12,26,29,32,59,67-69,122,123,133,152,162,187,202,207,208,243,250,258,305,317,330,333,337,341,342,373,379,385,391,408,422,442,454,457,488,491,507,511,514,515,526,532-534,542,549,572,585,588,589,606,612,614,626,643,646,661,662,676,677,683,703,704,707,715,727,732,742,773,791,810,811,835,841,842,848,851-853,907,919,923,942] among thousands of others. Many of these diseases are caused by problems in the construction of channels, or the insertion of channels in the wrong places in the wrong cells, or in the regulation and control of channels. This review is not focused on such biological problems, because we do not know enough yet to write physical models of the problematic biological systems.

This article is written to show the interactions of physical chemistry and molecular biology in channels, in theory, simulations, experiments, and mathematics, as well as in the text itself. Channels are defined, along with enough discussion to show how they are used in biology, without (I hope) overwhelming the reader with complexity. The selectivity properties of channels are discussed at length because in some cases these can be understood quite completely with simple ideas from the primitive (implicit solvent) model of electrolyte solutions of classical physical chemistry. Selectivity implies interactions. In the world of ideal point particles $K^+$ and $Na^+$ are identical! The need to analyze finite size particles that interact because of their size and electric field is a recurrent theme. The need to analyze flows is also a recurrent theme, although most of that analysis is yet to be done. Flow must be analyzed because ion channels, like most devices, work far from equilibrium.

Equilibrium is death to biology. A variational method *EnVarA* is introduced that allows automatic extension of equilibrium analysis to nonequilibrium. *EnVarA* was developed to deal with complex fluids, with flowing interacting subelements far more complex than hard sphere ions. I argue that electrolytes can be viewed more realistically as a complex fluid than as a simple fluid of classical theory. The variational method is then applied to a few cases of interest. The article tries to go full circle: describing ion channels, using classical physical chemistry to deal with an important biological property of channels, introducing the new variational approach to deal with flows through channels, and finally arguing that this variational approach provides a new perspective on ions in solution as well as channels. I propose that ions in solutions are complex fluids in which interactions dominate: 'everything interacts with everything else'.

**Physical Chemistry and Biological Problems**. Many scientists want to apply physical chemistry to biological problems. The question is, what problems? The challenge is how to do it. The answers to these questions are hard for human reasons, I believe.

Physical chemists and biologists usually have different aims. Physical chemists want to know everything.[76] Biologists want to understand so they can control. Biologists want to understand how machines, systems, and devices work well enough to make life better, in health and disease.

Biologists have little interest in how living systems work under 'non-physiological conditions'. Only anatomists study 'fixed' (dead) material and they do so because fixed material is easier to view than moving, living systems. Structure is important, but it is important mostly because it can move and do something.

Biologists are much like engineers. Chemical engineers are as much physical chemists as they are engineers. Physical chemistry is linked to important parts of biology much as it is linked to chemical engineering. Physiologists and physical chemists dealt with the same issues until molecular biology came along and focused physiological attention on proteins, rather than the ions that surround them.

Engineers have a particular approach to problems shared by biologists. Engineers study devices as they function in a particular case. Engineers want to study an amplifier as it amplifies. (I use electrical examples because of my limited knowledge of chemical engineering.) Engineers are not eager to study amplifiers when they cannot amplify, when they are 'dead'.

Little work is done on amplifiers at equilibrium, with power leads soldered together and held at ground (zero) potential. Engineers (like biologists) usually study systems in a limited set of conditions in which the systems actually work. Few systems actually work at equilibrium. Most systems require specific 'power supplies'. Most systems are tolerant to some changes in conditions, but fail to work at all outside a certain range.

Biological systems only function when gradients of ions are in a certain limited range. Gradients of chemical and electric potential are the power supplies of biological systems. Gradients of ionic solutions drive signaling in the nervous system, the control of muscle contraction, the secretion of hormones, enzymes, and urine. Biologists are interested in ions because they power so much biology. For them, this is a universe. For a physical chemist, it is not even a solar system in the universe of all ionic phenomena.

Physical chemists have a broader view than engineers and biologists. They are interested in everything that ionic solutions can do in any temperature or pressure, in solutions made of many types of ions. They study the general properties of ionic solutions. The special properties of ions important in biology— $Na^{+}$, $K^{+}$, $Ca^{2+}$ and $Cl^{-}$ in ~150 mM aqueous solutions around 300K—strike them as a particular, perhaps boring, special case, while biologists (and physicians) call that case life!

Biologists and engineers, however, do not find special cases boring. Both know that their machines only work in special cases. Engineers know that locomotion is not a general consequence of burning gasoline. Special structures and particular ingredients and conditions are needed to make that happen. Biologists know that animals live in only a narrow range of conditions.[405] Biologists and engineers are interested in the special conditions in which their systems function.

Machines are robust in some ways and delicate in other. Rather small changes of ingredients or structure will stop the machine, and may in fact 'gum it up' so it will not ever work. Think of kerosene in the gas tank of a car.

Electrochemical devices, like electrical devices in general, require particular power supplies. With the wrong voltages, they cannot work and can in fact be damaged irreversibly. Similarly, ions in channel proteins perform useful functions only under special conditions. They need certain concentrations and gradients of electrical and chemical potential to function. They need certain concentrations of control ions that regulate channels much as the accelerator of a car regulates the speed of the car. The wrong ions or wrong concentrations of ions can irreversibly denature proteins just as the wrong voltages applied to an amplifier will irreversibly 'denature' it.

Biological cells and molecules only function under restricted conditions. Animals are the same. We are all too familiar with the fragility of life. Outside a narrow range of temperature, we are uncomfortable. Outside a slightly wider range of temperature we die. The properties of biological cells and proteins have the same sensitivity. Indeed, one of the roles of a biological organism is to maintain the special chemical and physical conditions that its cells, tissues, and proteins need to function. Homeostasis and 'fitness of the environment' are main themes in the classical physiology of the 19$^{th}$ and 20$^{th}$ century. The organism buffers the cell and its molecules from the outside world much as our houses and clothes buffer humans from what other animals experience.

Biologists and engineers know that their machines require power and specialized conditions to perform their function. Biologists and engineers know the importance of structure. They know that the essence of their machines are the special structures that use power to convert inputs into outputs of general use. Biologists and engineers are trained to study systems that are alive and performing their 'design' function. Biologists and engineers only study systems in the range that they actually function. It seems obvious to them that an amplifier should not be studied without its power supply, or with its power inputs soldered together. It seems obvious that function of living systems cannot be reproduced in death.

It seems obvious to biologists and engineers that no general analysis is possible. It seems obvious to them that the desire of physical chemists for a general approach cannot be satisfied. Biologists and engineers think a truly general approach is impossible.

In fact, a powerful general approach is indeed possible, but only in a special sense. A general approach is possible of the machine as it functions. Analysis of functioning machines will almost always reveal special structures that use special conditions to perform an important function. The combinations of special structures and conditions form motifs, design themes in engineering and adaptations in evolutionary biology used again and again because they are successful.

Machines are designed to use complex structures and specific power supplies to execute simple functions. An amplifier can often be described by a single number, its gain. Many transistors arranged in a complex circuit are needed to produce the simple behavior of an amplifier. Each transistor is described by complex field equations—coupled partial differential equations in fact not so different from the equations of ions in solutions and channels[267,270,284,286]. The transistors are connected in a circuit of some complexity. The physical layout of the transistors is a structure much more complex than the circuit diagram. All of that complexity is needed to produce a simple property, in this case the gain of an amplifier.

The general behavior of the machine—for example, the gain of the amplifier—can be simple and powerfully described by simple equations, often much simpler than the general equations needed to describe the underlying physics or structure of the machine. But that simplification is possible only because of the complexity of structures involved, and the restricted set of conditions under which the machine operates. A very complicated circuit is needed to make a linear amplifier, and that circuit only works when given the right power supply. But the resulting linear function can be described (for most purposes) by a single number, the gain.

**Complexity in structure and physics produces simplicity in biological function**. Molecular biology illustrates these facts very well. Molecular biology shows how complex structures and physics are used to make simple function. The revolution of molecular biology is so important because it has revealed some universals of life. These universals are complex structures that use physics to write the code of life's molecules. All life is inherited but almost the only thing inherited is a blueprint to make proteins. The only things the blueprint (DNA) describes are proteins. Attention is thus focused in biology on a very narrow world of proteins and nucleic acids. Proteins and nucleic acids exist in a narrow range of temperatures and pressures. These are the only conditions compatible with life. The study of life is not general. It is the study of these systems of complex structure at those temperatures and pressures.

The study of life is also the study of ionic solutions because life occurs in a mixture of ions and water. Attention must be focused on ions in solution because proteins and DNA require ions in water. Proteins and DNA 'come with' ions, in the sense that sodium comes with chloride in table salt or sea water. Ions are always present to balance the electric charge of the rest of the DNA or protein molecule. Indeed, the requirement for ions is usually more specific than that. Without quite particular ions, most proteins and nucleic acids cannot maintain their structure, cannot function. Indeed, without ions, many proteins simply denature, damaged as irreversibly as when an amplifier input is connected to a voltage beyond its design limit. Neither protein nor amplifier can survive too strong an electric field.

Ions set the necessary milieu for proteins and nucleic acids that shields protein permanent charge, so electric fields and potentials are small. Without the shielding of charge, electric fields in proteins would be very large, likely to damage the very structure that allows them to function. Ion channels illustrate this generality: without proper concentrations and gradients of ions, most channel proteins simply do not function.

Ions have another role. Ions act as controllers for most intracellular proteins. Many intracellular proteins are controlled by the concentration of ions, much as a gas pedal controls a car, or a dimmer (or rheostat) controls a light. Quite often proteins are controlled by the concentration of $Ca^{2+}$. The controlling concentrations are very small making the physical chemistry of trace solutions of particular biological interest. $Ca^{2+}$ concentrations of $10^{-7}$ M are typical; many controllers of proteins work in lower concentrations. Some 'hormones' function at concentrations of $10^{-11}$ M.

The role of ions in life is too much to review in general. But the specific role of ions in channel proteins is nearly manageable and so I write about them.

**Ion channels.** Ion channels are proteins with a hole down their middle that control the flow of ions through otherwise impermeable membranes (Fig. 1 & 2). Ion channels are the nano (nearly pico) valves of life with as general a role as transistors in integrated circuits. Ion channels form a useful path into biology for physical chemists, one that carries familiar ions over potential barriers. Ion channels can help carry physical chemists over the social and intellectual barriers formed by exhaustive descriptive biology. The study of ion channels involves physical chemists in a biological problem of great generality and importance, one that nonetheless can be attacked (in large measure) by physics (and chemistry) as usual, without invoking vital new principles of organization or complexity. We present a brief description of ion channels now so the gate to physical understanding is open and not organically blocked by unknown structures.

I focus attention on that part of ion channel function that involves ions most clearly and directly. Ion channels must conduct different ions if they are to function biologically because different ions carry different signals. Thus, the specificity of channels to different types of ions is of great biological importance. It is that specificity we will concentrate on in this paper. It is also that specificity that makes physical chemistry essential to biology. Fortunately, the physical chemistry of specificity can be dealt with by physics as usual, at least in the cases studied here. Specificity of ion channels is then a subject of mutual interest to biologists and physical chemists in which their interests and analyses overlap. The biologist needs the knowledge of the physical chemist. The physical chemist benefits by the focus on a specific, well defined, reasonably robust system that uses a definite set of forces and structures to maintain a definite set of properties, namely its input output relation. The physical chemist finds it much easier to study ionic solutions around 200 mM ionic strength made of alkali metal ions and alkali earth ions than to study all solutions of all elements at all concentrations. The physical chemist (or at least the chemical engineer) can find the problem of determining the transfer function of an electrochemical cell more approachable than the problem of determining all properties of the solutions inside the cell. Biology provides 'the existence theorem' that guarantees that a reduced model can describe an ion channel quite well.

The selectivity of ion channels is an easy subject to study in the physical tradition. The role of structural change (in the channel protein) is minimal and stereotyped, so elaborate descriptions of different types of conformational change are not needed. The physics of selectivity is also quite stereotyped and so only a few mechanisms are likely to be involved, although the balance between the energies of different ions (and processes) is likely to be quite different in different channels.

Physical scientists often dislike descriptive detail. The selectivity of ion channels requires less descriptive detail than many other properties of channels (or proteins), making ion channels a (relatively) easy biological object to study for physical scientists.

Biologists on the other hand relish descriptive detail and many find selectivity boring for that reason. Selectivity requires description in numbers, not names and biologists are often glad to leave that to their physically oriented colleagues. The common themes of all ion channels seem too common—and boring—to some biologists. I hope they do not seem too challenging for physical chemists.

**Channels open and close.** Not all properties of ion channels are as simple as selectivity. Channels open and close and open again and that process involves conformational changes of the protein and of the electrical and chemical potential fields within the channel. When channels are open, ions flow through a single structure that does not change on the biological time scale (slower than say 100 μsec). Ions flow through a 'hole in the wall' following the laws of electrodiffusion, at a fixed temperature, with a simple contribution from hydrodynamics. Ion channels also share themes with each other and with enzymes in general[285]. A general physical analysis of some functions of channels is possible. Ion channels are much less complex objects to analyze physically than amplifiers, particularly once the channel is open.

Ion channels are diverse biologically at the same time they are unified physically. The biological functions of channels require extensive description because they are so diverse. The underlying physical plan of channel's selectivity (probably) does not require so much description (at least as far as we can tell so far). The open channel so far falls into two types, one in which side chains of the protein mix with ions and water in the pore of the protein (calcium and sodium channels) and one in which the side chains face away from the pore and the protein forms a (relatively) strong 'smooth' surface (as far as we can tell) (potassium channels).

**Ion channels** are nanovalves that control most biological systems by controlling the flow of ions, water, and electric current across otherwise impermeable biological membranes. The membranes are lipid bilayers that define cells and structures within cells, like the cell nucleus or mitochondria as well as the cell itself. Evolution—like an engineer—isolates subsystems from their neighbors by insulators. Engineers use silicon dioxide. Evolution uses lipid bilayers. Subsystems are connected through 'channels' that cross the insulation.

Ion channels exist so ions can cross otherwise impermeable membranes in disjoint physically separate pathways ('parallel resistors' if one likes engineering images). Ion channels have internal structure that allows ions to move and external structure that allows them to exist in lipid membranes.

Physical chemists need to be reminded of the importance of structures. Complex structures are found in nearly all devices and biological systems. Complex structures must be understood as much as physics must be understood if we are to control and build devices and biological systems. The complex structures need description. No engineer would consider studying a machine without its parts list. A great deal of biology has been devoted to constructing that parts list, for centuries at the macroscopic level, and now at the molecular level. It takes a great deal of work to separate and identify the parts of a complex system, if you do not have a blueprint. Molecular biology learned part of the blueprint when it learned of the existence of genes and their biochemical basis, DNA and RNA. The blueprint provides (most of) the information needed to make proteins. It determines the sequence of amino acids joined into a linear chain to make the protein. That is the only function of the blueprint. But the linear chain is like a chain of beads that folds to make the devices and machines of life. We do not know the rules that fold the chain into devices and indeed sometimes additional information is needed in the form of structural templates. That is why so much effort has been spent on the protein folding problem. Much progress has been made but the truth remains that we must catalog the structures of proteins as we catalog their sequence of amino acids if we are to have a reasonably complete parts list. Our parts list must include three dimensional structures as well as sequences. Biology thus necessarily involves a great deal of description. That description is needed before the role of structure can be understood in biological function.

The structural complexity is achieved in many cases by combining subsystems used in other systems, and so already found in the parts catalog. Keeping subsystems separate and (reasonably) independent allows construction of complex systems from common building blocks. Linking (more or less) independent systems is an obvious strategy for building complexity from simpler building blocks. Valves are the physical building blocks that control flow from system to system. Ion channels are the nano (nearly pico) valves of life. Vacuum tubes (‘valves’ to 20$^{th}$ century speakers of UK English), and Field Effect Transistors (FETs) are a great deal easier to use than bipolar transistors because they are much more independent. The input characteristics of a FET are far more independent of output characteristics (of itself or its connecting devices) than bipolar transistors and this makes design much easier and more robust.

A large fraction of the proteins in an organism are channel proteins that use ions to carry a current of particles, just as a large fraction of the devices in a computer are field effect transistors (FETs) that use their own channels as nanovalves to carry the current of (quasi)-particles, holes and ‘electrons’. The biologist and computer designers need to know about channels and transistors because they are such a large fraction of their systems. The physical chemists and physicists need to know about channels and transistors because they are so interesting, providing mechanisms by which tiny structures and powers can control large movements and flows. Channels and transistors are valves. And valves use interesting physics.

The analogy between channels and FETs is useful and productive[270,283,284,286,820] because they both follow similar mathematics even though the charged objects are quite different. The charged objects of channels are ions, atoms with permanent charge that do not change during the great majority of biological processes.

The electrons of semiconductors are more ephemeral. They are mathematical constructs with quite distinct properties from the electrons in a vacuum tube. Surprisingly few scientists are aware that the negative quasi-particles of silicon/germanium have little in common with the isolated electrons of physics textbooks and cathode ray tubes. I wish the ‘electrons’ of silicon/germanium were named (something like) ‘semi-electrons’ (for semi[*conductor*]-electrons) so they are not confused with isolated electrons that flow in a vacuum. A mathematical construct and a physical electron are really quite different things. It is a naughty convenience to use the same word ‘electron’ for both, at least in my view.[270]

**Sigworth’s Equation: single channel currents**. The best way to become acquainted with ion channels is to think of them one molecule at a time, corresponding to the way they can be studied experimentally when reconstituted into lipid bilayers or measured in patch clamp set ups. Of course, ion channels are not used one at a time in biology, where large numbers of several types are involved in almost any function. The question of how to account for the properties of large numbers of channels is not a major subject of this review, and in fact is too general for anything less than a book. I concentrate on ion channels one at a time, dealing with macroscopic behavior of ensembles of channels only a little. I mention enough of the properties of ensembles so the reader will know the main additional features that occur when channels work together. It would not be good for the reader to be entirely divorced from the biological reality of cells and tissues.[2,3,47,309,319,486,487,489,494,530,547,580,701,762,904,905]

Consider an ensemble of $N$ independent channels of the same type. Ion channels control current flow in a simple way, summarized by Sigworth's equation[819]. The equation describes the current flow $I$ through $N$ single channels measured in a voltage clamp experiment in which all single channels have the same (open channel) conductance, voltage and concentration across them and all single channels open with probability $p$ to an identical mean current level $<i>$. (Historical note: I choose the name for this equation for personal and historical reasons. Most channologists learned of the equation from Sigworth[819]. I certainly did, in a seminar he gave us early in his career, while still a graduate student, if I remember correctly. I do not know if there was a significant previous history.)

$$I = N < i > p = \underbrace{N(\text{seconds})}_{\substack{\textbf{\textit{N}}\text{umber}\\ \text{of Channels Available}}} \quad \cdot \quad \underbrace{< i(\mu\text{sec}) >}_{\substack{\text{Amplitude}\\ \text{of Single Channel}}} \quad \cdot \quad \underbrace{p(\text{msec})}_{\substack{\textbf{\textit{p}}\text{robability}\\ \text{of opening}\\ \text{of one channel}}} \qquad (1)$$

Much of the lore of studying channels ('channology' if you will) is the application of Sigworth's equation to the particular case where there is just one channel protein being recorded at a time. I try to make explicit what experimentalists learn implicitly 'in the lab'. This knowledge is needed by physical scientists, if their work is to deal with real data. Theory and simulations are most useful when they produce results indistinguishable in form and format from experiments.

Sigworth's equation is used in the lab for the case where only one channel protein of one type can open. It is used when there is negligible probability of more than one channel being open. Experiments with single channels are often designed to show openings of only one type of channel to only one level. Indeed, one of the most important features of the single channel revolution of Neher and Sakmann[384,655,656,774] is often unremarked. If an experiment records a single channel molecule, it records a single type of molecule. One molecule can only be of one type. Currents measured from that one molecule do not have to be 'deconvolved' into the currents of a multiplicity of channels.

The multiplicity of channels that plagued earlier recordings from membranes was replaced with singular clarity with a single suck (i.e., the suction needed to create the gigaseal that is the electrical key[31,384,555-557,658,710,711,713,769,770,774,816,847] to single channel recording). An old saying used to be "never do clean experiments on dirty enzymes", i.e., be sure to purify an enzyme before you study it. Until single channel measurements were done, almost all experiments on channels were dirty, i.e., they almost all involved more than one channel type, and often unknown channel types as well.

Single channel experiments are designed to study one channel protein of one type. That is appropriate when the goal is to understand that channel protein. That is the goal of much of molecular biophysics.

But channel proteins function in physiological systems made of many types of proteins. In those systems many types of channels are present that interact to make the system function. Physiology as a profession tried to link these scales for several centuries, culminating in the successful analysis of the neuron from molecule to membrane to cell, even to cells meters long.

The physiological tradition is not taught to most molecular biologists, however, in my experience. Most molecular biologists need to be taught what steps are needed to connect channels and cell function. They need to be taught the enormous importance of intermediate scale models. No engineer would think of making an all atom model of an amplifier. Indeed, most engineers would not make a model of the physical layout of an amplifier. Many would avoid the complete circuit of the amplifier. They would rather use the simplest equivalent circuit that illustrated the properties they needed to know. Classical physiologists used the same approach to the nerve cell. Thus, the linkage of atoms (e.g., ions), molecules (e.g., channel proteins), cell structures (e.g., membranes), cells (e.g., nerve fibers) into a single system is very well understood in classical neurobiology. Neurobiology has learned to use a multiscale approach just as an engineer would, with different resolution descriptions of ions, channels, membranes, axons, and nerve cells. A traditional[920] and a modern textbook[334] illustrate this material very well, much better than I can here.

More complex cases of more than one channel, or channel type, or of many superposed channel openings—repeated openings of the same channel molecule, or overlapping openings of several different channel molecules—can be handled in theory by multiple convolutions. This is straightforward when individual openings are independent and uncorrelated.[177,774] Openings of different channel proteins are likely to be uncorrelated if the proteins are widely separated. On the other hand, multiple openings of the same channel molecule are likely to be correlated. Nearby channel proteins that interact ‘directly’ will have correlated openings, by definition. In these cases, theoretical methods must include every way a channel can open, for example, by using enough nested convolutions to describe all interactions that can open a channel.

Eq. (1) has an explicit separation of scales that is important in its use. If the time scales are not widely separated, the distinctions I am about to make do not hold, and a more elaborate analysis is needed, and it is likely to be different for each type of channel and each situation.

The time scale (seconds) indicated in parentheses in eq. (1) means that the number of channels available for current is typically constant on a time scale shorter than seconds. Regulatory and metabolic processes can change the number on the time scale of minutes, and (mechanistically obscure) processes labeled ‘slow inactivation’ can change the number of available channels on the time scale of (say) seconds. The longer time scales are thought to reflect the construction of the channel and the regulation of its biochemical state by covalent bond changes like phosphorylation.[909] Those time scales are supposed to be captured by the $N$ variable in eq. (1) although that description is obviously an oversimplification. The inactivation, regulatory, and metabolic processes might change. Description by a constant $N$ would be misleading in that case. The properties of $N$ need to be checked by experiment in each case.

In eq. (1) each channel protein conducts a current of $i(\mu\text{sec})$ amps, where the $(\mu\text{sec})$ indicates that this current reaches steady state in less than a microsecond[634,856]. The current through a single channel in eq. (1) follows a rectangular time course. The current (in the mean, averaging out noise, by averaging over an ensemble of measurements) is zero for a stochastic time after a voltage change (or other perturbation) is applied. The current then suddenly switches to a new level, the open channel current. The new level is maintained without drift or internal correlations (once noise is averaged out[381,556-558,713,714,817,818]) for a stochastic duration, and then the channel suddenly closes, usually to the same zero level it opened from. The channel may reopen. If it does, the channel will suddenly open to the same level, within measurement error. Fig. 3 and 4 show some classical sudden openings of single channels and illustrate the difference between gating and permeation.

**Opening time course of channels.** Everyone wants to know how fast a channel opens. The time course of a single channel opening has been studied by Miodowinik-Aisenberg and Nonner[634] and independently by Tang et al[437]. Neither group chose to publish a full length paper (personal communication) because their results were so complex. On a time scale in which a single channel opening could be resolved (say 100 nsec), an enormous range of behaviors was seen, that could not be easily summarized by either group. Some openings were sudden, some were very gradual, some were gradual and then sudden. It was not even possible to define a clear 10-90% risetime. What is possible is to say that after a few microseconds, this complex opening behavior seems to have no further effect on the current through an open single channel. We do not know what questions to ask of the opening of the channel. We do know what to ask of the open channel. So we study the open channel from now on, and ignore the opening process. It is important for scientists not to ask questions they know they cannot answer, particularly if they do not even know what question to ask.

What is relevant is the current through the open single channel. The open channel current is independent of the duration of opening or the latency, within experimental error. Channels open and close stochastically, all to the same level, if no perturbation is applied. The duration of opening is stochastic, usually exponentially distributed, often distributed as the sum of a few exponentials.[597,621,622] The fraction of time the channel is open is a reflection of the gating process that opens the channel, widely assumed to be a change in conformation of the protein, although the relevant conformation in general is that of the (electrochemical) potential profile along the channel, and not the anatomical shape of the channel protein.[271,283,284,286,288]

The current through the open channel is a reflection of the permeation and selectivity processes in channels and corresponds more or less to the ‘instantaneous current voltage relation’ postulated by Hodgkin and Huxley[424,433-437]. Permeation through channels occurs by electrodiffusion in a structure of fixed dimensions (on the 10 μsec and slower time scale, see Fig. 3 & 4). Flexibility seen in simulations is on a much faster time scale[16-18,756].

Selectivity is produced by physical forces that depend on the ion charge and size and chemical interactions with parts of the channel protein that form what is called ‘the selectivity filter’[102,103,106,273,274,352,357,664,668,669,672,675,752,896,936]. Those chemical interactions can arise from physical interactions of ions with side chains of the protein, physical interactions with the permanent electrical charges in the protein, physical interactions with the induced polarization (i.e., dielectric) charge on the protein, chemical interactions involving changes in the distribution of charge within the molecular and atomic orbitals of the protein. (If the changes in distribution of charge in the orbitals are proportional to the local electric field, the change in charge is identified as the induced polarization charge of a dielectric. If they are not proportional, I call them ‘chemical’.) In any case, those interactions do not vary significantly on the time scale of 10 μsec or longer: if they did, the single channel current level would not be ‘constant’ independent of duration of opening, or time after an opening.

It is important to realize the sensitivity of this argument. Changes in (average) open channel current of 5% are easily detected, since the (averaged) signal to noise ratio of single channel measurements is often >40. Charged groups in the protein (whether partial charges in a carbonyl or full charges in a carboxylate) are within an angstrom of permeating ions. The change in the electric field from Coulomb’s law is large if these groups move. The electric force has an exponential effect on current flow. Thus tiny changes of structure (< 0.5 Å) would have an easily measured effect on open channel current, if the structural change lasts long enough, longer than (say) 10 μsec.

Sigworth's equation is applied to cases in which the electrical potential across all the channels is the same, as in voltage clamp experiments. Sigworth's equation is not applied to nerve or muscle cells when they are carrying action potential signals because the voltage is then changing. The potential in voltage clamp experiments (but not in real cells) is usually controlled in steps that change from one value to another at one time or another (in the tradition of Hodgkin, Huxley, and Katz[424,434]). The chemical potential (e.g., the concentration of ions) on each side of the channel is assumed not to change significantly with current flow or time in the classical tradition, although significant exceptions to this rule are common and indeed inevitable in important cases like calcium channels, where the concentration of one of the permeating ions is very low on one side of the channel or the other. Experiments and theory should always be wary of the assumption of 'constant' concentration (independent of time and current amplitude) because it is an approximation that cannot be generally true.[833]

Classical voltage clamp analysis makes the tacit assumption that all important properties of channels can be investigated by steps in potential. This is obviously not true in general, since a channel property that depended on the time derivative of potential would not be well studied by a voltage clamp command potential that was a step function in time. A step function has time derivative that is zero, or $\pm\infty$, and so it does not explore the range of dependence on the time derivative of potential, at all. This fact was taught me by Julian Cole[37,501] as he learned and questioned traditional electrophysiology. The inadequacies of step functions was in fact well known to the founders of electrophysiology, as I found out when I asked Prof. Hodgkin "Why do you think voltage steps allow a complete analysis?" (around 1972 in a personal communication). and was the reason for the experiments reported Hodgkin's answer to my question was that he did not, and never had thought that responses to voltage steps was enough to analyze any system. He had understood the need to analyze the response to a continuous voltage waveform, not just discontinuous steps. He actually had dealt with timing varying voltage early on (in Fig. 10 of reference[437]). He chided me that I had not read his papers as carefully as I should have if I had missed the importance of that experiment. Indeed, Hodgkin and Huxley both made clear to me in repeated personal communications that this issue was a main motivation for their calculations—by hand! without even a calculator that could multiply—of the action potential itself. Huxley and Hodgkin showed they could reconstruct a full waveform[436] of an action potential using data taken only from step functions of voltage, using an intermediate scale theory that described the flow of current in transmission lines, submarine cables, and long cylindrical cells, called cable theory.[293,334,476,920]

We return from history to the modern study of single channels. In the idealized case, the individual channel behaves very much like a pore with a gate. The pore conducts current carried by ions, with a diameter about half that of the pore. The flow is controlled by a gating process that turns the current through the pore on and off but does not otherwise affect that current. At least this is the classical view, and a most useful initial approximation.

**Open Probability.** The probability that a single channel opens varies on a msec time scale and so is written as $p(\text{msec})$ in our version of Sigworth's equation (1). This $p(\text{msec})$ is the probability of any one of the $N$ channels opening. It is not the probability of a particular channel opening. If one wishes to write the probability of any channel opening as a functional of the properties of one channel opening, one must deal with the variable latency and duration of individual (i.e., one channel) openings. One must perform multiple nested convolutions, being sure to include the possibility of multiple openings, of the same or of different channels, and durations that depend on the number of openings, as well as the duration of previous openings and so on. The convolutions must sum all the possible openings one would see if one made a very large number of measurements of the response of a single channel protein to repeated steps of voltage in a voltage/patch clamp experiment. One must be sure to include all the $N$ channels that can conduct current. Usually those channels are assumed to open independently (if the voltage is maintained constant), but such is not expected to be the case if the concentration change produced by channel openings correlates their properties, or if the channels interact through their internal charge movements (i.e., through electric fields within the protein and membrane), or by mechanical interaction, or by mysterious allosteric forces of unspecified physical origin.

**Relation to nerve function.** We turn now to the properties of ensembles of currents in one important case, the action potential of nerve fibers, the natural activity of nerve axons, which are the ‘wires’ that conduct signals long distances in the nervous system. This detour from physical chemistry seems necessary to lend biological and medical reality to our discussion.

The electrical currents flowing in voltage clamp experiments are related to the natural activity of nerve fibers in a complex way that is in danger of being forgotten when channels are studied in the voltage clamp tradition on the atomic scale of molecular and cell biology, and of molecular dynamics simulations. Thus, I include the following description of textbook material[334,920] meant mostly for physical scientists new to channel research.

Channel proteins are typically some 10 nm in diameter, while the pores through which ions move are typically 0.6 to 1 nm in diameter. Channel proteins are typically (but not always: important exceptions include the node of Ranvier and parts of the endoplasmic and sarcoplasmic reticulum) separated by substantial amounts of lipid membrane. Channel proteins are thought to be built to operate independently so it is natural to assume no interactions between channel proteins if they are all held subject to the same electrical and chemical potential gradients in a voltage clamp experiment. Many channel proteins are modulated by nearby proteins however, that themselves are not channels, for example, the neurologically important GABA receptor complex.[264]

Channels are **not** independent when functioning in a natural unconstrained way. Most channels are independent in a voltage clamp experimental apparatus but not in nature in their natural setting in cell and organelle membranes. When experiments are done without voltage clamp, the responses to stimuli and drugs are complex, obviously not just the sums of components, in many cases. When experiments are done with a voltage (and ‘space’) clamp system, responses to stimuli are much simpler, often just the sums of components.[433,434]

Channels in nerve cells interact with each other strongly through the electric field generated in part by the current through the channel. Channels control the electric field by regulating the current flow through the membrane. The current flow through the membrane in turn changes the potential. In the classical case of the voltage clamped squid axon, **channels interact only through the electric field** created by their current because the currents flow only for a few milliseconds. Even in squid axons, however, current flow of potassium over somewhat longer time scales than initially studied[424,437] produces significant concentration changes that produce interactions of channels.[321] The accumulation of potassium after a series of action potentials in fact may be of crucial importance in a number of biological functions (memory) and medical disorders (epilepsy).[635,836]

Confusion can be easy here if the experimental situation is not precisely described. The movement of current through a single channel rarely if ever is large enough to change the concentration of ions significantly, unless the concentration on one side of the channel is very very small (as with $Ca^{2+}$ ions[365]). But the movement of current through a macroscopic number of the same channel may change the concentration quite significantly. (One must never forget that the concentration of messenger molecules like $Ca^{2+}$ is in fact very very small, e.g., $10^{-7}$M. Thus, calculations of $Ca^{2+}$ flows must not assume that the intracellular concentration of $Ca^{2+}$ is constant, or independent of current flow or experimental conditions. The same is the case for calculations of other messenger molecules.)

Biophysicists, like traditional electrochemists, go to some length to separate the properties of channels (that they are interested in) from properties of the surrounding solutions that seem (traditionally) to be uninteresting. Thus, they work hard to remove the effects of gradients of electrical potential outside the channel in the baths (by 'series resistance compensation', starting with[437] and then developing into a substantial literature[776,808]) and to remove the effects of gradients of chemical potential (i.e., concentration) in the baths, by judicious restriction of time scales, choice of ionic solutions, and even selection of experimental records. 'Concentration polarization' — as such effects were once quaintly called — are particularly difficult to deal with because they necessarily involve water flow and convection (at least in general and probably nearly always). The rotating disk electrode method of electrochemistry[10,178,553,726] was developed to control such effects by establishing a known and controllable convection field that dominates the spatial distribution of concentration. Biologists seem not yet to have used rotating disks or other methods to establish convection gradients and control concentrations outside channels, membranes, or cells. Such methods might be particularly helpful in studying water and solution flow, which are plagued by controversy, perhaps arising from their sensitivity to bath conditions. It is interesting to note that controversy decreased dramatically in the fields of polarography and electrodics (i.e., electrochemistry involving chemical reactions at electrodes) once the rotating disk electrode was introduced.

Time dependent concentrations involving water flow are also particularly hard to deal with in theories or simulations. Simulations on atomic scales ignore such flows typically because they occur on time scales so remote from the time scales of the simulations. (Water flows take msec to second; atomic motions take femtoseconds.) Theories have difficulty knowing how to mix different types of flow (i.e., macroscopic pressure volume flows, diffusion, and electrical migration) and how to deal with interactions of different solutes and ions.

Recently, variational analysis has been applied to this problem[280] and progress seems possible. Energies and dissipations of different components are combined in the energy variational approach and Euler Lagrange equations are then derived, as we show later in this paper. These partial differential equations are the unique consequence of the contributions of individual components. The form and parameters of the partial differential equations are determined by algebra without additional physical content or assumptions. The partial differential equations of mixtures automatically combine physical properties of individual (unmixed) components without arbitrary parameters. It will be interesting to see how far this approach is able to simplify the formidably complex descriptions[601,648,802] of water flow in complex ionic solutions with many components.

**Action Potential is a Cooperative Phenomenon**. One of the most important interactions of channels is the cooperative behavior that produces the main signal of the nervous system, the action potential. The interaction of sodium channel opening, channel current, and potassium channel opening at one location with the same openings at another location produces the action potential that carries information long distances in nerve fibers[425,429,430,436]. Hodgkin proved[428,429] (when he was an undergraduate) that the cooperative behavior that produces propagation of the action potential is entirely electrical and does not involve any ‘propagating wave’ of chemical activity in the nerve membrane itself. Hodgkin’s papers are beautiful examples, easy to read today, of what can be done with minimal apparatus and maximal thought.

Hodgkin once told me that he had quite purposely written the papers to show they could have been performed with apparatus available around 1900. Modern readers of Hodgkin simply need to know that ‘electrotonus’ is an old way of stating, more or less, the change in membrane potential as a function of time and distance. (Readers of the “There will always be an England” features—found in the New Yorker magazine for many decades—need to know that Hodgkin did in fact use an oscilloscope and not an electromechanical apparatus to record potential: being aggressively archaic does not mean being foolish.)

Later work showed that the action potential itself, as well as its propagation, is a passive process, that does not involve the chemical interaction of different parts of the membrane (i.e., that does not involve chemical interaction of different channel proteins, in modern language). Channels cooperate to make an action potential ***only*** through the electric field that the channel currents create (in the classical case[436]). In the less classical case[322], concentration changes produced by these currents also produce interactions. In no case are covalent bond changes or ATP hydrolysis involved in the action potential.

Allosteric interactions between different channel proteins are not involved in the main signaling phenomena of the nervous system, although it is certainly true that the effect of membrane potential on the opening of a single channel might be called allosteric. Binding proteins near ion channels can importantly modify the properties of ion channels[264] but they are not involved in the voltage or agonist responses of the classical sodium and acetylcholine channels of neurobiology.

**Allosteric mechanisms**. The physics that produces allosteric interactions is not widely discussed, perhaps because the inventors of the phrase ‘allosteric’ thought physical mechanisms uninteresting or unapproachable on the atomic scale.

Indeed, some scientists have avoided the atomic scale altogether. Some suggested that an important allosteric interaction is produced by rigid rods[131] that reach across an intracellular space to join two membranes. Such models seem to me more appropriate for the macroscopic deterministic scale of laboratories than the atomic Brownian scale of atoms.

When considering atomic or molecular scales, one must always remember that friction dominates all motion[64,708] because every atom reverses direction an enormous number of times in the shortest biological time scale (say 10 μsec) as it moves more or less at the speed of sound[75] in a system with almost no empty space. Indeed, friction limits all motion[64,708] to the overdamped case[294] in condensed phases like ionic solutions. (It is important to remember that ice floats on water, even salt water, thus showing that the density of liquid ionic solutions is less than that of the relevant solid. Thus, the empty space in liquid ionic solutions is ***less*** than the empty space in solid solvent. Gas phase models of biochemical reactions are more or less worthless, in this view, because the density of ideal gases is nearly zero.)

My view[271,284,286,288] is that local (i.e., incompletely shielded) changes in electric field within the membrane, within a Debye length of proteins, are likely to be involved in these long range allosteric effects as well as conformation changes. These local electric fields are likely to be focused by regions of low dielectric coefficient such as found on the inside of the KcsA channel[240]. The focusing of electric fields by dielectrics can create an enormous range of machines in the hands of engineers.[313] It would be surprising if evolution did not do some of what engineers do so much.

**Conformation Changes**. A macroscopic description of the forces driving conformation changes seems necessary in both physical and biological models. The gap between atomic scales and macroscopic scales seems too large to deal with in the foreseeable future, as we discuss later in this paper.

Conformation changes and allosteric forces arise from the random motions of atoms computed on a $10^{-16}$ sec time scale in molecular dynamics but the resulting conformation changes of channels are so slow (from $10^{-5}$ to $10^{1}$ sec) that computing them directly is likely to be difficult. It will be even more difficult to do the computations, given the size of the system (involving $10^{9}$ memory locations so positions can be specified in three dimensions to 0.1% of a 1000Å computational box) and the ratio of time scales ($10^{11}$ to $10^{17}$) and number densities involved. (The number density of water is some $10^{8}\times$ larger than the number density of intracellular $Ca^{2+}$.) Doing the calculations in a reliable way with known numerical and mathematical error bounds requires calibration against known macroscopic properties of physical systems.[275,705]

The requirement for calibration cannot be avoided in biological applications by redefining the problem to an atomic scale. Biological function occurs on a macroscopic time and concentration scale (even though the function is controlled by atomic scale structures inside molecular scale proteins). Thus, the output of calculations must be on the macroscopic scale of concentration and time with structural inputs on the atomic scale. Calibrations of this sort will be difficult even when the calculations can be done. It is important to note that as of now, as far as I know, no calibrations of biological (i.e., mixed) solutions have been made successfully. Attempts to calibrate NaCl solutions are beginning [2,3,47,309,319,486,487,489,494,530,547,580,701,900,904,905,943] and are reasonably successful when dealing with properties of the neutral combination NaCl, but are not so successful when dealing with cations and anions separately. Cations go through channels and act in enzymes separately and so calculations involving them must calibrate ions separately. Indeed, sodium channels are called sodium channels and not sodium chloride channels precisely because sodium—not sodium chloride—goes through them. Calibrations of solutions with divalents have not been completed as far as I know. Mixtures of ions have not been attempted, even though mixtures are invariably present and important in biological solutions. Thus, atomic scale simulations of conformation changes in realistic ionic conditions with calibrated results are not on the immediate horizon.

In my view, uncalibrated analysis is dangerous, not really part of a proper scientific process. Uncalibrated analysis can go on and on, sucking resources from other kinds of science that might lead to reproducible results and answers to questions.

In my view, multiscale analysis is needed, in which parts of the problem are handled on atomic scale and parts on macroscopic scale. The coupling of scales is part of the model in any multiscale analysis. Multiscale analysis requires treatment of reduced models, and the interactions of reduced models. What is important is that the reduced models can be checked against experiments in a range of conditions relevant to their use. In my view, physical chemistry can be of enormous help in dealing with biological problems if it addresses multiscale problems explicitly and insists that structural biologists and simulators deal with the macroscopic reality of ionic solutions. Studying individual trajectories, whether imagined or calculated, does not deal with the macroscopic reality of trajectories and ionic interactions. Trajectories and ionic interactions are incredibly complex, involving a more than astronomical number of terms if all ions are coupled by the electric field.

The models of structural biologists and molecular dynamics must produce results comparable with experiments if they are to be compared to experiments. Graphs must have the same variables whether they come from experiment or theory. To replace graphs with verbal discussions, or individual trajectories, is to replace science with metaphor. However beautiful, metaphors are not helpful in actually building machines and devices. They are hardly likely to be more helpful in understanding and building biological machines and devices. There is no engineering without numbers; there should be little molecular biology or biophysics without numbers, for the same reasons.

One of the most important targets of multiscale analysis should be the conformation changes of proteins. These occur on time scales from femtoseconds to seconds, and length scales from atoms to whole proteins.

Conformation changes have a physical origin, partially electrical. Identifying conformation changes as arrows in models is only the first step in understanding them. Naming them as allosteric might in fact be a step away from understanding if newcomers or structural biologists felt that the adjective ‘allosteric’ had a specific physical meaning. The arrows of allosteric models might seem to be disguises for the law of mass action. Indeed, the arrows of allosteric models do become specific statements involving the law of mass action once the rate constants of that law are identified with physical processes. But that identification is not one of the main topics of discussion in the usual treatments of allostery. Physical chemistry is needed to replace the metaphor of allostery with the reality of numbers.

Later we will discuss the law of mass action and the physical origin of its ‘constants’. Briefly stated, we will show that the law of mass action is usually used in a misleading way. The law is most helpful if reactants are ideal. It is most helpful if reactants are uncharged solutes, and rate constants are identified as a property of unidirectional fluxes into absorbing boundaries. But reactants are often not ideal. Indeed, reactants are present in enormous concentrations in channels and close to active sites of enzymes where they are important. Reactants in those locations more resemble an ionic liquid[5,516,519,778,838] (e.g., melted NaCl) than a solution of infinitely dilute uncharged solutes. Indeed, ions in solution always come ‘in pairs’ (more precisely in groups of anions and cations that are exactly balanced in charge so the systems are neutral taken as a whole). Ions come in pairs because anions and cations interact so strongly. Anions and cations cannot be independent solutes and also form solutions that are exactly electrically neutral. Ionic solutions are electrically neutral because their ions interact. If the reactants in the law of mass action are not ideal, the rate constants must be variables, and rather strongly varying variables at that. In fact, the variables often change exponentially with potential.[308,316,386,485,524,928] The law of mass action with constant rate constants seems a poor initial guess at a transport law for ions in multi-molar concentrations near active sites or in channels.

The forces that drive conformation changes (and determine the rate constants of models of conformation change) are not known, but are likely to be electrical, and mechanical arising from excluded volume of ions and side chains.[103] The forces driving conformation changes are closely related to the mechanical forces in classical Donnan systems[399,669] that produce osmotic pressure.

**Cooperative behavior produced by current flow**. In this multiscale spirit, we change perspective (and scales) now from intermolecular forces (femtoseconds and 0.1 Å) back to the cooperative behavior of channel proteins in a normal nerve signal (0.1 to 10 msec and 100 Å to 1 cm). We present in more detail a specific biological system that extends across scales from atomic to macroscopic and is well understood. The issues that arise need to be considered in general. They are likely to arise whenever physical analysis is applied to biological systems.

The ionic current that flows through channels in an action potential flows down the length of the fiber a long distance (think mm in myelinated nerve fibers of vertebrates and cm in squid nerves) to a neighboring region. This longitudinal current (that happens to be conducted by different ions from those that cross the membrane to make the action potential at the original location) changes the potential in that neighboring region. That change in potential changes the probability of a channel being open quite dramatically, which in turn produces an action potential in this new region, and the process continues down the nerve fiber, creating a propagating action potential. The process is rather like that in an underwater cable with repeater stations some distance apart that restore a digital signal to its original specification.

**Computation of the Action Potential**. A complete mathematical theory of the propagating action potential has been available for a long time.[8,415,419,640] Measurements of channel currents in voltage clamped axons allow computation of the propagating action potentials[436] without the introduction of any new physics. Thus, a complete mathematical theory of a major biological function is possible and was essentially complete (although without molecular explanation) in 1952. Molecular explanation followed in the work of many channologists,[8,415,419,640] culminating in the single channel measurements[774] and Nobel Prizes to Sakmann and Neher with structural insight provided later by Mackinnon.[241,550,596] A complete mathematical theory of a major biological function starting from atoms and finishing with nerves of the length of a giraffe leg has been available for several decades. Sadly, we do not have room here (or time in our lives) to write a full description of this remarkable accomplishment of physiologists and biophysicists. Good accounts can be found in [29,334,415,425,427,432,920].

In thinking of this theory of channels and their function, it is important to remember the distinctions between experiments and natural behavior, so the single channel molecular biophysics of one protein[415,774] is not confused with the physiology of conductance and gating[425,431] of billions of channels. Experiments are designed—or anyway, should be designed—to remove as much of the complexity of nature as possible so they can reveal mechanism. Experiments like the voltage clamp were designed to simplify the complex process of the action potential and its propagation. In mathematical language, experiments should be designed to remove as much of the ill posedness of inverse problems as possible.

In particular, voltage clamp experiments are designed to remove correlations arising from macroscopic properties of current flow in nerve and muscle fibers. Voltage clamp experiments are designed to remove the (electrically produced) cooperativity that makes the action potential a propagating all-or-none phenomenon. Electrically produced correlations are removed by 'short circuiting' both the spatial and temporal dependence of the natural propagating action potential. The spatial dependence is removed ('short circuited by a space clamp', in lab jargon) by inserting a wire down a cylindrical nerve fiber. The wire keeps the potential at all locations (hopefully) the same at all times.[859-861,864] The spatial dependence can be removed by studying spherical cells that do not allow propagation.[290,302]

The temporal variation of electrical potential is prevented by electronic circuitry and the experimental setup, the voltage clamp apparatus, see[459] for history, and reference[774] for recent implementation in 'whole cell clamp'. A variation called the patch clamp is used if the system includes only one or two channels.[555-558,710-713,774,920]

**Changes in ion concentration**. The voltage clamp experimental apparatus is designed to isolate single types of channels or individual channel molecules so they can be studied independently, as we have just discussed. In nature, channels interact by passing current that changes the electrical potential across the membrane of the cell and the thus across neighboring channels. The current through the channels is carried by one type of ion depending on the selectivity of the channels. The current inside the cell is carried by a different type of ion depending on the types of ions inside the cell. The potential inside the cell changes the potential across the cell membrane and is usually nearly equal to the membrane potential. (See [37,290,302,695] for an experimentally important exception.) The experimental apparatus is designed to prevent this natural interaction so the underlying mechanisms can be isolated and studied. The apparatus does this by controlling electrical potential with an electronic feedback apparatus (and a system of four electrodes to link amplifiers with ion currents) that prevents the natural interactions between channels.

In nature, channels may also interact by changing the concentration of ions near channels (called accumulation or depletion, or 'concentration polarization' in the older literature). Such effects are particularly important in calcium channels, because the calcium concentrations inside cells (on the intracellular side of the channel protein) are small $[\mathrm{Ca}^{2+}]_{\mathrm{in}} \cong 1\times10^{-7}\,\mathrm{M}$ compared to the intracellular concentrations of potassium $[\mathrm{K}^{+}]_{\mathrm{in}} \cong 3\times10^{-1}\,\mathrm{M}$ or sodium $[\mathrm{Na}^{+}]_{\mathrm{in}} \cong 3\times10^{-2}\,\mathrm{M}$. The amount of charge actually carried through a channel is often set functionally by the capacitance of the nearby cell membrane (typically $0.8\,\mu\mathrm{F/cm}^2$), producing a capacitance of some 63 pF for the $7.85\times10^{-5}\,\mathrm{cm}^2$ of membrane in spherical or cylindrical cell of diameter $5\times10^{-3}\,\mathrm{cm}$ with typical electrical properties.

An important function of current through the channel is to change the voltage by a given amount. The concentration of ions near the channel must also change, when current flows through the channel. Electrical forces are much stronger than chemical forces (see the first paragraphs of Feynman[312]) so, in some vague sense, we expect the charge carried by current to be more important than the mass, i.e., change in concentration, produced by the ions that carry the current.

The importance of charge vs. concentration depends on the concentration of ions present before the charge moves. If the electrical charge and current are carried by sodium or potassium ions through the membrane, the concentration of ions is changed by a much smaller fraction than if the current is carried by calcium, because there is so much less calcium present to begin with inside cells. Calcium concentration inside cells can be easily changed by calcium current through channels while potassium and sodium concentration cannot because there is so little calcium inside cells (some $10^{-7}$M $Ca^{2+}$ vs $10^{-2}$M $Na^{+}$ or $K^{+}$). In chemical and biological language, calcium is hardly buffered by the 'background' concentration inside cells but sodium and potassium ions are buffered, because the concentration of sodium and potassium inside cells is so much higher than calcium.[320,321,833] (The reader should be warned not to think of this buffering process as simply as this. Elaborate subcellular machinery is present to control these concentrations and that machinery involves transporters that use chemical energy to control concentrations.) A given amount of current (i.e., charge) passing through a channel changes the concentration of any ion (with the same valence) carrying the charge by the same amount. The relative importance of this change in concentration depends on the background concentration (before the current flow). If the concentration is high, like $K^{+}$ inside a cell, the relative effect of the inward charge movement is small. If the concentration is tiny, like $Ca^{2+}$ inside a cell, the relative effect of the inward charge movement is large.

Changes in calcium concentration inside cells are often used by biological systems as signals ('messengers') of channel activity for this reason. In a similar, but less dramatic way, changes in potassium concentration outside cells can be used by biological systems as signals of channel activity. The background resting potassium concentration outside cells is typically $[K^{+}]_{\text{out}} \cong 2\times10^{-3}\text{M}$ comparable to the calcium concentration outside cells, but much less than the sodium concentration there. Of course, the concentrations of ions inside and outside cells are also regulated by diffusion and convection and 'active' properties of nearby biological structures designed to control concentration.

The sketchy calculations made in the last paragraphs are only a motivational starting point for serious work. Models of the flow of ions through channels need to be coupled to models of the flow of ions inside cells and to descriptions of the active systems that help control these ions. An elaborate mixture of biophysics, structural biology, physical chemistry and mathematics is needed to describe and understand these processes. They are much harder to deal with than the selectivity issues I focus on later in this paper.

**Gating Processes.** This review does not deal further with the gating process that opens channels because simulations of gating are not quite in our grasp and the physical basis of gating has not yet been described by reduced models. (Historical note: despite their numbers, models based on arrows[412,415,420] instead of physics have not proven useful and make little connection to the physical properties of channel properties or ions. A British physiologist, who in fact had written a number of arrow models, once told me "You can tell how much is known by how few papers are written on a subject. When it is understood, little more needs to be said." He was referring to his early work on the (mesoscopic scale) sliding filament 'hypothesis'[388,460,461] not to his own later arrow models on the atomic scale.)

Some of the many states postulated by arrow based gating models are likely to exist in a well defined sense, even if most of them cannot be unambiguously identified or measured, but the problem is that the traditional arrow models of biochemistry, biophysics and channology (compare [411,412,415] and [425,431]) do not let you know which states are significant and which are not. It seems unlikely that enough data could possibly be measured to determine the arrangement of the arrows in the models containing tens or even hundreds of reactions. Indeed, the amount of information needed is much larger than the number of rate constants themselves, because rate constants vary with conditions and cannot be constants. It seems likely that deviations from expected behavior would always be explained[234,415] not by questioning the models themselves, but by invoking still more states and rate constants that could themselves be measured only with great difficulty. This process seems unlikely to be an efficient way to understand how channels work, particularly given that the underlying models themselves are not appropriate, at least in my view.[144,186,271,276,284,286,288]

I believe one needs a physical model that shows specifically how proteins move from state to state, under what forces, and with what dependence on electrical and chemical potentials, physical properties of the channel protein, and so on, if one is to make progress. One needs a physical model that shows how the conformation changes of the channel protein reveal themselves as an ionic conductance or as the nonlinear capacitive current called gating current.[77,81-85,128,129,694,777,807,815,843,844] Physical analysis of gating in this spirit is beginning[92,815,899,671] but it is not mature enough for me to review.

Direct atomic simulations would be another path to understanding gating. The conformation and charge storage processes involved in gating and gating current are too slow (from $10^{-5}$ to $10^{1}$ sec) to easily simulate in simulations of the molecular dynamics of all atoms (that occur on a $10^{-16}$ sec time scale). Reduced models are not yet available to deal convincingly with protein conformation changes. We look forward to the successful simulation and analysis of gating in the next few years, as the exponential progress of Moore's law[590,638,639] provides tools for the actual calibration and checking of simulations in atomic detail as well as simulations of long enough duration.

**Selectivity and Permeation**. Physical models of selectivity and permeation are easier to construct than physical models of gating because the structure of the open channel that produces selectivity and permeation does not change on the biological time scale $>10\mu\text{sec}$. It is important to realize that ions in channels are often next to the charged atoms of the channel protein. (Remember many atoms of proteins have significant partial charges—i.e., are polar—even if they are not ionized acids or bases. Something like half of the atoms in molecular dynamics programs have significant charge, i.e., more than 0.1 proton charges.) Coulombic and excluded volume forces are immense on this scale. Even a small change in the separation of such atoms that lasted more than $10\mu\text{sec}$ would produce a large change in the forces on the permeating ion (e.g., on the electrical and on the chemical potential) that in turn would change the current observed to flow through the channel. The resolution of single channel measurements is great,[80,774] and changes of 5% (that would occur for example from a 2% change in diameter in a macroscopic system) are easily seen because signal to noise ratios are >40. Random changes in current are in fact seen in experiments and complex, more deterministic phenomena such as subconductance states,[774,947] but the classical approximation is a good first step to understanding the important features of reality. The conductance of a single open channel is (nearly) constant unchanging in the mean on a scale that can resolve at least 5% changes in amplitude, no matter what the duration of the opening, showing no correlation with the time from the channel opening.

The open channel behaves as if it has a single structure that hardly changes on the biological time scale. Selectivity and permeation can then be analyzed with reduced models that do not involve conformation changes of the protein slower than some $10\,\mu\text{sec}$. Of course, how to construct such models is another question altogether, and it may even be the case that no useful reduced models can be built, and all atom simulations are needed to understand the physics of permeation and selectivity in some channels.

**Physiological models of permeation and selectivity**. Physiologists[415] continue to use state models of permeation (as they have for a long time[396,411-414]) that assume specific states of ions in channels and simple rate models for the transition between these states, although the traditional use of these rate models has long been attacked[70,184-186] and discredited in the biological[35,106,144,268,269,271,277,284,286,288,294,666] and physical[316,386] literature (see Appendix). (References[144,288,666] summarize the main criticisms. Reference[70] shows simulations designed to reveal the diffusive nature of ionic motion through channels, compare reference[949]. References[102,103,106,351,352,357] show the success of diffusion based models.)

**Rate models have their place.** Rate models (see Appendix) are useful when:

(1) the energy profiles and landscapes that define the states are well defined;

(2) states are well defined and separated by large barriers[598], as discussed in the next paragraphs, and the number of states is not increased arbitrarily[234] whenever new conditions are studied;

(3) transitions between states in fact follow well-defined one-dimensional paths independent of conditions,

(4) physically appropriate equations describe motion along those paths,

(5) fluxes predicted by the models are comparable to experiments.

It is important to realize that these prerequisites/assumptions are not easily or automatically satisfied. We consider them one by one.

**(1) Energy profiles are well defined.** It is difficult to define energy profiles well and specifically. The potential profiles and landscapes must describe free energy, and specify how they deal with the entropy production and dissipation that inevitably occur[647] when atoms move in a condensed phase without empty space, like solutions[76] or proteins or channels. (Any simulation of molecular dynamics shows that proteins and channels, like solutions, have no empty space on time scales of nsec, let alone the biological time scale > 10μsec.) Does the potential profile include entropy? How does it deal with entropy production that always accompanies flow? Does potential profile use the dissipation principle[128] of Rayleigh[724] (eq. 26) and Onsager[680,681] to compute entropy production? A well-defined theory using energy profiles has clear answers to these questions.

Ion channels are open systems, in which matter, charge, and energy cross boundaries. Rate models need to deal with potential profiles in these open systems. How is potential described in these open systems? Open systems always involve boundaries and electrodes to maintain and supply concentrations and electrical and chemical potentials. How are the boundary conditions involved in the computation of the electrical potential, or the potential of mean force used in these models or in the energy landscape in general? Indeed, is it meaningful to define a single energy for a channel system as is classically done, given that the profile of concentration, electrical, and chemical potential within a channel must depend on the concentrations of ions (because of screening[135] as reflected in the sum rules[403,610] of equilibrium statistical mechanics)? These profiles depend sensitively on the same variables that determine biological function,[103,106,107,273,274] including the electrical potential across the channel, the charge distribution and structure of the channel protein itself, and concentration of controlling ions (like $Ca^{2+}$, hormones, and second messengers). Crudely speaking, 'potentials of mean force', like biological function, must be expected to vary with a host of significant parameters. The rate constants of reaction models must then also vary with these parameters, often exponentially.

A single potential of mean force or energy landscape cannot capture even the main properties of channels or proteins for these reasons.[324,325] The classical constant field[362,415,438] model is unlikely to describe the field corresponding to anything that actually determines biological function or current flow through channels, no matter how widely it is taught or used. Constant field models[644] made the understanding of semiconductor rectifiers much harder[734,814], and (in my imagination of historical events[175]) prevented their authors from inventing transistors. Constant field models have had nearly as serious an effect on the study of ion channels, I fear.

A model that imposes a single unchanging energy landscape on a system, even if conditions change, can do so only by injecting energy, charge, and or matter into the system. If the model did not inject energy, charge or matter, the chemical or electrical potential would change. A model of this sort must not be used to describe a system that does not in fact have access to energy.

If the system does not have access to energy, charge or matter, the energy landscape must change when conditions change. Thus, describing a protein or channel as an unchanging energy landscape[184,294,298,324,910] is likely to seriously misrepresent[118,273,274,284,286,288,299,798] the system. After all, most systems, and almost all channels have no access to sources of energy, charge, or matter as was established long ago[115] in the history of ion channels.[425,426,431,459]

This misuse of energy landscapes and constant field theory is an important example of the 'Dirichlet disaster' that undermines the law of mass action and the usual treatment of Brownian motion[276] that we will discuss later in this paper. These disasters are automatically avoided by a variational approach that deals self-consistently with all interactions of the energies and dissipations specified in a model.

The energy variational approach[11,151,154,468,551,552,574,582,806] we call *EnVarA* ***derives*** partial differential equations from the energies and dissipations built into a model, instead of assuming those differential equations. The differential equations are the results of algebraic operations on the energies and dissipations and involve no physical assumptions (beyond those built into the energies and dissipations themselves). The equations change when different energies are involved or different conditions are present. Extra energy cannot be inadvertently introduced into the system because the variational process itself constrains the energy before the partial differential equations are derived. *EnVarA* does not assume an energy profile. *EnVarA* produces boundary value problems in which energy can be injected as it actually is injected in the real experimental or biological system.

**(2) Existence of a large barrier.** It is not obvious that large barriers exist in channels while they conduct current.

There are real biological arguments suggesting that large barriers should not exist. Flow through the classical voltage-activated channels of nerve fiber immediately determines the conduction velocity of an action potential and the conduction velocity of an action potential is an important determinant of survival in most animals, particularly an animal like a squid, one imagines. Thus it is reasonable to assume that ion channels are built by evolution to maximize flow, by having as small barriers as possible.

It is interesting to note that traditional models of current flow through channels, like the Goldman-Hodgkin-Katz[362,415,438] constant field[644] equation, assume no barriers, because they describe ion movement through channels as electrodiffusion through an uncharged system without barriers, much like the (electrically neutral) salt water in a macroscopic pipe, or more precisely like a solution of KCl in a pipe (KCl cannot support substantial diffusion potentials because the diffusion coefficient of $K^+$ and $Cl^-$ are nearly the same). It is surprising that so much of the channel literature simultaneously uses constant field equations that assume no barriers and rate models of selectivity and permeation that assume large barriers, following standard texts.[415]

**(3) Assumption of a single path.** A single reaction path, independent of conditions, cannot usually be assumed as discussed in references.[63,512,729] Even two-dimensional reaction surfaces have anisotropic behaviors that cannot be described as a simple single invariant reaction path. This fact undermines[62,512,729] one dimensional models of chemical reactions and rate models of ion permeation.[132,294,308,325,462,463,524,832,926,928]

The failure of the idea of a single path is easy to understand. In a high dimensional space, like that of a chemical reaction, which may involve $10^{20}$ coordinates, it is much harder to define an optimal path uniquely than it is to define an optimal path when walking through a three dimensional mountain range. (Even that task is hard in three dimensions. Only one dimensional paths are easy to define uniquely in my experience.) This difficulty arises in the mathematical treatment of steepest descent in even the simplest textbook cases of one or two dimensions (see p. 265 of reference[126]). Crudely speaking, error terms in steepest descent treatments appear inside integrals and so the full integration must be done before one can be sure of the size of the total (integrated) error. The error can be very different in apparently similar expressions. A term that is small inside the integral—in the integrand—may have a large effect once the integral is performed. Singularities in the integrand must be studied in detail in each specific case to evaluate the size of the total integrated error. This mathematical reality defeats general treatments based on steepest descent methods in chemical reactions.[132,412-414] To put it baldly, a mathematical derivation of a reaction path that does not deal with the specific properties of a system is not possible.

In fact, the singular nature of reaction paths is apparent in direct calculations of ions moving through a bacterial protein porin.[227,787] Different ions follow different paths. The idea of a single path fails ('is singular' in mathspeak) because anions and cations follow different paths. The electric field perpendicular to the path of the ions (i.e., parallel to the plane of the membrane) is large because the permanent charges on one side of the channel are acid (negative) and on the other side are basic (positive).[200,227,482,587,659,699,771,786,787,866,888,892]

These calculations show that a single potential of mean force cannot describe the movement of ions through a channel. The path depends on the charge of the ion, and presumably (judging from other work) on its diameter, chemical interactions with the channel protein, concentration, and electrical and chemical potentials on both sides of the channel. It is not clear that much is gained by introducing the idea of a potential of mean force or pathway for current flow. It may be just as easy to compute the current flow itself.

In any case, one must calculate the current flow separately for each type of ion under each set of conditions and not presume that they all follow the same reaction path. A complete treatment shows that different ions follow different paths and these paths depend on the charge and concentration of the ions. They are expected to vary with the transmembrane potential, temperature, and so on.

**(4) Motion along the reaction path.** Traditional models[411-415] also use an unrealistic description of flux along a reaction path. Flux along a reaction path is diffusive, as we have discussed previously. Flux occurs in a condensed phase without significant empty space and so always involves friction. Flux must be treated as Kramers did[316,386] even in idealized cases where a single path is guaranteed by assumption. The importance of friction was clearly stated by Eyring[928] in his original paper on condensed phases, although it seems to have progressively[308] been lost sight of in later years.[419,420,485]

Rate equations of the ideal gas phase[415,416] that ignore friction can never be used (in my opinion) to describe the motion of ions in condensed phases under biological conditions.[41,144,186,271,276,284,286,288] Rate equations of the gas phase assume no collisions (because ideal gases[127,761] have vanishing density). Condensed phases of ions in proteins always have friction and rarely allow potential profiles or rate constants independent of conditions. In my view, traditional rate models do not apply. Thus,

**(5) Rate equations predict too little current.** Rate equations that do include friction[415,420] among their variables cannot predict anything like the currents actually observed in open channels.[144,186,271,284,286,288]. Rate models that leave out friction describe ions moving in ballistic trajectories, without collisions. They thus predict much larger currents than ions that move in condensed phases. Currents predicted by rate models with friction appropriate for a condensed phase cannot exceed a few tenths of a picoamp. Currents measured in single channels are usually tens of picoamps, often much larger. Currents are reduced by the friction that must exist whenever ions move in a condensed phase with little empty space. (Remember: ice floats on water so water contains less empty space as a liquid than it does as a solid.)

This issue is so central to the view of channels and proteins that I will describe the underlying situation at some length. Motions of atoms in a condensed phase with little empty space, like a protein, cannot occur without collisions. Empty space is a small fraction of a liquid. (Remember ice floats on water so it has much space between atoms than water does.) Atoms move more or less at the speed of sound[75] 1551 m/s or ~1.6 nm (16 Å) every **pico**second, or 1.6 **milli**meters every microsecond (!). If the mean first passage time of an ion through a channel 1.5 nm long is 100 nsec[35]—one ion crossing a channel every 160 nsec produces 1 pA of current—the ion must travel 100,000 lengths of the channel before it first reaches the other side. The ion travels a total distance of 0.15 mm as it moves a net distance 1.5 nm. In fact, calculations and simulations of several models[35] show that the great majority of trajectories of ions go back and forth 'innumerable' numbers of times and almost always (but not always) wind up on the same side—the so called '*cis*' side—of the channel they begin on (if the potential barrier is not negligible). Mathematical analysis confirms[294,513] what the simulations show: the doubly conditioned ionic trajectories—conditioned by the location of their source and also of their sink—that cross the channel are often a vanishingly small fraction of all the trajectories present. These *'trans'* trajectories that cross the channel are the biologically important ones. Because they are such a tiny fraction of all trajectories, their calculation (with known error) poses special problems. These back and forth, *cis* vs *trans* properties of trajectories in channels cannot be approximated by rate models of the gas phase which do not include collisions at all.

The reality of diffusion and diffusive paths mean that classical rate models of channels[415,416,641] do not deal with these fundamental properties of the physics of condensed phases and proteins. It is no surprise that they have not proven very helpful[415] in understanding how the function of proteins arises from their structures and physical laws.

Despite these negative words, I have been taught[452,466,468,575,581,584,766,806,939,944] that it is possible create a field theory of ion permeation in an open system, involving dissipation and interactions of everything with everything else, if the proper mathematical apparatus is used, of energy variational methods *EnVarA* that produce boundary value problems appropriate for the energies and dissipations of the model, and boundary conditions imposed by the experiment and biological structure.[11,151,154,468,551,552,574,582,806] This theory is just past conception,[642,292] so its success cannot be judged, despite the enthusiasm of its parents.

None of this discussion means that high barriers are unimportant. Whenever distinct states exist, barriers exist that define those states. Distinct states clearly exist in the substrates and reactants of enzymes. Distinct states clearly exist in the gating of most types of channels. Motion over the barriers that define these states must be computed in such cases. What is crucial is that the barrier be computed selfconsistently from the physical model, preferably with a variational approach that guarantees selfconsistency. It may in fact be possible to make elegant approximations of high barriers computed selfconsistently, e.g., if the high barrier is one of permanent charge.

However, this computation is trickier than it might seem. It should be noted that flows actually computed over high barriers are surprisingly sensitive to assumed details of the shape of the high barriers, e.g., the shape and symmetry of even very high barriers, as we found to our surprise.[35] As in so many other cases, actual computations show properties not present in simplified discussions. Flow over high barriers is so sensitive to details that even when high barriers exist, and are computed selfconsistently it may be safer and wiser to compute the flow over general shaped barriers, where incorrect assumptions are harder to make inadvertently. Simple expressions can be derived for flux over arbitrarily shaped barriers (see eq. (13) below) and it is not clear why these cannot be used in general instead of the slightly simpler classical exponential expressions.

**Structural models of permeation and selectivity**. Analyses of mechanisms—including selectivity and permeation—of proteins with newly discovered structures are found in most issues of the widely read journals **Science** and **Nature** as well as the more specialized journals of molecular and structural biology and biophysics. These analyses of mechanism[240] follow the traditional practice of biochemistry textbooks and are nearly always verbal, without quantitative specification (however, see[641] that cites Kramers[524] but uses arbitrary rate constants and prefactors without physical discussion). These discussions of the mechanism of protein function are entirely in words, without reference to measurements of function, or graphs or numbers at all. Poetic license has its place but this is not it, at least in my view.

The mechanisms of structural biology usually depend on arbitrary choices of impossible ionic trajectories—impossible because the trajectories never reverse direction, unlike trajectories of real atoms that reverse so often that they travel 0.15 mm before they reach the end of a 1.5 nm channel, as we have just seen. (Indeed, in the Brownian approximation, widely used in simulations of channel motion,[35,60,184,185,198,212,294,470,471,480,535,571,604,673,784,787,798,886] trajectories reverse an infinite number of times in any finite time, no matter how small. That is a fundamental property of the stochastic processes mathematicians call Brownian.)

The functional models of structural biology ignore the statistical reality of atomic physics known at least since the time of Maxwell (see history in reference[117]) and so are even less helpful than the arrow models of physiologists. They resemble Kekule's molecular dreams more than physical reality. The simulations of molecular dynamics available now for decades[116] should have provided a visual vaccination against the idea that ions move in smooth slightly curved paths. Roux[70] addresses the issue most directly, in contrast with MacKinnon[641], in the same issue of the journal **Nature**. Sadly, a glance at the literature of structural biology shows essentially no 'back and forth' paths of ions like those that actually characterize atomic motion. Verbal models of smooth paths are nearly always used to describe the 'mechanisms' of molecular biology, just as in biochemistry textbooks for what seems to be forever.

The reader may think that these smooth paths should be thought of as average paths. Average paths of course can be smooth, but renaming the paths of structural models begs the question. How are the smooth paths chosen? It is not at all clear how one should average the astronomical number of atomic motions that determine the motion of an ion as it crosses a channel without calculation and theory. Statistical physics and molecular simulations were in fact developed to do that averaging. Modern simulations of ions in channels are beginning to average trajectories successfully and this work may eventually succeed in reaching biological time scales. The other issues of scales are so large, however, that atomic simulations of biological systems seem likely to remain out of reach for a long time, as we discuss at length later in this paper. The key idea is that all the gaps of scales (see Table 1, much later in the paper) must be dealt with simultaneously in a fully atomic simulation, because all those scales exist and are significant to the natural function of ion channels.

Calculations of selectivity or functional properties involve much more than the average paths, of course. They also involve the driving forces (of concentration, electrical and chemical potential) that send ions through these paths and the 'resistances' of various forms of friction that result from the motion along these paths. The friction involves collisions, with water, ions, and atoms of the channel protein and even more importantly (I imagine) dissipative electrostatic interactions (dielectric friction) with charged atoms of the protein, water, and solutes within a Debye length or two of the ion itself. The average paths change with experimental conditions and so the friction must as well.

In the real biological case, the number of atomic motions and interactions involved are far larger than astronomical. The current flow through a sodium channel during a propagating action potential depends on the electrical potential over a distance of millimeters. All the ions in that region interact and are significant in producing the propagation and waveform of the electrical potential. The number of interactions of some $10^{19}$ ions is very large indeed.

Verbal descriptions do not deal with these issues at all. But the issues exist whether or not structural biologists choose to discuss them. Statistical physics has been developed over the centuries because words alone cannot account for the properties of inorganic solids and liquids. Indeed, it is hard to see how verbal models can be falsified, or verified, even in principle. They have more characteristics of metaphor and poetry than of science and engineering.

Poetry and metaphor have important places in scientific thinking, as motivators of the guesses that start most scientific work. But the checking that makes the metaphorical guess into science needs to be objective and quantitative if at all possible. Quantitative analysis addressing the known properties of atoms is needed to deal with ions in channels. The molecular mythology of smooth reaction paths in traditional biochemistry textbooks is not useful if we wish to compute and control biological systems the way engineers compute and control inorganic systems. An objective method of computing those smooth paths is needed, that includes the ionic conditions, and boundary conditions, as well as the structure of the protein. Engineers do not use verbal models to design or build things, any more than building contractors do.

Verbal structures fall of their own weight unless buttressed by numbers. But verbal structures can take a long time to fall if words are allowed to replace actual theories that confront specific experimental data.

**Reduced models are the essence of biology and engineering**. Despite their misuse, verbal descriptions of structure and mechanism have a crucial role, even if they cannot be the endpoint of analysis. Verbal descriptions and qualitative discussions are starting points for quantitative investigation.

Biological systems are like engineered systems. One needs a general description, a list of parts, and a knowledge of function (and power supplies) before one knows how to write equations. Biological systems cannot be analyzed without some knowledge of structure, just as engineering systems cannot be analyzed without some knowledge of their structure. One must know which wires are the inputs and outputs (and power supplies and ground wires) of amplifiers, before they can be analyzed. But one does not need to know the full (logical) circuit diagram, let alone the physical layout, let alone the location of all the atoms of the amplifier. One needs enough knowledge to specify a reduced model, an input-output relation that allows us to summarize quantitatively what we need to know to control and improve the amplifier.

Of course, what we need to know is not unique. It depends. An engineer working for NASA needs to be concerned about power consumption of amplifiers (as does a designer of memory or CPU chips nowadays). But for most of us power consumption is not so important. A designer of audio amplifiers needs to worry about distortion. How much she or he worries depends on the type of music. Folk songs and Mozart arias are one thing; rap music is another. A designer of patch clamp amplifiers needs to worry about input currents. But even in these cases, reduced models are used. Only a few elements are needed to describe the input and output impedance of amplifiers, including their input currents and even their output slew rates. These extra elements make reduced models more complex, but they do not involve all the circuit elements let alone all the atoms of the device. A handful of extra elements are involved, not the astronomical numbers in molecular dynamics, or the tens to hundreds in rate models or the uncountable number in metaphorical treatments of smoothed trajectories.

Complete descriptions are rarely needed or wanted in biology and engineering. The magnificent molecular architecture of proteins are cathedrals of structural knowledge. We can admire their beauty but we do not need to know all the details of that beauty to know what the cathedrals do, or how to enter them, and even how to make them work. The doors and altars are often easy to find. The architecture can be reduced to a simple plan if we have limited needs and goals. We do not need to know everything.

Reduced models are what are needed to understand protein and channel function, in my view. The problem with the reduced rate models of physiology[214,418] and biochemistry[411-414] is not that they are reduced but that they are based on the properties of uncharged ideal gases, which do not resemble the properties of ions and proteins in condensed phases. The problem with the verbal models of structural biology is that they do not involve numbers and equations and so cannot deal with experiments: reduced models cannot be used unless they involve numbers and equations.

Reduced models of proteins and channels are of great help if they start with recognizable properties of condensed phases and then can successfully calculate properties that can be measured in a range of conditions. Such reduced models can be tested and a sequence of models constructed that will allow understanding and control of biological and engineering systems.

**The scientific method and channels.** Guess a model; check it; fix it and add more if needed. One should start by assuming that the usual models and methods of physics and engineering can deal with biological complexity. The usual procedures of physics and engineering can then be employed to understand and control channels. If those procedures prove inadequate, new principles can be introduced, special to biology, if they are specific and quantitative, and testable.

We follow standard procedures of physical analysis here as we use reduced physical models and try to understand the selectivity of calcium and sodium channels. We guess what physics may be involved. We derive as carefully as we can the consequences of that guess. We check the consequences, and modify the model as needed to improve the guess. "Guess and check" is the name of the game.

**Reduced Models of Calcium and Sodium Channels**. Reduced models of channels are built in this tradition of guess and check. The hope is that these models capture the essential physics used by biology to create the selectivity important for biological function. The models are simple enough so that the physics they contain can be calculated with some accuracy. They are justified by their fit to data and by the robustness of their results: methods from MSA[664], to SPM[669], to Monte Carlo[99,100], to DFT[748] and PNP-DFT[353] give essentially similar results, often quantitatively[99,100,664,881] as well.

We concentrate on reduced models of calcium and sodium channels because they have been quite successful in dealing with the selectivity of these channels as measured over a wide range of conditions. These models represent the side chains of proteins as spheres of charge, that occupy volume and interact with mobile ions ($K^+$, $Na^+$, $Ca^{2+}$, and $Cl^-$) through volume exclusion and electrostatics much as the mobile ions interact with each other. The ions and 'side chains' mingle together in the selectivity filter of the channel, typically a region 10 Å long and 6 Å in diameter to which the side chains are confined. The solvent and protein are represented implicitly as dielectrics.

We hope that the principles of a general approach to channel permeation and selectivity emerge from this specific analysis, along with some general physics that may be present in most problems of channel and protein function. The general approach assumes that understanding of selectivity requires measurements in a wide range of solutions of different concentration and types of ions. Computations of a 'free energy of binding' in a single solution are not helpful for two reasons. Properties in a single solution are too easy to explain. It is difficult then to separate one model of selectivity from another. Secondly, it is clear that the 'free energy of binding' is not constant, but depends on ionic conditions. The ions in the baths and channel are not ideal. Everything interacts with everything. The free energy of binding depends on all concentrations. Thus, calculations done under only one set of conditions are not very useful, for our purposes. They do not permit comparison with a range of data; they are vague enough that calculations with different models cannot be compared.

This reduced model of selectivity is in the long traditional of primitive implicit solvent models of ionic solutions. Models of bulk solutions using implicit solvents have a long and

successful history in physical chemistry,[868,2,3,47,136,256,257,309,390,494,502,547,548,700,701,735,762,904] and have been particularly investigated and compared with experiment in Turq's group.[206,252-254,256,257,736,743,745,825] Implicit solvent models of proteins are also widely used in the study of protein function. Indeed, the literature of implicit solvent models of proteins is too large to review.[25,33,116,213,239,259,358-361,394,395,448-450,469,510,637,663,722,751,763,800,803-805,826,850,915-918,950 452,466,468,575,581,584,766,806,939]

The lack of detail in implicit solvent models is primitive, as the name implies. The treatment of polarization as a dielectric is actually embarrassing to those of us who are aware of the complex dielectric properties of ionic solutions[45,46,593,678,679] and electrochemical systems in general[45,593, 272,273,356,460,532,615,616,808]. I spent many years making impedance measurements of the complex dielectric properties of biological systems to determine their electrical structure[287,289,295,554,611,879] and so have measured the dielectric properties of cells, tissues, or ionic solutions that cannot be described by dispersion, or a single dielectric constant.

The dielectric 'constant' of ionic solutions, in particular, is nothing like a constant. It varies from 80 to 2 in the time range of atomic motions relevant to molecular and Brownian dynamics (i.e., from $10^{-13}$ sec to $10^{-7}$ sec). The likelihood of nonlinear field dependence in the region close to an ion (particularly a multivalent ion) cannot be denied. Indeed, electron orbital delocalization may occur in some cases, and then solvation has some of the characteristics of a classical chemical reaction involving (partial) covalent bond formation. The fact is, however, that so far the most successful treatments of ionic solutions are primitive despite the impressive progress of a number of laboratories[2,136,486,494,502,904]. Only the primitive model has allowed calculations of the fundamental properties of ionic solutions, namely their free energy per mole or chemical potential. Primitive models are a good place to start. They are also surprisingly successful. Perhaps the most important properties of ionic solutions depend mostly on integrals of the dielectric properties over all frequencies because of the Kramers Kronig relations and this integral property is captured by implicit solvent models well enough.

The fundamental property of any ionic solution is its free energy per mole, its activity, or electrochemical potential, all nearly the same thing, differently normalized, written in logarithmic, exponential, or linear scales, see [47,75,76,309,547,660,701,735,762,790]. Almost all solutions have excess chemical potential, activity coefficients, or osmotic coefficients not equal to unity because few solutions are ideal.

The central fact of electrochemistry is that the excess chemical potential of an ion is not zero.[530 319] The excess chemical potential in fact varies as the square root of its concentration (speaking loosely for 1-1 electrolytes like NaCl) and not linearly[217,309,547,735]. Ions are not independent in ionic solutions.

Ions are not independent in ionic solids, where we take for granted the fact that there are exactly equal numbers of $Na^+$ and $Cl^-$ ions (or we would be electrocuted each time we salt our food[829]), and ions are not independent in solution[217] because of the screening[135] reflected in the fundamental sum rules[403,610] describing ionic fluids. The requirement of electroneutrality in bulk solutions guarantees that ions in solutions have highly correlated behavior not found in ideal infinitely dilute gases of point particles without charge.

Ionic solutions are not ideal. Their extensive properties are not proportional to number density. The 'independence principle' that Hodgkin and Huxley[430,433-437] used so brilliantly to understand the properties of nerve membranes does not apply to bulk solutions. The independence principle correctly describes the movement of different species of ions through different (and independent) protein channels in a membrane, if they are perfectly selective to those species. The independence principle describes almost nothing else.

**Different ions carry different signals.** Selectivity is an important, even fundamental property of channels and proteins. Indeed, one of the founders of molecular biology (Nobelist A. Klug) said[691] "There is only one word that matters in biology, and that is specificity. The truth is in the details, not the broad sweeps." While this might be mild (but understandable) hyperbole, the historical fact is that specificity of channels is so important that it is widely used to name them. The specificity was unprecedented in the physical world until quite recently. Even now, such specificity in chemistry is usually found only in biomimetic applications.

The true significance of specificity and selectivity in biology is their biological importance. Biology uses the concentration—really, the activity—of individual types of molecules as distinct signals, the way a computer uses the voltage in an individual wire (or transistor) as a distinct signal. If the wires (or voltages) get mixed up, the computer stops functioning. If animals lose their ability to distinguish ions (e.g., even the commonplace $Na^+$, $K^+$, $Ca^{2+}$, and $Cl^-$ ions), the animals die. Selectivity of channels and proteins allows ions to carry specific signals.

Selectivity of channels allows ions to provide energy for biological function. Animals use gradients of $Na^+$, $K^+$, $Ca^{2+}$ and $Cl^-$ to provide energy for many of their most fundamental functions, from signaling in the nervous system, to coordination of contraction in muscle (including the coordination of contraction of cardiac muscle that allows the heart to function as a pump), to the transport of nearly every foodstuff that provides nutrition. Animals use gradients of these ions to maintain the osmotic integrity of cells. Failure of selectivity between $Na^+$ and $K^+$, for example in the nerve terminals of the human brain, quickly leads to irreversible swelling of the terminals (i.e., they burst) and death. The more important the nervous system is to the animal, the more dense the network of nerve terminals, the more it must cope with the finite selectivity of channels to ions, the more likely the animal is to die if it cannot cope (by active transport requiring oxygen) with the 'leakage' of $Na^+$ (and accompanying water) through imperfectly selective channels.

Nothing is more important than selectivity in my view, although I would not say it is the only word that matters, as did Klug[691]. As we shall see, details are in fact crucial in the selectivity of calcium and sodium channels, as Klug said they would be in general, but the details can be computed[102,103,106,273,277,664-666,668,669] from the broad sweeps, much to the surprise of the founders of the reduced model (Wolfgang Nonner and Bob Eisenberg, soon joined by Douglas Henderson, Dezső Boda, and Dirk Gillespie).

**Selectivity of the Calcium Channel**. We turn now from generalities to a specific model and its analysis, a reduced model of the L type calcium channel of skeletal and cardiac muscle described extensively in the literature.[20,21,61,153,274,300,346,397,410,474,517,520-522,541,545,546,616-620,737,785,789,849,871,872,931,945,946] This channel coordinates contraction in the heart, allowing the heart to function as a pump. Its biological importance is difficult to exaggerate. Its medical and clinical importance is as great as its biological importance. The L-type calcium channel is the target of calcium channel blockers of considerable pharmacological significance. The easiest way to document the medical, clinical and biological importance is to do a Google search on 'calcium channels'.

The selectivity of calcium channels has been well described[617-619,785] and so we need not describe it again here. It is important to say, however, that one class of experiments on calcium channels, and calcium binding systems in general, has been misinterpreted because of a technical difficulty in the construction of solutions buffered to known activity of calcium.[106] Standard methods[608,609] of calculating the activities of calcium buffers (like EGTA) involve the ionic strength of the solutions; they treat $Na^+$ and $K^+$ the same way, for example, see p.224 of reference[608]. This treatment is, however, an unfortunate oversimplification.

$Na^+$ and $K^+$ have different effects on the activities of $Ca^{2+}$ when $Ca^{2+}$ is at high concentration because such solutions are not ideal[309,547,548,700], not even approximately ideal. $Ca^{2+}$ is at very high concentration close to the EGTA molecules used to buffer its concentration. The standard methods of computing the activities of ions in calcium buffers introduce corrections based on ionic strength, and so treat $Na^+$ and $K^+$ the same way, for example. They are in error because $Na^+$ and $K^+$ do not have the same effect on $Ca^{2+}$ when ions are highly concentrated. This common method of computing activity produces incorrect interpretations of experiments, as discussed in detail in one case in reference.[106] Sadly, this problem is likely to be important wherever calcium buffers have been used and other ion concentrations have been varied. $Na^+$ and $K^+$ concentrations have been varied in many such experiments.

The question then is how do we calculate the properties of the calcium channel? What kind of model should we use? Later in this paper (see Table 1) we argue that a full atomic scale calculation is likely to stay out of reach for sometime: the gaps in scales, for length, volume, time, and concentration are too large, particularly when one has to deal with all the gaps at once, as the channel itself does. The channel 'knows' how to use atomic scale structure to control macroscopic flows controlled by trace concentrations of ions. But we do not. Our task is to try to determine how evolution has chosen to make such a system.

We proceed by guessing a reduced model using the insight that channels are very crowded places with enormous densities of charge. We will try the simplest representation of the ions we can think of (as hard spheres) and the wild guess that side chains of the protein can be represented as spheres just like mobile ions. The spheres are free to move anywhere within the channel without constraints, but they are not allowed out of the channel. It is not clear that anyone—particularly well trained and (we hope) well bred classical physiologists like Wolfgang Nonner and myself—would have guessed *a priori* that such a system would work. But we wanted to understand publications of Turq's group[256,257] on the mean spherical approximation[821-825] and we knew of the role of the EEEE (glutamate glutamate glutamate glutamate) side chains in the L type calcium channel (see more recent review[785]). John Edsall (in his last conversation with me, at his 95th birthday celebration symposium around 1998) had guided me once again "Bob, can you include the size of ions in your nice work on electrostatics? You know, van der Waals doesn't do too badly." Perhaps he knew of Widom's paper[377,586] reworked with such clarity by Henderson.[400]

Early calculations[664,668] showed promise and motivated Douglas Henderson to show us how to extend them[95,96,99,100] using more appropriate and convincing methods—Monte Carlo simulations of the primitive model of ionic solutions with an implicit solvent—he had developed with Dezső Boda.[96,97,101] Before we knew it, high quality simulations[95,96,99,100,669] showed that the EEEE system had the main selectivity properties of a calcium channel. Later work[95,96,99,100,103,106,107,274,347-352,600,669,764] suggests that this simple model captures the essential features of the biological adaptation used by evolution to create calcium selectivity.

**Reduced Model of the Calcium Channel**. The reduced model of the calcium channel we analyze (Fig. 5) considers only the signature amino acids of the channel EEEE that confer selectivity on the channel according to experiments[300,785,927,931] discussed in more detail below. The side chains are represented as movable spheres, so they have excluded volume, which in fact fills a substantial fraction of the selectivity filter. The selectivity filter is 6 or 7 Å in diameter[81,80,84] and 10 Å long. The number density of ions (called 'concentration' for short) varies with location and so plots of concentration vs. distance are always examined. Occupancy is defined to be specific as the total number of ions in the central 5Å of the filter. (Other definitions could have been used. This choice seemed sensible but needs further investigation.) The ions in the solution are treated as hard spheres with 'crystal' radii of $Ca^{2+}$ 0.99 Å; $Ba^{2+}$ 1.35Å; $Li^{+}$ 0.6 Å; $Na^{+}$ 0.95 Å; $K^{+}$ 1.33 Å; $Cl^{-}$ 1.81 Å; $O^{-1/2}$ 1.4 Å. Precise numbers are given here to avoid ambiguity. The carboxyl groups are treated as two half charged oxygens because the oxygens of glutamates are symmetrical in bulk solution and so the acid making electron of the carboxyl is expected to be shared equally. The side chain oxygens are called 'structural ions', 'oxygens', 'glutamates', or 'side chains' in different places in the literature. The side chains are allowed to move within the channel according to the same rule as the ions themselves except the side chains cannot be outside the channel. Monte Carlo (attempted) moves that place oxygens outside are rejected.

The diameters of the ions and side chains are never changed. They are the same in all our publications (except for inadvertent errors). Indeed, a sensitivity study of the effects of small diameter changes is probably needed.

The constancy of ion diameters is of great importance because it is an important distinction between our work and that of workers on K channels who often 'back engineer' diameters of ions (and other parameters), for good reasons, no doubt. We avoid such procedures because we think they would make it too easy to fit data. It would seem quite easy to adjust diameters to get reasonable results with a wide range of disparate models, although we cannot be sure, not having done that ourselves.

To reiterate, our goal is to fit existing experimental selectivity data with a model with unchanging diameters and parameters in a wide range of ions (e.g., $Li^{+}$, $Na^{+}$, $K^{+}$, $Rb^{+}$, $Cs^{+}$, $Ca^{2+}$, $Mg^{2+}$, $Ba^{2+}$) in a wide range of concentrations including the biologically crucial range of $10^{-7}$M to 0.1 M for $Ca^{2+}$. We believe this much data with this much constraint on our parameter estimation is needed. Estimates of free energy of binding made in one condition (i.e., one solution of one concentration) are not helpful in our experience because the free energy of binding is a highly nonideal quantity depending on all concentrations in the baths. This is what we expect from a system containing some 20M salt in the channel, in which everything interacts with everything else in a highly nonideal way.

There are serious problems with our model. For example, the choice of ionic diameters can quite rightly be criticized because it ignores hydration shells and other specialized interactions between ions and water or channel. These have not been ignored out of ignorance—they were explained clearly to me as an undergraduate in 1959 by John Edsall.[172,262] Rather, our goal is to see how well we could do without such shells and interactions, whose energy and nonideal properties are so difficult to calculate reliably. In fact, it seems that representation of solvation and hydration by the implicit model is good enough. Properties estimated from a dielectric structure and spherical ions fit experiments quite well.

**The ionic environment of the channel is remarkably crowded.** In the reduced model of the calcium channel, 8 half charged oxygens are located in a channel of volume $2.8\times10^{-28}\,\mathrm{m}^3$ at number density of $\sim 2.8\times10^{22}\,\mathrm{cm}^{-3}$ . The structural oxygen ions are present at a concentration of about 46 molar! For comparison, the concentration of NaCl in a solid is 37M and the concentration of oxygens in liquid water is 55 molar.

The volume of the oxygens in our model is $\sim 8.8\times10^{-29}\,\mathrm{m}^3$ which means the oxygens occupy about 31% of the volume of the channel. The four permanent negative charge of the oxygens can be balanced by four sodiums that have a concentration of 46 molar and occupy another $1.67\times10^{-29}\mathrm{m}^3$ or 5.9% of the volume of the channel. The two calciums needed to balance the negative charge of the oxygens would have half that density, namely 23 molar, and occupy half that volume. It is clear that the two calciums would be much easier to pack than the four sodiums and the difference in the energy of repulsion is an important contributor to the selectivity of the calcium channel.[664]

Concentrations of this size are far beyond saturation in a bulk environment. This environment inside a channel is that of an ionic liquid,[519,921] in which nearly all atoms are charged. This environment inside a channel has no resemblance to the infinitely dilute environment envisioned in a perfect gas because the environment is so concentrated. Perfect gases are perfectly dilute. It is not surprising that theories based on properties of ideal gases are not very useful. Solutions in channels or active sites are not small perturbations of an infinitely dilute solution. For that reason, simulations using force fields calibrated in infinitely dilute environments should be expected to face difficulties in such concentrated environments. Even so, nearly all force fields used in molecular dynamics simulations are calibrated in infinitely dilute environments, or in environments without definite ion concentrations. In the first generation of molecular dynamics force fields, the problems of calibration in biologically relevant mixtures of ions were understandably avoided. Now that molecular dynamics is established as an indispensable tool of structural, if not molecular biology, more realistic calibrations will be made, no doubt.

It is important to realize intuition is no guide in a concentrated environment like that inside an ion channel with large forces of opposite signs. The excluded volume forces are enormous. The electric fields are enormous, often larger than 0.1 v in 1 nm, or $10^8$ v/m. The electric forces are long range. The excluded volume forces extend at least the length of the channel. The electric forces can have either sign and thus can balance the excluded volume forces or add to them. In a mixture like this, the only thing that can be certain is that reasoning based on the properties of uncharged point ions is useless. Sadly, the traditional theories of channel permeation and selectivity all use chemical reaction models in which electric charge and excluded volume play no role. (The Appendix discusses some of the difficulties with these models.) It is not surprising that such traditional analysis has not been helpful.

**History of Reduced Models of $Ca^{2+}$ channels**. The reduced model of calcium channels used here was originally motivated[664,669] more by its simplicity than anything else. Neither Wolfgang Nonner nor I imagined that a handful of glutamates—represented as spheres—would be very successful in dealing with selectivity in calcium channels, because selectivity was such a complex phenomenon, involving properties in tens of solutions, particularly compared to the stark simplicity of the model.

The first calculations of the model[664,669] showed striking calcium selectivity, however; and a long series of papers [98-100,102-104,108,273,274,349,351,353,354,356,629,630,664,667,669,748,906] has shown that the model works remarkably well for both calcium and sodium channels, despite its obvious defects and lack of structural detail.

In particular[102,103,106], a single model, with two parameters, and one set of diameters of ions (chosen to be crystal radii, not hydrated radii) deals well with the selectivity properties of two different types of channels (calcium and sodium) with the same unchanging parameters (dielectric coefficient and volume of the selectivity filter), in a wide variety of solutions of different composition and concentration.

Only the side chains differ in the models of the sodium and calcium channel. The difference in side chains is enough to produce the strikingly different properties of the channels, including the biologically important selectivity of the sodium channel for $Na^+$ over $K^+$.

No additional 'chemical' energies are needed to reproduce the behavior sodium and calcium channels over a wide range of conditions. The binding free energies are outputs of this model and vary with conditions. Simulations in atomic detail are not needed. Indeed, simulations of selectivity that ignore ionic conditions, or that produce a **single** free energy of binding as an output are obviously irrelevant in this case, because the free energy of binding varies so dramatically with ionic conditions. Amazingly, the reduced model calculates the free energy of binding correctly (i.e., well enough to fit experimental data), over a wide range of concentrations of different types of ions. The reduced model of a calcium and sodium channel does far better than any atomic detail model of any channel we know of.

**Reduced Model Properties**. The reduced model[106] represents the channel protein as a dielectric surrounding a cylindrical pore some 10Å long and 6Å in diameter. The pore contains spherical ions $Na^+$, $K^+$, $Ca^{2+}$ or $Cl^-$ (and sometimes other ions) in amounts and locations that are determined as outputs of the calculation. Water in the pore is represented as a uniform dielectric. The bulk solutions are represented as spherical ions in a uniform dielectric. This is the primitive model of ionic solutions using an implicit model of the water solvent. In much published work, the dielectric constant of the pore is taken as equal to that of the surrounding bulk solution, although obviously one should use a smaller value. Preliminary work shows that using a more realistic value in a channel model containing three dielectric regions does not change important conclusions of our work.

The pore contains spheres meant to represent the side chains of the amino acids that make up the wall of the channel protein. The side chains are charged spheres and the charge is taken as the charge of the fully ionized side chains. The model corresponds to experiments at say pH 8.0 where acidic groups are likely to remain fully ionized at all membrane potentials and all solutions of interest. Treatment of cases with variable ionization would add complexity to our analysis but can be done simply by including the ionization energy (and dissipation if desired) into the *EnVarA* theory described later in this paper.

The side chains (as we will call the spheres from now on for the sake of simplicity but not realism) are treated as mobile ions that are confined to the selectivity filter of the channel (as we call the 10×6Å cylindrical pillbox) making it a classical ion exchanger[399] and a typical Donnan system described in physiology and biochemistry texts for more than 100 years.

What is different from the classical ion exchanger is that in our model the ions and side chains are mixed into an 'electric stew' (as Ed McCleskey[618] so aptly called it) in which the ions and side chains have finite volume. This stew corresponds to the view of the calcium channel of many experimental papers[20,21,153,300,410,474,541,785,872,931] that study the accessibility of side chains in a variety of experiments using mutations and cysteine scanning.[927] That work is reviewed in reference[785] where an important conclusion is reached (on p.134). "The Ca channel field is convinced that the EEEE carboxyl side chains project into the pore lumen". This structure is in contrast to the better known K channels where side chains face away from the pore lumen and the protein presents a wall of carbonyls to permeating ions and water.

What is different from the classical ion exchanger is that the Donnan system is analyzed with modern methods and molecular insights: Monte Carlo simulations are used to deal with the important nonideal properties of the ion exchanger arising from the finite diameter and electrostatic interactions of the solid spherical ions.

The spatial locations of ions and side chains are an output of the calculations using the reduced model. The content of side chains (i.e., the total number of side chains in the selectivity filter) is fixed, but the content of ions is not fixed. The content of ions varies with conditions as does the location (but not number) of side chains. The energy of the selectivity filter is taken as the sum of the electrostatic energies of all the spheres, assuming that the spheres have the dielectric constant of the bulk, with all their charge at their center. Spheres are not allowed to overlap, so an important determinant of their locations and free energy are excluded volume effects. The most realistic calculations are done by Monte Carlo simulations (Metropolis algorithm) with methods discussed at length in the literature[96,98,99,102-108,349,350,599]. The reader is sent to that literature, and standard references for more details of the Metropolis Monte Carlo Method.[13,326,380,539,721,788,901] Suffice it to say here that the method is remarkably robust and able to produce good estimates of the distribution of particles in a wide variety of systems.

Monte Carlo simulations of the primitive model used here are quite efficient when ions are treated as hard spheres. If the spheres overlap in one of the trial moves of the Monte Carlo simulation, the energy is taken as infinite, and the trial is rejected. If the spheres do not overlap, the energy of the ion (injected in the trial) is computed from the electrostatics of its central charge, and the trial move is accepted or rejected according to the Metropolis criteria. The Metropolis procedure guarantees that the computed distribution of particles is in a Boltzmann equilibrium once computed distribution has aged and lost its dependence on initial conditions. It is easy to examine a calculation to see if it has aged sufficiently.

**Crowded Ions: Properties of the Model of Calcium Channels.** The most striking property of this model of the calcium channel is the density of ions in the channel, as mentioned previously. The number of ions within the ionic atmosphere of the selectivity site must equal the net charge of the protein within the selectivity site if the system is electrically neutral. The system must be neutral because biological systems are destroyed by voltages more than 200 mV. So the net charge cannot create a potential more than 200 mV. The net charge can be estimated by Gauss' law applied to a sphere, say 10 Å in diameter, in a dielectric of 78 extending to infinity. The net charge that produces a potential of 200 mV is very small indeed, a negligible fraction of the ions present.

The charge of the side chains themselves (without the mobile ions) is enormously high in the channel. The charge is 4 charges in $2.8\times10^{-29}\,\mathrm{m}^3$ or a number density of $1.4\times10^{22}$

charges/cm$^3$, in chemical units 23 molar! For comparison, the density of solid NaCl is 37 molar. It seems clear that active sites of channels are extraordinarily highly charged. The contents of a selectivity filter or active site are an ionic liquid[519,921] with constrained components.

Of course, the mobile charge—that balances the side chain charge—does not all have to be within the channel itself: the ionic atmosphere must extend outside the channel, perhaps quite significantly. The effects of the ionic atmosphere outside the channel can be modeled in a simple way by representing that atmosphere as a spherical capacitor, with a potential equal to the potential on the end of the channel. The equations of the capacitor and channel can be solved simultaneously and the effects of the spillover charge can be computed this way. These effects are significant quantitatively in many cases, particularly when the ionic charge cannot balance the fixed charge (e.g., $Ca^{2+}$ cannot balance an odd number of fixed charges), as Douglas Henderson pointed out to me long ago. Nonetheless, the qualitative properties of the model survive. Interestingly, $Ca^{2+}$ channels have much lower single channel conductance than $Na^+$ channels, suggesting that the mobility of $Ca^{2+}$ is much lower than $Na^+$ inside a channel. Perhaps this fact is connected to the charge balance between $Ca^{2+}$ and the permanent charge of the channel. Perhaps it is not: $Ca^{2+}$ conductance of a $Na^+$ channel is low too. The physical origin of the conductivity of ions inside channels is an important subject to address with high resolution methods of molecular dynamics.

Our model of a channel as a rigid body containing spherical side chains probably works so well because channels are so crowded with charges and side chains free to move. In the special case, exploited biology, the energies of the crowded charged spheres probably resemble those of the much more complex structures they crudely approximate. Future models should clearly include more complex and realistic models of side chains. Of course, this is a guess, written provocatively on purpose. This guess needs checking by simulations with higher resolution models. We need much more work on the physical chemistry of systems of crowded charge to make clear just what properties are well described by our simple model, and what are not.

The enormous concentration of ions in a selectivity filter reflects a general property of proteins. Proteins are highly charged objects. Early workers on proteins often were trained as physical chemists and immediately noticed the large charge density of proteins. As Tanford says[855,172,262,578], proteins 'bristle with charge.' These workers were referring to the charge density of the outer surface of a protein. Imagine what they would have said if they had known that the charge density of active sites was 10–50 times larger than that!

The enormous concentration of ions in an active site or selectivity filter has not been widely discussed even though the argument about electrical neutrality is not difficult. Indeed, there are a large number of enzymes, and a few channel proteins in which x-ray crystallography can actually resolve these counter ions. We (Liang, Jimenez-Morales, and I) are currently involved in searching databases to establish the density of acidic and basic amino acids in active sites and document this important fact in an objective way. It seems likely that the density is very high in most enzymes, binding proteins, and channels, suggesting that the crowded charges may play a central role in protein function in general. Proteins may be designed to exploit the nonideal properties of highly concentrated ionic environments, stews of charge.

It seems natural that any ion exchanger used for a specific purpose will have as large a concentration of ions as possible, just as electrodes in electrochemical systems or batteries will have as large a density of ions nearby, or transistors will have as large a density of holes and electrons as possible so they can conduct large currents.

It is clear that understanding why a simple model works so well will be very important in making reduced models of other channels, enzymes, proteins, and nucleic acids. Atomic scale simulations of these wide range of molecules do not yet include ionic solutions of varied composition realistically, if at all. Perhaps it will be easier to make reduced models of these biological systems than to make realistic simulations of the ions and atoms of the solutions and proteins of the systems.

**Limitations in analysis**. One might object, of course, that the ions in these models are at such density that in some sense they are not able to move. One might imagine they are 'salted out' as a local precipitate. However, that precipitate kind of binding cannot be present in ion channels. Ions in channels carry current through the channel in a reasonably normal way, albeit with a mobility probably some 100× less than in bulk solutions (see p. 1182 of reference[102] for discussion and references). The properties of the trajectories that carry current are interesting but problematic. So few trajectories succeed in crossing a channel that one must be very careful. Individual trajectories as usual are uninterpretable, even individual successful trajectories, because they reverse direction so very often, as we have discussed.

The set of successful trajectories that would be computed by a simple Langevin version of the primitive model[5,783,888] might be easily characterized by its flux, mean first passage time, doubly conditioned probability functions, and so on as in stochastic analyses[294,513] of related problems. Or the successful trajectories might have very special properties—e.g., cooperative motion of (neutral) ion pairs over potential barriers—quite different from those of typical trajectories. **In that case, the special properties would need to be identified and dealt with separately. In other words, the stochastic analysis would need a reduced model to interpret its results.** The set of successful trajectories computed from a full molecular dynamics simulation is more likely to have such special properties than those computed from a Langevin simulations, because so many more forces and effects are dealt with. The set of trajectories selected by biological structures and evolutionary pressures to perform a specific biological function may be very unusual. The useful trajectories may be a tiny subset of all trajectories, yet those may be the 'only' ones that determine biological function. The complex structures of technological devices and biological systems are not random. They are designed to make certain inherently implausible things happen all the time. Reduced models are built to describe these useful properties of a system. Reduced models can be viewed as models built to describe a highly conditioned system, with trajectories chosen to perform a particular function. Estimators are needed to select the useful trajectories, to take the staggering number of atomic trajectories and select from them those which perform a useful biological or technological purpose.

The estimators needed to make sense of a complete stochastic analysis have many of the properties of a reduced model. Indeed, one systematic strategy for finding reduced models is to use the theory of inverse problems[118,304,493] to find estimators of biological properties (e.g., unidirectional flux) with useful sensitivity properties and robustness. Physical chemists are needed to study the actual set of trajectories that cross channels in high resolution simulations and determine their special properties. Mathematicians with expertise in inverse problems may be needed to help identify these special trajectories efficiently.

In the case of proteins in general, ions are also probably reasonably free to move in active sites, for example in enzymes. Precipitates are probably not present because the ions in question are often the substrate of the enzyme, and thus obviously participate in a reasonably normal way in the chemical reaction catalyzed by the enzyme.

I imagine that if ions were so tightly bound that they were precipitates, they would always be visible in x-ray structure along with the protein structures that bind them. After all, binding means that the positions of the bound molecules are fixed and the motions of the bound molecules co-vary the way atoms co-vary in a solid. If the parts of the protein that do the binding move with enough order to be seen in crystallography, the ions that they bind as precipitates should (often) be seen as well. In fact, ions are not often seen in x-ray crystallography, implying that the ions are not precipitates.

It seems to me that a single number—the density of ions in a channel, selectivity filter, or active site—tells us where to start our thinking. It provides an important starting point for any discussion of selectivity in channels, or indeed for a discussion of the physical chemistry of channels and enzymes, and binding proteins.

For example, enormous number densities in active sites and ion channels immediately imply that traditional treatments of enzyme kinetics[234] and channel permeation[415,420] are suspect. It is clear that number densities in channels exceeding those of solid salts—46 molar for oxygens or $Na^+$ in an EEEE calcium channel vs. 37 m for solid NaCl—cannot be approximated—in any normal sense of that word—as ideal chemical species. They do not resemble uncharged, infinitely dilute perfect gases of uncharged atoms with negligible interactions because they have negligible densities. Traditional treatments of enzyme kinetics and channel permeation do not deal with the finite volume of ions and hardly deal with their electric field. Force fields of molecular dynamics simulations are almost always calculated and calibrated in ideal environments very different from the ionic liquids found in channels or near active sites. Errors must result.

The active sites of channels and enzymes and binding proteins are very special, very highly charged places, in which ions and atoms are severely crowded. They are a kind of ionic liquid, not an ideal gas.[134,519,775,921] They have little room for much water, and certainly have excess free energy (beyond the ideal) that dominates their properties. Ions in active sites resemble an ionic liquid, like a molten salt at room temperature (with constrained counterions, namely side chains) more than they resemble a perfect gas.

**Crowded charge as a biological adaptation**. The above physical and chemical view of active sites can be supplemented by a powerful biological perspective. When a biologist, particularly an evolutionary biologist, sees a structure or physical property that is rarely or never seen in the physical world, she/he knows that it is likely to be part of the adaptation evolution has made to solve a problem of natural selection, as documented in many examples in the essays of SJ Gould, for example,[366-369]

Many non-biologists are skeptical of teleological arguments of this sort, because they seem like a meta-biology, more akin to metaphysics than to hard science. Reading the extensive literature of evolutionary science should change this view. There should be little barrier to that literature, given the magnificent series of essays available in the semi-popular scientific press[366-369]. Even without such reading, skeptics would admit, I think, that the initial guess—the working hypothesis for investigation—should start with the unusual feature of the biological system, rather than assume its opposite. One should start a treatment of active sites and ion channels from their unusually high density of charge. One should guess that the special properties of ions in high density have been used by evolution to make the special properties of the channel or protein. One should obviously not start by assuming properties that are the opposite extreme from what are found. Close packed ions do not resemble infinitely dilute gases. Calculations without definite ion concentrations seem a poor way to approximate a system which has a specialized adaptation of enormous ion concentrations at its active sites.

The efficiency of the scientific process, like that of any investigation of the unknown, depends on the quality of the initial guess. If the guess about the mechanism is entirely wrong, it will be hard to correct, particularly given the human tendency to elaborate the initial guess, rather than replace it. If the initial guess is good enough, the scientific process rapidly converges, much as the solution of an inverse problem proceeds rapidly if one knows where to start. So starting with the idea of crowded ions seems a good idea to me.

I believe that investigations of mechanisms of selectivity, and active sites of proteins in general, should begin with the investigation of ions at enormous density mixed with side chains. That is what our reduced models of the selectivity filter do, although I am not sure Wolfgang Nonner and I were fully conscious of that when we started. That is where I think we should begin, adding structural details as they prove important in determining the forces and energetics of the system and model.

**Necessity of Calibration**. When we add structural details, however, we must be sure that we retain the essential features of the reduced model that ensure its success. We must be sure that more detailed calculations actually reproduce the properties of simple models that are important for the success of the simple models. The necessity of such calibration is obvious in the laboratory. Before one does new experiments, one must show one can repeat the old ones, if they are relevant. This habit of calibration and step by step extension of results is essential to the success of experimental science. It will be essential to the future success of computational biology, I believe, a view shared by many in computer science[705] if not by students used to running simulations without calibrating the programs that run them.

**Balanced Forces and Structures in Crowded Systems**. The key properties of systems with so much crowded charge is the competition between charge and space. The electrical forces are enormous when charges are at this density; but so are the repulsion forces that arise because ions and atoms cannot overlap. (More precisely their electron orbitals cannot overlap in the kind of systems we are dealing with here because of electrostatic repulsion and Pauli exclusion, as long as covalent bond formation is not involved. The systems of interest here do not involve formation of covalent bonds.)

The initial guess in the "guess and check" process should be that the properties of crowded ions dominate the properties of active sites and selectivity filters.

In such systems, forces and potentials will be very sensitive to conditions. The smallest change in position of a charge has a huge effect on the forces on a charge 1Å away, and charges within a selectivity filter, or active site are nearly this close. Systems like this with balanced large forces are well designed for control. Small changes in the balance can make large changes in the results. Thus, systems like this are widely used in engineering, where control is often the key requirement, more important than efficiency or even performance. It is the same in biology, I suspect.

**The structure of forces in proteins**. Systems with large balanced forces depend on details of the forces and the structure supporting the forces. Systems like this depend on details of boundary conditions and have very different behaviors for different balances of forces on the boundary or for different structures of boundaries. Systems like this do not have many general behaviors that are true in a range of conditions. Thus systems like this are unattractive to mathematicians and theoretical physicists seeking generality. Even so, we must dive into biological detail with the enthusiasm of a zoologist and structural biologist so we can behave like a bio-engineer and understand how these systems work. The potential profiles in these systems are outputs sensitive to the balance of forces. They must be computed, not assumed. The spatial distribution of atoms in these systems are outputs sensitive to the balance of forces. They must be computed not assumed. Both the spatial profiles of (chemical and electrical) potentials—that determine the forces on mobile ions—and the spatial profiles of locations of atoms are needed to determine the flux through the channel. The biological function depends on both the structure of the forces and the structure of the atoms. Both have important spatial structure. In ideal cases, the spatial distribution of location of atoms can be seen by x-ray crystallography. The spatial distribution of forces cannot be observed by any method I know of. Structural biology concerns itself with only the spatial distribution of location. Biophysics, like physiology and engineering, must concern itself with both the spatial distribution of location and the spatial distribution of forces. **The conformation of forces is as important as the conformation of the atoms of a protein**.

Models in which potential profiles are assumed or assumed to be independent of experimental conditions are not likely to be useful because actual profiles vary in the real world. Keeping the profiles constant in models requires introducing artificial sources of charge and energy not present in the real world. Such models are unlikely to deal correctly with the real world. The scientist using them is likely to need to introduce new complexities in the potential profile (i.e., new states and rate constants) every time she/he changes experimental conditions. (In my view, this is the likely origin of the enormous number of states used in models of channel gating.) Or she or he is likely to simply ignore the experimental conditions in the hope that others will forget their importance. (In my view, this is the likely origin of the practice of ignoring experimental conditions when calibrating force fields in molecular dynamics.)

We will return to this issue again under the name 'the Dirichlet disaster' because it is the key, in my view, to why so many classical models fail. Classical models often inadvertently specify potential within a system, and keep them constant as conditions change, when those potentials in the real world change as conditions change. By keeping potentials artificially constant, classical models introduce severe artifacts that often prevent the understanding of systems, I fear. The disaster would be much less common I think if we realize that when we talk of potentials in systems, or potentials of mean force, we are really specifying the forces in the system. I shudder at the idea of keeping forces fixed as conditions change. I find the idea of keeping a potential fixed as conditions change much more abstract. Both are of course actually equivalent.

Rates of chemical reactions, or diffusive crossing of large barriers, often depend exponentially on energy. In that exponential scale, the effects of the visually dominant details of structure of 1 $k_BT$ or less would be hard to resolve. Perhaps that is why reduced models with so little structural detail are reasonably successful. They deal with the large free energies $(5-10\ k_BT)$ of crowded ions quite well and the structural details they do not include have free energies that are much smaller than that.

**Importance of structures.** The availability of crystal structures of proteins has importance that cannot be over emphasized. The structures observed in x-ray crystallography represent the basic architecture of the systems. Without knowing the basic structure, one can often not make intelligent guesses about mechanism so the scientific process of guess and check does not converge to a useful result in such cases. It is the crystal structure that reveals the basic architecture of the system, not the sequence of amino acids. The sequence of amino acids may reveal the evolutionary control of the system (e.g., four glutamates producing calcium selectivity) and even the evolutionary control of the crystal structure (the characteristic sequences that produce alpha helices).

But observing the basic structure is not the same as observing the energetics. Indeed, if the basic structure reveals enormous densities of permanent charge, as in active sites and selectivity filters containing acid and basic side chains, the structural analysis implies that electrical and excluded volume forces of crowded ions will dominate the system. These are more or less invisible in the structure, so in this case ***the structural analysis points to the irrelevance of its own structural details***—i.e., the irrelevance of the details seen in x-ray structures—while providing evidence about what is important, namely the free energy of the unseen ions and the (often seen) charged side chains.

The impressive detail of magnificent structures, however, does not automatically reveal the free energies and forces that determine mechanism. In crowded systems, energies and forces determined will be typically $5-10$ times the thermal energy[102,664,665,668,669], i.e., $5-10\ k_BT$ where $k_B$ is the Boltzmann constant and $T$ is the absolute temperature. The entropic effects associated with particular locations of atoms and side chains are unlikely to have free energies this large, except in very special cases. They are more likely to be $0.1-1.0\ k_BT$. The structural details which are so evident in x-ray structures of proteins are likely to be associated with much smaller free energies and forces than the structures of crowded charge. The free energies of crowded ions are likely to be much larger than the free energies associated with the details of protein structure.

Indeed, the structures important to function may be so sensitive to conditions that they are more or less impossible to observe by available structural methods. We have examples of this every day in our technology. The structure of an amplifier is complex and impressive even on the macroscopic scale. The circuit diagrams are immensely complex, and every detail is worked on by an army of engineers to optimize performance, reliability and profit. The amplifiers and digital circuits of our technology all are built in crystals of silicon, with insulating layers of $SiO_2$ (that have a role like that of lipid membranes), punctuated by the channels of FETs (field effect transistors) in which charged particles (holes and electrons) flow across an otherwise impermeable barrier. The crystal structure of an amplifier or digital circuit is extraordinarily complex and beautiful and important, just as is the structure of a protein. **Yet the electrons and holes that make an amplifier work are more or less invisible to structural analysis.[237] The only relevant structure is the distribution of doping (permanent charge).** Of course, proteins are not semiconductors; their charge carriers have more reality and solidity than the quasi-particles, holes and electrons. The charge carriers of life are ions with diameter and permanence. The diffusion terms in proteins are very much more important than in semiconductors (at least at the frequencies we use them). Thus, the structure of proteins must be known in detail to have full knowledge of their function. But knowing the structure is not enough. Knowing the forces is also needed. Neither knowing the structure, nor knowing the forces is enough. One must have a reduced model that deals with both structure and forces to be able to calculate and thus understand the function of proteins, in my view.

Reduced models are the essence of engineering. Most of the properties of an amplifier can in fact be described by a single number (the gain), and almost all its properties can be described by a handful of numbers, that summarize its input impedance and output properties, its output impedance and ability to deliver current (e.g., slew rate, fan out, or some such).

In a similar way, the structures of side chains and ions that form the active sites in our models of calcium channels would be invisible to ordinary structural analysis. But they probably can be calculated. Calculations with Metropolis Monte Carlo methods can estimate the structure of the side chains in reduced models as outputs. And similar methods may be applied successfully to fully resolved structures of enzymes and binding proteins[65,66,225,226,505,506,628,809] some day. So much is known of some structures, and some are so important to biology and medicine (e.g., thrombin[120,125,228-230,338,607,684-686]), that creating successful reduced models of these soluble proteins would have dramatic practical effects, I believe.

**Some Structural Properties are so Important that they Cannot be Observed.** In crowded conditions, small changes in the location of the ions in this structure produce large changes in forces and energies, because the ions are so crowded. The ions cannot move without 'hitting' each other. This means excluded volume forces are enormous. Any movement of charge produces large changes in electric forces. Thus, the structure of these systems is a sensitive function of the conditions of the system.

The side chains (like the other ions) in Monte Carlo calculation of a reduced model are always in a Boltzmann distribution with a distribution of location (both mean and distribution of location) that is an output of the calculation. In other words, the side chains form a self-organized structure. That is to say details of the structure are determined by the forces between the atoms that create the Boltzmann distribution of their locations and velocities, and their energies and entropies. **These structures may be so sensitive that they have to be computed, and cannot be observed, just as the structure of forces in a protein must be computed because they cannot be observed.**

**Self-organized structures**. The structure of the protein is self-organized by the Boltzmann distribution. Self-organized means that all the atoms are in a Boltzmann distribution with both entropy and energy at their 'optimal' values.

Locations are not fixed in proteins, no matter how fixed they seem in the beautiful images of crystals published weekly in **Science** and **Nature**. Locations of atoms are not held at the positions seen in crystallography. There are no magic forces to hold atoms at the locations seen in crystallography. Atoms are in the locations seen in crystallography because those locations minimize the free energy ***under those conditions.*** Under different conditions, different locations minimize the free energy. The conditions of crystallization and x-ray crystallography are rather extreme. Temperatures are very low, to reduce entropy and to make the crystal 'strong enough' (i.e., with little enough entropy I suspect) to withstand radiation. The 'mother liquor' from which crystals are formed is not a physiological solution, but in fact is a peculiar cocktail of ingredients not always stated in full detail in publications. The ionic environment around the proteins themselves is not the same as the mother liquor. Thus, the locations of atoms that minimize free energy in x-ray crystallography are different from the locations of atoms in the real protein; the entropy of the atoms is likely to be even more different. When the ions surrounding a protein are changed, on one side of a channel, or the other, or the electrical potential across the channel is changed, the free energy profiles are changed, and all atoms will be in different positions with different entropies. In one phrase, the structure is self-organized and different in different conditions.

For these reasons, it is useful to think of the structure of the protein side chains as induced. The structure of the protein side chains and the location of the ions are both induced by each other. The binding site—of ions and protein side chains—has a structure induced by the experimental conditions including the concentrations of ions in the surrounding baths. The structure of the binding site is self-organized and is induced in the same sense that the structure of the ions are induced. All are in a Boltzmann distribution.

The Metropolis Monte Carlo method produces an ensemble of structures which form the self organized systems that so many physicists discussed as they turned to biology soon after the second world war. The Metropolis Monte Carlo method lends quantitative specificity to the qualitative idea of induced fit and self-organized structure. Indeed, both the dispersion and the mean location of side chains and ions in these structures are important.

Self-organized structures automatically have an induced fit between their components. The side chains fit to the substrates, and the substrates fit to the side chains, with mean locations and dispersions of the Boltzmann distribution. The self organized structures of these models are outputs of the Monte Carlo simulations. They vary with conditions. The variation in energy and entropy (roughly speaking mean and dispersion of location) are both important.

**Limitations of crystallography.** Structures of proteins and ions change with conditions so that measurements of them would have to be made under the conditions in which they function if they are to show the structures that perform the function. This is a difficult requirement, more or less incompatible with x-ray crystallography as we know it. Crystal formation requires special conditions, with special salts, detergents, and other conditions. The free energy of ions near the protein active sites are unknown in crystals. The number density and free energy of ions near the protein active sites are usually unknown in crystals. It is not reasonable to expect structures that are sensitive to experimental condition in simulations to lose that sensitivity in the laboratory even when they are in crystals at 100K.

None of this is to deny the enormous importance of structure for what it is: the average distribution of ions under special conditions. Models must certainly be compatible with those structures ***when the models are computed under the conditions in which the structures are measured.*** But structures measured in one (unknown) ionic condition cannot show how the structure changes under other conditions, let alone what the energetics and forces are in either case.

A further point needs making, judging from experience with students and colleagues, even if it seems obvious. Most measurements of structure are made at 100K, where disorder is much less than at biological temperatures. The entropic contribution to free energy is proportional to the absolute temperature: it is roughly speaking $T\Delta S$. The difference in the entropic contribution at 100K and 300K is substantial[382] and would have enormous effects in any calculation of rates that depend on the exponent of the free energy (see[383] and p. 1991 of ref [664] and Fig. 11 of ref[103]). Of course, the structures observed at 100K are not artifacts. Of course, many of the structures are visible at 300K. Of course, cold temperatures are used to minimize experimental artifacts (like crystal damage). Of course, it is also true that ions found in specific locations at 100K cannot be assumed to be there at 300K.[383] Ions found at specific locations at 100K may be there simply because the entropy at that temperature is too low to move them to where they normally reside and function. They may be 'salted out' by the cold, to use some old fashioned language.

In my view, detailed physical analysis is required along with structural analysis, if selectivity and protein function are to be understood. Computations in atomic detail starting with the dynamics of all atoms may someday provide such physical analysis, but they can only do that after the computations are calibrated to be sure they estimate macroscopic variables correctly. We know that macroscopic 'thermodynamic' variables describe many biological functions essential to life. Concentrations of ions and average electrical potentials determine whether nerves conduct, muscles contract, and patients live. Computations in atomic detail must be able to compute these variables if they are to deal with the biological functions they describe.

Perhaps it is best to view the functionally relevant 'structure of a protein', as well as the structure of forces in a protein, as the computed consequence of a model, constrained by conventional crystallographic and functional data.

**Inverse Methods and Selectivity Models.** Determining the mechanism of selectivity is an example of "reverse engineering". We wish to determine what is inside the black box of the selectivity filter from measurements taken outside. Inverse problems of this sort are notoriously difficult, and involve a set of mathematical problems[304,376,493] quite different from forward problems. In general, the central issue in reverse engineering or inverse problems is doing the right measurement. Inverse problems are usually ill-posed, without enough data present to give unique solutions, and often show extreme sensitivity to error. When a desired result (e.g., the energetics of calcium ions) is insensitive to the measurement, the measurement is not worth doing. When the desired result is too sensitive to the measurement, the measurement cannot be made well enough (with little enough systematic and random error) to give reliable estimates of the parameters of interest. It is the goal of the experimental scientist to make measurements of variables that allow the appropriate sensitivity, so the parameters of interest can be determined.

Despite their difficulties, inverse problems can be solved and are done so routinely in industry,[303] where they are an important tool in process design. The key is always to have enough data of the right kind. The inverse problem for the distribution of charge within a channel has been solved with these industrial methods,[118] using a PNP-DFT theory applied to the reduced model described here. No *ad hoc* assumptions were needed. Calculations done with added noise and systematic error were quite robust because of the large amount of data available from a range of ionic conditions, and the high signal to noise ratio of the measurements of current voltage relations. In fact, the numerical problem was that of (far) too much data (for computers at that time) and not too little! Thus, one can have confidence that reduced models can be built that can be tested experimentally. This is one of the few problems in channel biology or perhaps molecular biology in which a formal mathematical treatment of the inverse problem has been made.

The key to solving the inverse problem in channels was to have a large number of measurements in many different solutions of different concentrations of different types of ions. Reduced models allow calculations of channel properties under this range of conditions. So far, high resolution simulations have not been done in a range of concentrations. Indeed, many are done with an ill defined concentration of ions with uncalibrated free energy. (That is to say, the activity of the ions is not known.) Reduced models are needed (so far) if calculations are to be done under the wide range of conditions needed to solve inverse problems.

The need to solve the inverse problem should not be viewed as a mathematical nicety. If the inverse problem cannot be solved, different investigators will not be able to distinguish their different models of how selectivity occurs. A search of the literature concerning the selectivity of $K^+$ channels will show an enormous diversity of explanations, sometimes with more than one explanation per scientist or research group. I only cite a fraction of the enormous literature here.[9,16,17,30,50,70,73,74,86,109,111,112,119,164,169,188-190,204,223,233,236-238,240,242,251,296,297,318,370-372,379,389,416,473,496,525,527,529,543,592,605,633,636,641,654,657,670,672,674,675,719,720,750-757,760,779,780,782,792,812,813,865,870,882,890,893-896,933,936-938,947822] Because these explanations do not actually deal with experimental data over a range of concentrations of different types of ions, they cannot be told apart (in my view). In my view, calculations producing results that cannot be distinguished have limited utility. Reduced models are needed so far to do calculations in a large enough set of conditions that one can tell one explanation from another.

I argue again that reduced models have a general role in biology. Without reduced models, discussion is ill-posed. Different models cannot be distinguished. One cannot tell one idea from another.

I argue that biology can be analyzed by a series of reduced models, each with a complexity appropriate for the biological question be asked, each appropriate for the scale on which evolution has built that biological function. I argue in fact that the range of scales involved in biology forces us to use a sequence of reduced models. Inverse methods can help us choose these models intelligently. Variational methods allow us to compound models on different scales into one overall theory, at least in principle. Both methods are discussed later in this paper in a general context.

Practical scientists are not impressed by the theory of inverse problems, understandably enough, since successful biologists are often people who wish to test all ideas themselves without the help of abstractions of mathematics or abstracted mathematicians. So now we move on to the specific case of calcium channels. Practical scientists can judge the utility of reduced models in everyday experimental work. Calcium channels can actually be built using the design principles of these models.[629,630,906]

**Building calcium channels.** Miedema and his colleagues at BioMade Corporation (Groningen, Netherlands and Rush University Chicago) took a bacterial protein OmpF[200,227,482,786,866,891] (in the outer membrane of *Escherichia coli)* and systematically mutated and modified it to make a calcium selective channel (Fig. 6).

Many biologists speak of selectivity as a global property arising from the entire structure of the channel protein. If that were the case for calcium selectivity, mutating a few amino acids in OmpF would not be expected to produce a calcium selective channel. A bacterial channel shares no properties with eukaryotic calcium channels. The bacterial protein is built on a different plan, using β barrels instead of α helices so it has no structural resemblance. Its gating is very different from eukaryotic channels. It is obviously built to survive the environmental stresses faced by *E. coli* that thrive in the intestinal tract, survive the acids of the stomach (of industrial strength, pH 1), and drying on the ground. On the other hand, calcium channels of the eukaryotes are in highly protected environments, maintained by the homeostatic mechanisms of mammalian life[110]. Small changes in the ionic environment can have large effects on eukaryotic channels, whereas large changes in environment have little effect on outer membrane proteins of *E. coli.*

Miedema found however that placing glutamates in the constriction zone of OmpF made a quite selective calcium channel.[629] If glutathione derivatives were used to fill the constriction zone further, following a suggestion of Professor George Robillard, very selective channels indeed were created.[630,906] Fig. 6 is redrawn from that work and shows that two different mutants combined with glutathione derivatives produce channels nearly perfectly selective to calcium over chloride and reasonably selective to calcium over sodium. The full selectivity of eukaryotic calcium channels was not reproduced, however, and this is not surprising because homology models of the channel show that the volume of the selectivity filter is much larger in the mutated OmpF than in the wild type eukaryotic calcium channel. The theory does not predict a fully selective calcium channel because the ions are not crowded enough.[906]

The success of these experiments suggest that the reduced model is 'on the right track'. But OmpF is a hard protein to work with. It comes in a "three pack" (a trimer of channels), and its gating is hard to control. Further work on other proteins closely related to OmpF is under way to see if a fully selective calcium channel can be built and if one can be modified into a sodium channel.

**Mutations of Channels**. We turn back now to computations of specific systems, in this case the selectivity of the sodium channel. It turns out that the selectivity of calcium and sodium channels is interchangeable, in some sense. Mutations can change one into the other by changing the amino acids in the active site. The question is, can the models deal with this reality without adding arbitrary complexity?

When mutation experiments of this sort were first proposed, many objected, on the logical grounds that a mutation of one amino acid into another is likely to produce confusing, un-interpretable results. The structure of the protein is likely to change and a variety of interacting physical forces will surely change. Comparing properties of mutated and un-mutated proteins would be comparing different proteins. Results were expected to be uninterpretable.

These objections made good physical sense. After all, changing one transistor type for another in a digital circuit is not generally a good strategy for modifying the circuit in a defined way. Confusion is likely to occur when trying to interpret the results of such a swap, if any results can be found at all: in most cases, switching transistors will stop function altogether, and there will be no properties of the computer to study. The computer will not compute, after most modifications, so the mechanism of its amplification cannot be studied at all.

Fortunately, these physical objections did not prevent biologists from forging ahead and doing the experiments anyway. Good experimentalists like to test everything they can, and then deal with the confusion. This strategy often succeeds in biology, when it would fail in an engineering system, because the design principles of biological systems are often quite simple. Biological systems use design principles chosen by evolution and these adaptations can be simple: biological systems have to work or they die.

Indeed, an enormous number of site directed mutations have been made, looking for specific signatures in the amino acid sequence of proteins that determined important biological function. Many were found. Many results were confusing but many were not. In a very real sense, these signatures are the genomic adaptation that form the blueprint for the physical adaptation. We see this in channels, where the function of many channels is known to be determined by small sequences of amino acids.[4,174,179-182,415,443]

In particular, calcium channels were found to be[300,785,927,931] defined by their signature sequence of amino acids EEEE Glu-Glu-Glu-Glu. The negatively charged carboxyl (COO-) side chains of these glutamates are known to extend into the selectivity filter region of the channel.[517] Two calciums (or four sodiums) can balance the permanent negative charge of this active site, and so the electric field is likely to be more or less neutralized inside the channel.

Sodium channels are defined by their signature sequence DEKA Asp-Glu-Lys-Ala with a very different charge distribution: the net charge is −1 but the DEKA active site is very 'salty': it has two negatives and one positive (two acidic amino-acids and one basic amino-acid). A calcium cannot balance the net permanent charge of this channel but a combination of anion and cations can. It seems likely that the permanent charge of DEKA is not neutralized strictly by the contents of the selectivity filter, and the electric field outside the selectivity filter is of more importance than in the calcium channel.

**Mutations of models**. Site directed mutagenesis is most helpful in comparing sodium and calcium channels. Amazingly, calcium channels can be mutated into decent sodium channels and vice versa simply by changing the key amino acids[397,789,849]. A sodium channel can be mutated to give the 'titration curve' typical of a calcium channel[20,21,210,300,397,409,410,523,541,545,546,579,615-620,785,789,849,871-873,931,946] although with reduced selectivity. There are several comprehensive reviews in the literature[579,785] that are most useful summaries of the mutation work.

The question then is what happens if the reduced model of the calcium channel we have just discussed is mutated into a sodium channel? Specifically, what happens if we switch the EEEE amino acid side chains for DEKA? The answer is rather remarkable. If the appropriate mutation is made in the active site in the model, the calcium channel becomes a sodium channel, ***without changing any parameters of the model at all***[103].

A little historical detail is needed so our latest simulations[103] are in proper context. The sodium channel DEKA was studied[95] as soon as simulations[96,100] of reduced models reproduced the main features of the calcium channel. (Previous calculations were done with crude analytical models[664,668,669].)

The results of the initial simulations[96,100] were disappointing. Indeed, the mutation reproduced the switch from $Ca^{2+}$ to $Na^+$ selectivity, but the other properties of the DEKA channel were not satisfactorily reproduced. The crucial biological properties of the $Na^+$ channel were not found in the initial simulations. It was easy to reproduce the switch in selectivity from $Ca^{2+}$ to $Na^+$ but it was not easy to make the simulated $Na^+$ channel selective for $Na^+$ vs $K^+$. This disappointment was hardly a surprise, since changes in active sites involving such a large change of charge seemed likely to change the structure and other properties of the channel.

The real sodium channel is quite selective for $Na^+$ vs. $K^+$ and this selectivity is of the greatest importance for its function[415,417,425,430,432,436,438] although $Na^+/K^+$ selectivity was not found in the simulations. One could have adjusted parameters, following the traditions of molecular dynamics and quantum chemical calculations of selectivity, or ignored the $Na^+/K^+$ issue, but my colleague Wolfgang Nonner and I felt that would miss the main biological point. The sodium channel is all about $Na^+/K^+$ selectivity, so we decided to estimate the extra (unsimulated) free energy needed to produce the selectivity[355] that was not found in the raw simulations. I think it fair to say that Nonner and I felt we could not solve the problem without knowing an x-ray structure.

We proved to be wrong thanks to the diligent work of our colleagues who were slowly improving the model of the calcium channel. It turns out that improved reduced models are able to reproduce all the important properties of both the calcium and sodium channels, without changing parameters and without knowing an x-ray structure. Evidently, a switch in the amino acids of the active site is enough to perform this 'miracle' even without changing the model, without even changing the diameter of the channel or the dielectric constant of the protein. This result is quite surprising since it seems hard to imagine such a drastic change in an active site that preserves the diameter. Nonetheless, the simulation results fit experimental results on both calcium and sodium channel properties, with the same parameters, the only change being the amino acid side chains themselves.

The improvement in the model that allowed this to happen was the treatment of dielectric boundary conditions. An important stumbling block had been the dielectric boundary condition needed to describe the different mobility of induced (i.e., polarization) charge inside the channel protein and in the pore of the channel itself. Guesstimates of the energies[649] produced by this dielectric boundary condition suggested that it would not dominate the properties of the channel, simply because the charge induced in the dielectric was smaller and further away from the ions than many of the charges on the side chains. The calculation itself was also difficult because the field needs to be computed accurately close to the dielectric boundary (where the field is very large) and also far away from the dielectric boundary (where the field is small but there are many ions). There are so many ions far away from the channel that the total energy is significant even though the field itself has much smaller magnitude at that distance.

Rosalind Allen, working with J.-P. Hansen[14], introduced a variational approach to this problem that showed how to solve this problem. Gillespie [98,108] substantially improved the Allen algorithm and showed it was quite accurate when programmed so that it could deal with curved boundaries. Much work has been done on this issue[22,39-41] and more remains to be done. Computation of the forces produced by polarization charge is a challenge at all levels of resolution, from macroscopic, to reduced models, to molecular dynamics, to quantum dynamics.

**Dielectric miracle: $Na^+$ vs $K^+$ selectivity**. When the dielectric boundary condition was applied to the original simulation, something quite remarkable happened. Even though the dielectric energy did not dominate the problem, introducing the dielectric boundary had a profound effect that we did not anticipate: when the dielectric energy was introduced, the same model with the same parameters accounted for the very different properties of the calcium and sodium channel in a wide range of solutions and concentrations. Indeed, substantially all of the selectivity properties of the L type calcium channel and the sodium channel were produced by one model, using crystal radii of ions, that were never changed, and one dielectric constant for the protein and one for the ionic solution, in a pore of 6 Å diameter.

This result was not anticipated for many reasons. One was that it differed so dramatically from results using molecular dynamics and quantum dynamics on the K channel, where artificial radii differing from crystal radii and other "tuning"[895]—easily justified by the complexity of the system and the ambiguity of the high resolution calculations—were needed to produce a reasonable free energy of binding.[70-72,74,111,113,238,318,379,674,675,752,759,893-896,936] It is not known how well these high resolution simulations deal with real experimental results because they are performed with force fields calibrated in ideal solutions (of zero ionic content) and are actually performed in solutions of one ionic composition. The activity of the ions in that one solution is not known. The inability of molecular and quantum dynamics to calculate selectivity of K channels (in a variety of realistic solutions) was not a surprise. The difficulties of simulating free energies of ions in bulk solution with molecular dynamics are well known to us (see the references in [943] and for example [2,47,309,319,328,329,390,453,487,490,492,494,502,518,530,547,580,693,698,700,735,764,904]). It does not seem constructive to go on at length with this criticism. The fundamental properties of ionic solutions are their free energies per mole. These cannot be simulated or computed in reasonably concentrated solutions or in mixtures at all, according to the leading workers in the field. See discussion and references in [47,309,319,530,531,547,904].

What is surprising is that a model as crude as ours could succeed then in computing the binding of ions in two types of channels. Similar models do even better and can compute current voltage relations as well in a closely related channel.[348-353,356] Our model does not have any preformed structure and we thought that details of the crystal structure would be very important. After all, the general view, from which we were certainly not immune, was that details of the atomic arrangement of ions and side chains would be important in determining selectivity and those details could not be determined without x-ray structure.

What is so striking is that the properties computed without knowledge of crystal structure could do so well.[103] Let us review these results before we seek to explain them.

1) The calculations were done for an EEEA and DEKA calcium and sodium channel to keep the model as close to experiments as possible. (Separate work shows that simulations of EEEE channel have the full selectivity expected.[102])

2) Crystal diameters were used $Ca^{2+}$ 1.98 Å, $Na^{+}$ 2.00 Å, $K^{+}$ 2.66 Å and side chains were represented as 2.8 Å spheres (glutamate and aspartate), 3 Å spheres (lysine), or ignored (alanine). The channel diameter was 6 Å. Note that the diameter of the channel is about twice that of the ions. The image of a tight single file is not appropriate. Sodium channels are known not to have characteristics of single file diffusion[716-718] that are prominent in the classical analysis of K channels[415,420,439] but have historically[415,416,420] been used as a presumptive signature of all channels[415], inappropriately in my view. The dimensions of the channel rather seem to guarantee large crowding effects between ions and side chains, as well as between ions and ions, and ions and the rest of the protein. Everything interacts with everything else. Everything is involved in structures of this type. No one effect dominates. Single file features of uncharged balls in narrow filled plastic tubes, that Hodgkin and Keynes[439] had in mind, along with others[415,416,420], are not an appropriate metaphor to analyze ions crowded into channels moving on time scales of femtoseconds that carry currents on time scales of micro to milli to seconds to even minutes (in measurements of unidirectional fluxes).

3) The water dielectric coefficient was around 78 and the protein around 10. It should clearly be understood that the dielectric coefficient in the pore and the bulk were the same in these calculations even though most biologists—including Wolfgang Nonner and I—would expect a different dielectric coefficient in the pore from that in the bath. Computational difficulties are serious in such a three dielectric model. Dezső Boda has recently overcome those and is in the process of writing a series of papers about the resulting effects. At this stage, it is clear that the conclusions discussed here are reinforced, not weakened, by his results to date.

4) The main selectivity sequences found in natural channels of monovalent and divalent cations are computed without changing any parameters for both the DEEA $Ca^{2+}$ channel and the DEKA $Na^{+}$ channel.

5) The qualitative properties of the DEEA binding and the DEKA binding are correctly simulated by the reduced model.

6) There is no simple physical explanation **for the range** of binding phenomena. No simple explanation would be expected because a number of terms are involved in the binding free energy and these are all of the same approximate size. Each term varies with concentration of all types of ions, because this system is highly nonideal. (A characteristic of all nonideal systems is that everything interacts with everything else, i.e., the free energy of any one component depends on the concentration of every other component, one at a time, see item 2 above). In a situation like this, one cannot expect to rely on a simple verbal explanation of the full range of behavior, because all the terms are about the same size and vary with all conditions.[104,106,348,350,357,664,748]. One can compute that behavior, but one must expect to need to compute, and then understand. This is a familiar reality in applied mathematics, physics and physical chemistry. It will become more and more common, I feel, in computational biology.

Despite the expectation of complexity, however, **simple qualitative explanations are possible for the biologically important** selectivity properties of the $Ca^{2+}$ and $Na^{+}$ channel. The selectivity of the calcium channel for $Ca^{2+}$ vs $Na^{+}$ has a simple explanation. (The selectivity of the sodium channel for $Na^{+}$ vs $K^{+}$ also has a simple explanation we will discuss a little later.)

**$Na^{+}$ vs $Ca^{2+}$ selectivity in the calcium channel**. The selectivity of the calcium channel for calcium over sodium is crucial for its function. If sodium displaces calcium in the calcium channel, the heart would stop, nerves would not work, death would result. Interestingly, in this case, the simulations show that only two components dominate the properties of the channel (of the several possible[664,669]). Sodium ions are much more crowded in the EEEE channel than calcium ions and so they are excluded. Four $Na^{+}$ are needed to balance the charge and four sodiums occupy about twice the volume of two calciums. The theory gives a convincing explanation of why our model produces selectivity of $Na^{+}$ vs $Ca^{2+}$. We cannot be sure the channel works this way, but the model fits a great deal of data, and has a simple physical explanation, so it is tempting to conclude that the explanation of the model is also the explanation of the channel.

The excluded volume term favors calcium over sodium for a simple reason. The charge of the structural oxygens—the glutamate side chains—pins the contents of the channel. The net charge of the channel must be close to zero to keep the voltage from being lethal. (Remember that ~250 mV is a lethal potential inside a channel.) So there must be four $Na^{+}$ in a sodium filled EEEE channel, or two $Ca^{2+}$. Sodium and calcium are nearly the same size. It is clearly energetically much more difficult to crowd four spheres into this space than two, so the number advantage of calcium is very large.

The electrostatics also play an important role here, as they almost always do when divalents and monovalents are both involved. Divalent calcium brings two charges (from its nucleus plus inner shell electrons) to more or less the same distance from a glutamate as monovalent sodium does. Thus, the electric field is much smaller when calcium is present. It is much more effectively screened. This electrostatic shielding term is nearly as large as the excluded volume number advantage caused by the extra crowding of sodium ions.

The crowding and electrostatic energies are so large that the entropic orientational free energies (corresponding to the details of the crystal structure) are unimportant. The structural details that so preoccupied me for many years just do not contribute much energy compared to charges packed into some 300 cubic Angstroms at a number density (in chemical units) of more than 20 molar!

Ions this crowded obviously are in a very special environment. The smallest change in their average location will produce huge changes in crowding (in the excluded volume energy) and in the size and shape of the electric field and thus in the electric energy term.

What was not clear for some time was how a reduced model could deal with this sensitivity. We have already presented the answer. (For the sake of motivation and clarity, I departed from the historical order of things.) The model is self-organized and the fit of ions to each other, and to the side chains and to the channel is determined by a Monte Carlo simulation that automatically adjusts the location so the ensemble of ions has the free energy (both energy and entropy) of a Boltzmann distribution.[274]

**$Na^+$ vs $K^+$ selectivity in sodium channels**. The selectivity of the sodium channel for $Na^+$ vs $K^+$ also has a simple explanation. It turns out that the selectivity arises because of the rejection of $K^+$, not the binding of $Na^+$.

Most scientists who study enzymes, proteins and channels have assumed that selectivity arises in binding because enzymatic chemical reactions usually include a binding step and it seemed natural to include different binding for different ions. Indeed, it is hard to find any other explanation of selectivity or specificity in textbooks or reviews of the biochemistry, channels, or binding enzymes. Our simulations do not include chemical free energies of binding. All binding in our models is the result of crowded charges and electric fields.

We were, then, very surprised to find in our simulations that $Na^+$ vs K selectivity in the DEKA sodium channel was produced by depletion of $K^+$, not binding for $Na^+$. (See Fig. 7.) Biology seems to have chosen to use simple phenomena to determine functionally important properties. $Na^+$ ions are rejected by calcium channels because they collectively occupy too much volume. $K^+$ ions are rejected by sodium channels because they are excluded from a specific depletion zone.

Indeed, binding sites were found in our results—as OUTPUTS of the simulation—but these binding sites were not selective. Rather, $K^+$ was altogether absent from the selectivity filter. The selectivity filter created a depletion zone without $K^+$.

It is important to reiterate again that binding in our model is an output of computation. There is no chemical energy involved. Binding in our model is simply a concentration of ions beyond that in the baths. The energy for binding comes from the balance of forces in the model, namely the hard sphere excluded volume effects and electrostatics. In the computation of these energies, the charges and excluded volumes of all types of ions interact with the charges and excluded volumes of all other ions, in a highly nonideal environment that resembles an ionic liquid more than the ideal gases of most of our educations. Of course, our explanation of binding may be incomplete. Other forces and energies may well be involved in other systems. But in the system we compute, no other forces or energies need be invoked to explain an enormous range of data. Indeed, the fits to data are good enough that there is not room for other energies. Adding additional energy terms would most likely disturb the existing fit to actual experimental data.

The depletion zones that produce the selectivity for $Na^+$ vs $K^+$ were a surprise to me in one sense but not another. I had not anticipated that depletion would be involved in selectivity, but I had anticipated that biology would use depletion zones in one way or another. My colleagues and I had spent a great deal of time in the 1990's[1,34,36,118,137-140,142-144,146,147,149,150,265-271,273,281,283,284,286,288,353,444-447,649,666,668,798,799,827,884,886-888] studying the electrostatic properties of ion channels and their resemblance to semiconductors: the holes of semiconductors resemble the cations of ionic solutions, and the semi-electrons (i.e., the quasi particles with 1 negative charge) of semiconductors resemble the anions of ionic channels. I thus knew of the importance of depletion zones in transistors: depletion zones govern most of the technologically important properties of transistors. And I knew why depletion zones were so useful: when concentrations of charge carriers are small, tiny changes in concentration can have huge effects. Thus, gain is easy to produce and control. On the other hand, in enhancement regions, large changes need to be made to change concentrations significantly, making gain hard to produce and harder to control.

I had expected to find biology using depletion zones, and said so in a number of papers, but I was thinking of electrical properties and not of selectivity. Indeed, in my work quoted on the Poisson Nernst Planck model, I had assumed ions were points (as were holes and electrons), missing the essential importance of the finite size of ions in determining all the nonideal behavior of ions in water and in and near proteins. I was not alone by any means[15,19,124,155-161,163,165-169,171,184-186,191-199,215,216,255,263,343,344,374,470-473,480,481,535,536,559-571,602-604,645,902,903] and even workers who explicitly compared Brownian dynamics of finite diameter particles and PNP of point particles[60,163,166,169,171,184-186,193,196-199,263,343,344,471,472,561,568,569,645,762,820,902] seemed unaware (along with me) of the importance of finite size in classical treatments of ionic solutions (a few of the references are[3,47,256,257,309,319,328,492,495,530,548,660,700,735,762,310,311,402,519,594,688,689,914]) and discrete ion effects[44]. Indeed, differences between Brownian dynamics simulations and PNP probably reflect the finite diameter of ions in the simulations rather than any more sophisticated difficulties in PNP. Thus, I had not anticipated that depletion zones caused by the interactions of the electric field and the competition for space inside a channel were the cause of $Na^+$ selectivity in DEKA sodium channels.

We see, then, that two of the crucial properties of calcium and sodium channels are produced in quite a simple way. The physics of interactions in these systems is very complex. Everything interacts with everything else. But biology seems to have chosen to use simple phenomena to determine functionally important properties. $Na^+$ is rejected by $Ca^{2+}$ channels because it occupies too much volume. $K^+$ is rejected by $Na^+$ channels because they have a specific depletion zone. (So far, we do not have a simple explanation of how the side chains and other features produce this depletion zone.)

It turns out that biology has also used simple parameters—diameter and dielectric coefficient—to control some crucial properties of the sodium channel, namely its contents and its selectivity. This simplification was not expected.

**Control parameters**. In general, when one is dealing with models of phenomena as complex as selectivity, particularly when one is using models as reduced as ours, effects of changing parameter are almost never 'clean'. Changing one parameter changes many outputs of the model.

Our model had only two parameters, dielectric coefficient and diameter, and not surprisingly, changing one or the other had complex effects on almost every property of a EEEE calcium channel.

We were amazed to find quite different behavior in the DEKA sodium channel. Here changing the diameter of the channel had no effect on the occupancy of the channel, i.e., on the number of ions in the channel. The diameter had a huge effect on selectivity, i.e., on the ***ratio*** of $Na^+$ to $K^+$ occupancy, but no effect we could measure on the total occupancy. (Fig. 7)

The dielectric constant (of the protein), on the other hand, had a huge effect on the occupancy of the channel but no effect on its selectivity. (Fig. 8)

In a nonequilibrium extension of the model, one would expect occupancy to determine conductance (to a first approximation under most conditions), and ratios of occupancy to determine selectivity. Thus, it seems that the model has 'orthogonal control parameters': the dielectric coefficient controls conductance; diameter controls selectivity.

It is as if the genome determines the phenotypes of selectivity and conductance by controlling the effective dielectric coefficient and effective diameter. We certainly have not proven this idea, nor is it clear how one would do that. But what is striking is that these orthogonal control parameters emerge as outputs of the simulations, when these properties have not been built into the simulation in any way.

It seems obvious that these control properties are more a result of biology than physics. By this I mean that only in a particular geometry and set of conditions would diameter and dielectric coefficient not interact. It seems as if biology has chosen to use simple strong energies (of crowding and electrostatics) to allow simple control of the biologically important properties of the sodium channel.

This behavior came as more of a surprise to me than to those of my colleagues who are practicing engineers. They know that devices are often designed to be robust and controllable first, and then to have good performance. The sodium channel behaves as if were designed to have robust and controllable selectivity and conductance. It is interesting that sodium channels in different locations in the heart and the brain are 'isozymes' (i.e., are closely related proteins) that have different selectivity and conductance.

**Reduced Model of Transport through a Channel**. A significant success of reduced models is their ability to deal with nonequilibrium phenomena involving fluxes and current flow through open channels. Currents through open channels change as the voltage is changed. Measurements of the current voltage relation are particularly revealing (and stressful) tests of models. It is easy to sweep voltages over a range of energies far wider than the range of energies that can be changed by changing concentrations and chemical potentials.

Voltages can be easily swept from –150 mV to +150 mV which is a range of approximately 12× the thermal energy, $12\,(k_B T / e)$. It is much more difficult experimentally to change concentrations than to change potentials—the experiment takes minutes or even hours compared to seconds, and the possibility of wrecking the experiment is very much larger. In addition, single channel measurements in the patch clamp with solution changing require special methods[857,858]: some of the methods reported in the literature from well known groups are hard to perform, hard to reproduce, and introduce dramatic noise and artifact. It is rarely possible to change concentrations more than a factor of 10× without encountering new phenomena irrelevant to the issues at hand: a 10× change in concentration is roughly a 2× change in thermal energy, $2\,(k_B T/e)$. It is far better to change voltage across channels than concentration when testing models.

The reader should be warned that most of the current voltage or current voltage time recordings in the literature are not directly relevant to the theories described here. The theories described here are for currents through a single channel protein molecule of a single type with controlled voltage and concentration across it. Such measurements are hard to make. Most recordings in the literature are from ensembles of channels, measured in what is often called the 'whole cell recording' using the patch clamp method[384,774] to measure current from whole cells, not from single channels. These measurements are of current through the 'conductances' of many channels, perhaps through many chemically different types of channels that have different structures, functions, and genes.

Most measurements are ensemble measurements that resemble those made with the classical 'space clamp' of Hodgkin and Huxley[437] and Cole's laboratory[864] or with microelectrode measurements from spherical or finite length cells[37,290,293,302,695-697]. Ensemble measurements record currents from channels that are opening and closing and so they involve both gating and permeation through an open channel.

The reader should also be warned that ensemble measurements sometimes, or even often, include currents from many types of channels and so can be nearly impossible to interpret in a unique way. Before the invention of single channel recording[384,632,655,656,774], measurements were almost always made from mixtures of different types of channels.

Measurements of single channels that we analyze with the theories of this paper have the enormous advantage that they separate properties of gating and open channel conductance. They also measure current from just one type of channels, not from a mixture of different types of channels. The reduced models of this paper are comparable only to measurements of currents from single channel molecules of one type. Measurements from ensembles almost always involve opening and closing properties of channels, not described in the theories used here.

Current voltage measurements have been interpreted with rate models for a long time,[415] starting even before single channel recording. These models are subject to serious criticisms discussed in this paper, and in the literature for some time as well[144,185,186,271,284,286,288]. A sufficient reason to reject these models is their inability to predict currents of more than a few tenths of a picoamp[144,186,271,284,286,288] when friction is included in the formulation of the prefactors of the rate model, as it must be in a condensed phase[386] like an ion in a channel[144,184]. Currents through channels are typically larger than 10 pA. Theories must fit experimental results as actually recorded in the laboratory, in the units recorded, not just in normalized units, or reported in ratios.

Physical models have been used to describe current voltage relations of channels for some time involving friction, starting with the work of David Levitt [559-570], as far as I know. Permanent charges in the channel protein were introduced[283] with PNP theory (see historical references that are sampled in[51,286,484]). Three dimensional versions of PNP theory soon followed[124,171,374,445,535,602,604] although numerical difficulties were not put entirely to rest until spectral elements were introduced[445] and tested extensively[444] by Uwe Hollerbach.[444,446,447,649,650]

Attempts to deal with selectivity with PNP were not very successful[34,141-143,231] because they did not include the finite size of ions. This early work[141-143] showed the way to include the excess chemical potential.[145,665] Finite size effects were introduced later using the excess chemical potential.[96,664,669] Recent references to finite size effects can be found in.[102,103,106,600]

The most successful treatment of selectivity and current voltage relations is that of Dirk Gillespie, working with Gerhard Meissner and his group[357,913,929] and Mike Fill.[351,352] Gillespie has described finite size effects in the tradition of the density functional theory DFT of Rosenfeld[307,740,748] (of molecules in fluids). The DFT of molecules in fluids should not be confused with the more widely known density functional theory of electrons[538] (in orbitals). Gillespie has extended Rosenfeld's DFT to deal with ions[354,356] and checked his approach carefully against Monte Carlo simulations of the primitive model.[353,354,356,880] We[353] and then he[348,349,351,352,357] took the excess free energy computed by DFT and added it into the ideal free energy of classical PNP (also see[142] that grew from the previous work of Eisenberg[284,286] and Chen[145]). Gillespie then developed a DFT+PNP theory that has proven remarkably successful in dealing with nonequilibrium data from the ryanodine receptor, as we shall see.

**Reduced Model of a 'Transport' Channel the Ryanodine Receptor**. The ryanodine receptor RyR is the final common pathway for the calcium movement that controls contraction in cardiac and skeletal muscle[301,314] and has been extensively investigated for many years.[91,483,537] We[282] first caught RyR in action in the electron microscope nearly thirty years ago, along with others no doubt.

The function of the ryanodine receptor is to transport large amounts of calcium so it can control the $Ca^{2+}$ concentration in the cytoplasm of the muscle fiber as quickly as possible. The calcium is stored in the large "sack" formed by a membrane called the sarcoplasmic reticulum in muscle fibers. (Indeed, the channel exists in neurons as Rosenbluth saw long ago[738] and is evident in thousands of papers in the modern literature. Many other cells use $Ca^{2+}$ as a controller of function, and the RyR is expected to exist and control function in all these cell types. Biology uses the same motif in many places to do many things, just as a computer designer uses transistors to do many things. The RyR may be involved in Alzheimer's disease as well[613]).

The RyR channel is specialized to pass fluxes of calcium and so it is natural[351] that the channel is not very selective.[351] If it passed only calcium ions, the electrical potential across the channel would rapidly change, the electrical potential in the sarcoplasmic reticulum would approach the chemical potential gradient for calcium. The difference between the electrical and chemical gradient would approach zero, and the flux of $Ca^{2+}$ would cease. Remember that membrane systems in intact functioning cells or organelles are not voltage clamped by external sources or apparatus. Their voltage is free to move and is in fact determined by the relative conductance of the open channels in their membranes.[334,920] The RyR channel is not only permeable to $Ca^{2+}$, but also other cations found inside the cell, most notably $K^+$ and $Mg^{2+}$ which are in fact in (near) equilibrium across the sarcoplasmic reticulum membrane and control the potential (for the most part) independent of calcium flow. The *net* current flow is small, with $K^+$ and $Mg^{2+}$ currents in one direction balancing the $Ca^{2+}$ current in the other direction. It would be interesting to examine the properties of the fluxes through the RyR system in the tradition of classical transporters, instead of the tradition of classical channels. To what extent do the unidirectional fluxes of the RyR system show coupling like the unidirectional fluxes through anti-porters or sym-porters of classical transporter theory? Does the RyR system behave differently from a classical transporter? Is it possible that the RyR channel is (more or less) a transporter in disguise? The Gillespie and Fill model[351] is a complete description so it can be used to compute unidirectional fluxes and answer these questions.

Gillespie and Fill[351] have shown with this discussion why the RyR is not a highly selective calcium channel like the L-type calcium channel discussed previously. In fact, RyR is so poorly $Ca^{2+}$ selective that the membrane potential across the channel is about ~2 mV, set almost entirely by other ions and maintained by large $K^{+}$ and $Mg^{2+}$ countercurrents. This interplay of selectivity, conductance, potential and current may seem strange to those not used to it, but this interplay has been at the classical heart of electrophysiology, and has been textbook material[214,260,261,334,920] since the selective permeability of membranes was recognized by Hodgkin, Huxley and Katz[425,438,458,459] The interplay determines, for example, the electrical signals[291,335] produced by different channels in different membranes[278,279] of skeletal muscle.

Gillespie has constructed a model of the RyR showing how its selectivity might arise from its known molecular properties. The reader is referred to the main reference,[348] including important supplementary information for a complete description of the model, along with the additional papers on the RyR[349,352,357] For our purposes, the model channel can be described as a DEDDE (glutamate, aspartate, glutamate, glutamate, aspartate) channel in which the seven nonequilibrium parameters (the diffusion coefficients for $Li^{+}$, $Na^{+}$, $K^{+}$, $Rb^{+}$, $Cs^{+}$, $Ca^{2+}$ and $Mg^{2+}$ ) were determined by nine data points out of the more than a thousand measured in more than one hundred solutions. The structural parameters and diffusion coefficients were never changed in [348] (see Fig. 10-11).

Gillespie's reduced model fits linear current voltage curves such as those in Fig. 10A, measured from a channel bathed in a single electrolyte. Producing straight lines such as these might be thought trivial, but such is not the case. Straight lines are in fact difficult to produce when the inherent functions and properties of the model are nonlinear. Classical barrier models with exponential flux relations cannot produce linear current voltage relations when large gradients of concentration are present.[415,420] Linear current voltage relations are found in the open channels of a wide variety of channels.[179-182] The first clue that classical barrier models are poor descriptions of channels came from the discrepancy between the curved behavior of their current voltage predictions and the (often) linear behavior of current voltage curves measured from real single channels.

Current voltage relations of the RyR are not always linear. When the RyR channel is placed in mixtures of ions (Fig. 10B), current voltage relations are nonlinear. Nothing needs to be changed to fit the very different properties of the channel in mixed solutions. These nonlinear current voltage relations are fit with the same reduced DEDDE model and same parameters that fit the linear current voltage relations in solutions of a single electrolyte.

Mutations have been made[913,929] that change the charge of the RyR selectivity filter a great deal. The permanent charge density is changed by 13M.[357]. Not surprisingly, the current voltage relations observed change a great deal when these mutations are made, because the permanent (fixed) charge changes a great deal (Fig. 10 and 11). Gillespie's model fits this data quantitatively without changing its parameters at all despite large changes in the permanent charge density of the channel. Evidently this change in charge density does not even change the diameter of the channel very much.

The fit is particularly striking because it has so little error. Remember that a horizontal displacement of some 12 mV in the graphs corresponds to a fraction of the thermal energy, about $0.5\,(k_B T/e)$. Many of the curves shown in the Supplementary Information of reference[348] fit much better than that; a substantial number of points are within 1 mV, $0.04\,(k_B T/e)$. This accuracy should be compared to the errors estimated by the authors of high resolution models of selectivity,[675,750,752,754,758,893,895,896] which are often worse than $2\,(k_B T/e)$ in one dilute solution. The errors are not specified in other concentrations or in mixed solutions of different types of ions at all. (Remember biological function, including that of the RyR channel occurs only in such mixed solutions.) Errors of $2\,(k_B T/e)$ would produce misfits in current voltage curves even in pure solution so large that the graphs would probably not be published. Current voltage relations measured in pure solutions of one type of ion would not be helpful in understanding the function of channels that only function in mixed solutions containing many types of ions.

Sensitivities and errors of this sort also occur in nonbiological contexts when all atom simulations are performed. Computations of the activity coefficients in a range of solutions have large errors, see[319,530,943] and the references cited there. The calibrations of simulations called for by numerical analysts[705]—and users[275] of simulations—are not easy to achieve in practice and has not yet been achieved[530] in molecular dynamics simulations of ions in water despite the progress and hard work of many laboratories.[2,319,453,487,494,502,547,580,904]

The biological channel system has anomalous properties not expected in ideal solutions, such as the anomalous mole fraction effect. The anomalous mole fraction effect AMFE is a nonlinear non-monotonic dependence of conductance on ionic composition. As one ion type is swapped for another—think[348] 125 mM NaCl and 125 mM CsCl being changed to 250 mM NaCl or 250 mM CsCl—the conductance of bulk solution changes monotonically. (In these experiments, the mole fraction of an ion type is changed but not the total concentration of all ions. Hence, the name.) The conductance of channels often changes non-monotonically in such experiments and this is called anomalous for that reason. See references in[665].

Gillespie's model[348,352,357] predicted the anomalous mole fraction effect of the RyR (before the experiments were done, writing to skeptical co-workers) without invoking any obligatory frictional interactions (see Fig. S7 of supplementary material to[348] and Fig. 1 of reference[352] , computed before the experiments were done), showing, as did previous authors,[665]—but now with much more work[349,350,352]—that the AMFE has been misinterpreted in the tens or hundreds of types of channels in which it has been studied.[179-182] The AMFE has been widely, if not universally used in the 20th century channel literature[415,420] as an indicator for single file behavior. This is incorrect. The AMFE can occur in channels without single file interactions, even in large diameter nonbiological pores.[350]

It is important to remind 21st century audiences that the criteria, even definition, of single filing is actually the competitive interaction of unidirectional fluxes,[418,869] not the AMFE. Unidirectional fluxes are also (understandably) confusing for people who have not made such measurements. Few measurements of unidirectional fluxes are made nowadays, and some of the original experiments[205,451] cannot be repeated legally within present laws and regulations, at least in the United States. The AMFE is much easier to understand and measure and is taught early and often to channel biologists. Thus, it is not easy to question and has come to replace the unidirectional flux ratio as an operational definition of a channel. This is unfortunate since the AMFE is not a marker of single file behavior.

In contrast to the AMFE, unidirectional flux rations are quite robust theoretical markers of single file behavior at least in diffusion models[48,49,623,624]. Competition in unidirectional fluxes seems to imply frictional interactions on the atomic scale (although selfconsistent theories of unidirectional fluxes have not yet been investigated, as far as I know). Concomitantly, the definition of transporters is the cooperative interaction of unidirectional fluxes. In this regard, it is important to remember that the classical voltage activated DEKA sodium channel does not show single file behavior in its unidirectional fluxes—measured with painstaking care by Rakowski and coworkers[716,717] in more than a decade of experiments on the squid axon. AMFE's are not correlated to selectivity or other behaviors of ion channels in any simple way.

Gillespie has not yet applied his analysis of the AMFE to the 'mixed alkali' effect of physical chemistry although an outsider like me imagines that the effects are quite similar, at least in some cases. A fraction of the literature of mixed alkali effects can be found by starting with references[23,183,328,329,378,690,735,867,874,925]

Mixed alkali effects seem likely to arise whenever interactions are strong and nonlinear. In my view, it will remain difficult to decide between competing physical explanations[23,167,304,305,352,642,682,805,812,855] of mixed alkali effects with classical approaches because they do not deal with interactions in a natural way or (self) consistent way. The variational approach *EnVarA* discussed later in this paper holds more promise in my view because it deals with interactions consistently without invoking many parameters that are hard to estimate.

**Conclusions and Implications of the Crowded Charge Reduced Model**. It seems clear that the reduced models channels as charged spheres in a small space are quite successful. These simple models using the simplest kind of physical chemistry deal with complex biology. It is also clear that this model is successful for two main reasons: one, it calculates the energies that biology seems to use to produce selectivity in these channel types; two, it calculates a self organized binding site, with an induced fit between ions and side chains and vice versa. The approach of the physicist—'guess and check', then add complexity—seems to have worked.

The implications of this success seem significant. It provides an alternative path to the common approach using molecular dynamics, best shown in the huge literature on the selectivity of the potassium channel already cited. The common approach tries to compute everything, despite the enormous gaps in scales that make this so difficult (see Table 1 below). Clearly, more resolution will be needed than we have used so far, as other channels and other properties of calcium and sodium channels are computed. All those features of the protein are there for a reason. The lysine of the DEKA channel does more than just contribute charge in all likelihood. But as we add resolution, it is important to preserve the features that make the low resolution model successful.

It seems that higher resolution models must share the important features of the reduced model if they are to share this success. Higher resolution models can deal with calcium channels (for example) only if they compute energies with similar properties to those computed by the reduced model under the range of conditions that the calcium channel model succeeds in fitting data. Specifically, a higher resolution model—applied to the same system of spheres—should give results nearly the same as the reduced model in a range of concentrations of $Ca^{2+}$ from $10^{-7}$M to 1M, and a wide range of $Na^{+}$, $K^{+}$, etc. concentrations, and a range of divalent concentrations as well. Just as importantly, higher resolution models should be shown to change selectivity when the side chains are changed from EEEE to DEKA.

So far higher resolution models of molecular dynamics, using elaborate force fields to describe interatomic forces, have not yet been shown to reproduce the results of simple models correctly. If high resolution simulations do not deal with the issues known to be important in the laboratory, it is not clear how higher resolution simulations can deal with real laboratory results. Higher resolution models need to include specific concentrations in the bath as inputs. Higher resolution models must estimate activities of individual ions with reasonable accuracy. Experimentalists know they must estimate activities reasonably accurately even to identify the type of channel they are studying. They know that most of the properties of proteins and channels depend on the concentration of ions. So far higher resolution models of molecular dynamics, using elaborate force fields to describe interatomic forces, have not yet been able to deal with a range of ionic concentrations in mixed solutions. They have not been able to calculate activities over a range of concentrations or in mixtures, or even in pure solutions of divalents like $CaCl_2$. Thus, at this stage we believe that high resolution models of selectivity are not yet ready to be compared with low resolution models, or with experimental data.

Despite all this discussion, it is obvious that higher resolution is needed than the crude reduced models that we have used so far. More details in the structure are certainly involved in some functions of the calcium and sodium channels we have dealt with. More details of the structure may be involved in the main selectivity or binding properties of other channels and proteins. Molecular dynamics with force fields is in fact a kind of reduced model, because the parameters of the model are determined (in large measure) from fits to macroscopic data. The advantage over the reduced models used here is that they include all atoms. The disadvantage is that they do not deal with ions very well in pure solutions, let alone in concentrated mixtures. Perhaps an important difficulty is the choice of macroscopic data. Perhaps when force fields are calibrated against measurements of the activity of ions they will do better.

Until higher resolution models are actually calibrated against simpler models, it will be difficult to compare the two classes of models, however. The tradition of the physical sciences assumes such calibrations are necessary in most cases. Certainly, engineers know that calibration is necessary if their devices and machines are to work or are to improve as complexity is added. Biological scientists perhaps need to learn here from our physical colleagues.

If the models cannot be compared, science as we know it cannot proceed, in my view. The scientific method requires us to be able to choose between models, at least in principle. If we cannot, because our computations are inadequate or our experiments are incomplete, we should do something else, until our technology advances to where we can do something useful. The scientific endeavor of ‘guess and check’ can be viewed as a social process, that is justified if it discovers something useful, or builds something that works. ‘Guess and check’ cannot converge to a useful result (to use mathspeak), if check is impossible. If different models cannot be distinguished, checking them is impossible and science, as I define, it does not work. The need for checking and the need for calibration are essential components of the scientific process I think.

The need for calibration of molecular dynamics thus seems self-evident to some of us[275,705,943]. The issues are not just the force fields, but the numerical difficulties themselves (which we shall describe later). Work actually comparing properties of ions in solution with ions in known physical systems is just beginning,[2,319,487,494,502,530,547,580,764,904], also see references in [453,518,943]. Extending this work to deal with experimental reality is a great challenge but certainly one that can be met as computational size and scientific wisdom increase, hopefully at comparable rates.

**What are the source of problems with high resolution models?** This review is not focused on issues of simulations, but on issues of channels. Nonetheless, it seems important to summarize the main difficulty in actually using the high resolution simulations that are so widely computed today. (References to high resolution simulations of K channel selectivity were given previously.) These simulations seek to compute biological function with atomic resolution. It is clear that biological function is controlled by individual amino acids and even by side chains of those amino acids. Computers are growing in power at an exponential rate. It is natural to try to exploit the growth of computing capability to deal directly with biological problems.

The main difficulty with this approach are problems of scale. These are formidable, as summarized in Table 1. There is little freedom in the choice of scales in Table 1. The reality of atomic motion is known. The reality of biological function is known. They both must be computed if atomic detail simulations of actual biological function are to succeed in their stated goal of being able to deal with experimental and biological reality.

Time scales are set by the time scale of atomic motion and the time scale of biological function. The fastest biological function is that of signaling in the nervous system (if the photon phenomena of vision and photosynthesis are excluded). Signaling occurs in $10^{-4}$s at the fastest. The time scale of atomic motion is set by the time scales of atomic vibrations which occur around $10^{-16}$ s. The ratio of time scales is $10^{12}$.

Spatial scales are more complex. The linear range of dimension starts with the resolution needed to define side chains $10^{-11}$m and reaches to the size of animals, say 100 m. Here we are concerned mostly with cellular processes so we choose a typical cell as our largest scale, say $10^{-5}$ m for a mammalian cell. The ratio of length scales is $10^{7}$. This spatial scale occurs in three dimensions so the ratio of volume scales is $10^{21}$.

Structures must be resolved with at least 0.1% resolution if side chains are to be represented in a given protein. Indeed, higher resolution than this would be desirable. Thus, spatial resolution in three dimensions requires scale ratios of $10^{9}$. It turns out that structures of this much detail cannot be handled in easily available packages of computer software in 2010. Memory bandwidth needs are enormous, since many bytes are needed to describe each of the billion numbers in a highly resolved structure. Graphical processing units may be able to deal with such bandwidths in the near future, with the required double precision arithmetic.

Concentrations must be resolved over an enormous range. Simulations must deal with concentrations of ~50 M if they are to deal with ions in active sites and channels as we have seen. Simulations must also deal with very small concentrations.

**Biological Control by Trace Concentrations of Ions.** Most biological proteins are controlled by trace concentrations of ions because biology uses trace concentrations of ions as signals. These trace concentrations are used as signals to control biological systems just as an accelerator is used to control the speed of a car. Most intracellular proteins are controlled by $10^{-7}$M $Ca^{2+}$ concentrations. Experiments show that $Ca^{2+}$ concentrations outside this range (by a factor of say 20) produce irreversible changes in many proteins. The rates of many enzymes and channels are controlled by the actual activity of $Ca^{2+}$ within this range. Thus simulations must accurately calculate the activity of $Ca^{2+}$ in this range. Many proteins are in fact controlled by much smaller concentrations of hormones, second messengers, cofactors and so on. The lower limit of concentration seems to be around $10^{-11}$M. It should be emphasized that the controls set by these concentrations are essential to most biological function. Failure of this control produces many diseases. Simulations must deal with $10^{-11}$ M if they are to deal with many proteins of great practical importance. Thus, the range of concentrations is $5\times10^{12}$.

The difficulties with small concentrations require a little more discussion. A $10^{-7}$M solution of $Ca^{2+}$ contains 55 moles of water for each $Ca^{2+}$ ion. If $10^3$ ions are needed to estimate properties of $Ca^{2+}$ properly, $55\times10^{10}$ molecules must be calculated, or $1.6\times10^{12}$ atoms. If $10^{-11}$M needs to be calculated, $1.6\times10^{16}$ atoms need to be computed. It seems unlikely that calculations of this size will be practical. They certainly seem unwise. If interactions must be directly computed between ions, the large numbers repeated multiply, because everything interacts with everything else. Multiplying numbers of numbers introduces exponentials and factorial functions to determine the number of states and combinations, as we remember from elementary statistical mechanics, and the numbers of energy terms in fully interacting systems quickly exceed astronomical.

**Table 1**
**Time, Space, and Concentration Scales**

| Variable | Computations | Biology | Ratio |
|---|---|---|---|
| **Time** | $10^{-16}$ sec *Vibrational modes* | $10^{-4}$ sec *Action Potential* | $\mathbf{10^{12}}$ |
| **Space** | $10^{-11}$ m *Side Chains* | $10^{-5}$ m Typical Cell | $\mathbf{10^{6}}$ |
| **Solute Concentration** | – | $10^{-11}$ to $5\times10^{1}$ M | $\mathbf{5\times10^{12}}$ |
| **Volume** | – | – | $\mathbf{10^{18}}$ |
| **Spatial Resolution** | – | – | $\mathbf{10^{9}}$ |
| | | | |

We do not try to estimate the resolution needed to compute the electric field. The electric field used in actual cellular function extends over meters in some long nerve axons, and includes cellular, and molecular, and even atomic scale fields. The ratio of scales is thus those of the spatial scales shown above. What is not clear is the extent to which periodic boundary conditions, etc., can deal with these scale issues of the electric field. For this reason, I do not use these scales of the electric field in further discussion. What is clear is that the electric field is dealt with in a very different way in simulations[209,477] and theories[484,591,801,889] of the properties of semiconductors. There, periodic boundary conditions are never used. I have speculated too many times[266-271,273,284,286,288] that periodic boundary conditions could be related to this difference between ionic and semiconductor calculations, but our attempts to deal with finite size ions without periodic boundary conditions[5,783,884,885] have not progressed enough to see if my speculations were right.

It is clear that periodic boundary conditions are a great help, and probably necessary when dealing with condensed phases of uncharged molecules, or even reduced models of ionic solutions in which the water is included as particles or even spheres. The hard sphere neutral particles have interactions that extend over a moderate range and can be accommodated with sufficient accuracy with the spatial periods possible today. What is not clear is how to deal with the long range nature of the electric field and the need to accommodate Gauss' law (and the Poisson equation). Semiconductor simulations involve different potentials at different 'far field' locations because those are the inputs and outputs of the system (corresponding to the electrodes in an electrochemical cell or physiological experiment). And it is clear that the strength of the electric field is such that one needs to have all the charge accounted for and consistent with all the forces in the calculation, lest strong and nasty artifacts result. It is not clear how well the standard periodic boundary conditions corrected by various Ewald (modern references are [203,398,456,497,834,912,948]) or reaction field[121,222,576,930] schemes accommodate these needs. It is not clear how well the particle mesh methods of the semiconductor community deal with excluded volume forces.

It might be wise to adopt different schemes for hard sphere, Lennard Jones, and van der Waals forces, on the one hand, and coulomb forces on the other. Periodic boundary conditions might do for the hard sphere type interactions. Periodic boundary conditions for the potential might do well for the screened component of the electric field at long times after all the charges are screened. Screening occurs after some nsec. The sum rules then apply[403,610] and everything is in the 'thermodynamic limit', with spatially uniform boundary conditions at infinity. It is also not clear how to deal with the long time component of potential needed to deal with far field boundary conditions present even in screened systems in the 'thermodynamic limit' if the systems have finite boundaries. Since that potential is the potential of the signal in a nerve action potential, or in a classical telegraph, dealing with it cannot be ignored in molecular dynamics simulations that seek to compute the properties of nerve or muscle fibers.[755] Computations the physiological properties of nerve or muscle fibers must extend millimeters (muscle fibers), centimeters (muscle fibers and nerve fibers), even meters (nerve fibers in large mammals) and last milliseconds (one action potential) to a second or so (trains of action potentials that initiate contractions for example). All atom simulations of nerve or muscle physiology must then extend into the (very) far field in time and space, compared to the femtoseconds and angstroms of atoms themselves.

What is also not clear is how to deal with the electric potential on the femtosecond time scale of molecular dynamics calculations, before screening is established. That potential can spread arbitrarily far because it is unscreened by the movement of mobile charge, i.e. ions. Ions take time to move, because of friction. In fact, that short time electrical potential is simply the solution of a classical dielectric problem. Potentials in a dielectric can spread arbitrarily far and have intricate properties that can be made into elaborate machines[313] depending on the details of the boundary conditions.

**All the scales need to be dealt with at once.** We return to the length, volume, time and concentration scales. The difficulty in dealing with scales shown in Table 1 is made much worse by the central biological reality that all those scales are used at once in a single functioning biological system. It is an experimental fact that the current through a single sodium channel, carried by atoms of $Na^+$, extends to macroscopic scales in time and space. That current produces nerve function over a range of meters. It is an experimental fact that the current through a single sodium channel is strongly affected by trace concentrations of $Ca^{2+}$ and other ions. Heavy metals are toxic in trace concentrations and this toxicity is of enormous medical importance. It is also an intensively studied environmental issue taking a significant fraction of the resources of governmental science agencies like the US Department of Energy. Simulations must deal with these ranges of times, space, and concentration simultaneously.

It is also clear that simulations must be calibrated over these ranges, as well as computed over these ranges. Simulations, like other numerical analysis, need calibrations to establish their validity. The rigorous tight error bounds of mathematics are rarely available.

These problems are daunting and if it were necessary to deal with the scales shown in Table 1 simultaneously, biological problems would probably be unsolvable. I argue, however, that it is not necessary to deal with atomic detail over all this range of scales. I argue that the set of questions interesting in biology are limited and are controlled by a limited range of processes and scales. I argue that analysis of biological systems is fundamentally an engineering analysis of what actually controls those systems. The scales that evolution uses are what need to be analyzed, not the range of all possible scales.

**Reduced models deal with the range of scales.** I extend now my argument for the need of reduced models. I argue that biology can be analyzed by **a series** of reduced models, each with a complexity appropriate for the biological question being asked, each appropriate for the scale on which evolution has built that biological function. Each system will require a separate analysis because each system is likely to have evolved separately. Thus the scales used to control nerve signaling clearly extend from atoms to arms, as we have said. But the scales used to build a hip joint are mostly macroscopic, except perhaps for the lubrication layers that involve cells, and even smaller structures.

This kind of analysis is fundamentally a type of reverse engineering: we seek to approach biological systems the way an engineer would approach an unknown amplifier, doing just enough to find out how it can be controlled, not trying to find out everything, certainly not trying to deal with it in atomic detail.

Little can be said in general about this approach since each biological system will need its own analysis, and general principles will be those of evolution as much as physical chemistry, not appropriate for this paper. There is a general issue that emerges, however, that needs further discussion. There is a need for a mathematical framework that can deal with the issues known to be present in the salt solutions of life.

**Interacting Ions in solutions and channels**. The mathematics used to analyze ions in channels must deal with interactions. Channels were defined for many years by the interactions of (unidirectional) flows of ions before channel proteins were known as definite molecules[430,869]. If the unidirectional fluxes were competing, the flows were said to occur in channels. If the flows were cooperative, the flows were said to be through transporters.

There is a huge literature precisely defining the words ‘competing’, ‘cooperative’, and the phrase ‘unidirectional flux’ that can be reached through references[110,327,375,418]. The Appendix of ref[148] tries to give a precise and comprehensible definition of unidirectional fluxes useful for physical scientists. Ref[294] defines unidirectional fluxes precisely in a stochastic model and shows how their properties can be exploited in a mathematically exact theory of transport. Singer’s work[828,830,831] extends this in important ways. Bass[48,49,623,624] shows the importance of unidirectional fluxes in classical diffusion models of transport. (Ratios of unidirectional fluxes in selfconsistent models of channels have not yet been studied as far as I know.) In my view, it is important to see how useful these ratios are in fully selfconsistent models and in models with atomistic detail, both with and without single file behavior of the type envisioned in classical work. These ratios may be useful definitions of transporters in general if the theory and analysis deal with the properties of charged channels approximately twice the diameter of ions. In classical work, attention was focused—understandably enough given the lack of computers—on a much simpler system. In the classical literature [415,416,418-420,439] single file meant the behavior of red and blue balls (without electric charge of course) in an (uncharged of course) circular cylinder only slightly larger than the diameter of the balls.

The mathematics used to deal with unidirectional fluxes in the classical membrane and channel literature was that of chemical kinetics, using the law of mass action with rate constants independent of conditions. This use naturally followed[285] the use of similar laws throughout biochemistry and enzymology, then and now[234,733]. I argue in the Appendix that the law of mass action is fundamentally incorrect when solutes have charge or size (i.e., in all electrolyte solutions). In that case, the rate constants must vary with concentration, often dramatically, and vary with the concentration of every species, and with every boundary condition in the system. Treating rate constants as necessarily independent of other ions implies that all solutions are ideal. Solutions in biology are rarely ideal. Solutions of the concentration found in channels and active sites of enzymes do not resemble ideal solutions at all. Assuming that ions of some 20 to 40 molar can be described as ideal solutions that follow the law of mass action (with constant rate constants) is likely to seriously distort the physics of the system and make progress in understanding mechanisms nearly impossible.

The analysis of ions in channels is not alone in this difficulty. Despite the work of generations of physical chemists—references[319,479,530,577,802] provide an entry into this enormous field of research— ions in solutions can be dealt with only with difficulty, or not at all, when the solutions are mixtures and highly concentrated. In those cases, everything interacts with everything else, and analysis that starts with individual atomic species cannot deal with the interactions without adding bewildering complexity. Even worse, that complexity changes in what seems an arbitrary way every time a new type of ion is added to the solution. Indeed, the complexity changes as one varies the concentration of one type of ion over the biological scale from $10^{-7}$M $Ca^{2+}$ (inside cells) to 20 M $Ca^{2+}$ (inside channels). A leading experimentalist Werner Kunz recently said “ … it is still a fact that over the last decades, it was easier to fly to the moon than to describe the free energy of even the simplest salt solutions beyond a concentration of 0.1M or so.” (p.10 of ref[531])

Ions in solutions can be analyzed in pure dilute solutions with some success, as Kunz’ quotation implies, but analysis fails when (1) solutions of one species are concentrated beyond say 0.1 M for NaCl (2) divalents are involved like $Ca^{2+}$ (2) solutions are mixtures of ions; or (3) flows of any type are involved. Clearly, concentrated solutions are involved in most cases where ions are important. Clearly, $Ca^{2+}$ is involved in much biological function. Clearly, mixtures are involved in many cases where ions are important, in all cases in biology and many in industry. Clearly, flows are involved in most cases where ions are important. Clearly, flows of mixtures of concentrated ions are crucial components of nearly every living process in animals and plants.

I have always been puzzled that mathematicians could compute the flow of air around an airplane, in sub, trans and supersonic domains, with more or less arbitrary accuracy, but cannot yet deal with the flow of $Na^+Cl^-$ mixed with $Ca^{2+}Cl_2^-$. Computational fluid dynamics deals quantitatively with problems of much greater complexity than anything I could imagine in ions in channels.

Ions in biology flow in a narrow range of temperature, by electrodiffusion, dominated by friction,[294,651,797-799,828,830,831] without waves or shocks or complex physical behavior. Shock waves are not involved. Fluid dynamics covers a much wider range of conditions and phenomena with all the complexity we see in water waves crashing on the beach and much more (water is incompressible; fluid dynamics deals successfully with compressible fluids, even with mixtures of different types of immiscible fluids like liquid crystals). Why could fluid mechanics succeed when computation of ions in channels, or bulk for that matter, fail (at least in comparison)?

An approach to the answer seems to have been recently provided by mathematicians working on fluid mechanics. Chun Liu, of Pennsylvania State University, has developed a variational principle equivalent to the full Navier Stokes equations, including dissipation, and has applied it to systems apparently far more complex than ions in channels. Recently, Chun's Energy Variational Principle *EnVarA* has been applied to ions in channels and solutions.[280,467]

It is certainly premature to review this work, which is just now appearing in print, but it is appropriate to state the principles involved since they or their equivalent are likely to be needed to deal with the fundamental properties of ions flowing in concentrated mixtures in biological and physical systems. Indeed, we argue in[280] that "The equations [of this new variational approach] must be tested against experiments in many applications, and then improved in a mathematically systematic and physically selfconsistent way …."

**Ionic Solutions as Complex Fluids.** The basic principle in a variational analysis of electrolyte solutions is to consider ionic solutions as (relatively) simple examples of complex fluids[280,467], using the theory originally developed for viscoelastic fluids [1,11,151,244,249,464,465,552,573,574,582-584,765,766,806,932,939-941], rather than as complex examples of simple fluids, as in the classical literature.[43,75,387,401,403,423,731] Simple fluids (in their original and ideal realization) have no interactions. Complex fluids[88,89,235] and viscoelastic fluids have complex nonlinear viscosity, large interactions, and sub-elements across many scales of size. I argue that ionic solutions always have large interactions and so should be considered as complex fluids. I argue that it will be constructive to view the classical literature of finite size[3,47,256,257,309,319,328,492,495,530,548,660,700,735,762,310,311,402,519,594,688,689,914] and discrete ion effects[44] in this light.

The physical reason is simple. The fundamental fact of ionic solutions is that ions come in pairs, i.e., in strictly neutral molecular combinations of charged atoms. One <u>**ALWAYS without exception**</u> dissolves a strictly neutral salt in water to make an ionic solution. The salt is neutral to at least one part in $10^{15}$ even in mesoscopic systems. The electric field is so strong that no violation of electrical neutrality is significant in chemical units (i.e., number density) even in systems as small as the pore of an ion channel. Yet the electric fields produced by these fantastically small violations of electroneutrality produce forces as large as diffusion forces. The electric field is strong (see the first paragraph of Feynman's text[312]). The electric field guarantees interactions.

The interactions of the electric field are in fact so strong that charges are completely screened. Any imbalance would lead to huge forces and flows that would quickly (in the relaxation time of the ionic atmosphere) produce screening. These screened idealized systems are described by sum rules[403,610] that apply for distances longer than a few Debye lengths, a few angstroms in many biological systems, crudely $3/\sqrt{I_s}$ in Å in a useful approximation where ionic strength $I_s$ is in molar units.

Screening[135] of course guarantees that ions are not independent and so should not be treated as simple fluids, in my opinion. Screening guarantees that ions do not follow the independence rule of classical electrophysiology[434,436]. The independence rules were developed to describe currents that flowed through independent conductances in membranes that we now know are distinct and independent ion channel proteins. The independence rules were applied to bulk solutions before the classical authors knew what they quickly later learned of screening and Debye-Hückel theory. (Historical note: I know this from personal discussions with the workers involved. Indeed, the independence principle *for bulk solutions* was being taught to students in Cambridge UK physiology courses into the 21st century, to my personal knowledge.)

Screening is sometimes thought to make ionic solutions behave essentially like uncharged systems but this is not the case. Screening takes many psec to develop because ions move slowly in response to a change of force. Ions experience enormous numbers of collisions and thus enormous friction. They thus respond to a change in force, for example insertion of charge, only after a time delay.

The time course of the response of a conductive dielectric is in fact described in textbooks of electricity and magnetism. The time course can be directly seen from measurements of the time course or frequency dependence of the dielectric coefficient[46]. Screening is complete only in systems in which boundary conditions and flows are unimportant. If boundary conditions and flows are important, screening is incomplete, and electrical potentials spread macroscopic distances. (This is the mode of operation of the classical telegraph, and of electrical properties of nerve and muscle fibers.[920]) Thus, interactions produced by the electric field are of great importance in all ionic solutions.

Interactions produced by the finite size of ions are important in these relaxation phenomena and in everything else, in most solutions. Even in solutions of 100 mM salt, the effects of the finite size of ions on the free energy per mole of ions are large. (See Fig. 3.6 of reference[309]; Fig. 4.2.1 of reference [547] and the general discussions in the textbooks[47,309,530,547]. Fraenkel[319] has a particularly useful discussion in a modern context.) Theories have had some success in dealing with finite size effects in pure solutions of ions in systems without flow and boundaries[47,252,256,257,309,547,548,736,825], but even there, theories are more ad hoc than one would like and are not mathematically well enough defined to be extended easily to new systems or ionic salts.

Interactions among ions of different types are much harder to deal with. Essentially all that can be done is to compile complicated equations of state with large numbers of parameters that need to be readjusted when conditions are changed or new components are added to the solutions.[479,577,648,802]

Nonequilibrium situations in which flows occur in mixtures of ions are hardly addressed by present workers on ionic solutions even though a great deal of technology and most of life are in that category. The flow of mixtures of ions is also more or less impossible to simulate with present day methods of molecular dynamics because of the enormous problems of scale we discuss near Table 1.

We are left with a remarkable situation worthy of the quotation from Kunz: we can send a man to the moon, but we can compute none of the properties of mixed ionic solutions. Theories of conductance of ionic solutions deal with pure solutions in a crude way[7,232,331,332,492]. Theories of mixtures[441,863] do not do better.

Density Functional theories (DFT) of mixed ionic solutions have promise. They were originally developed to deal with the properties of mixtures of uncharged spheres of finite volume in confined spaces.[118,211,307,353,354,356,401,739,741,746,748] In these situations interactions among spheres and boundaries are crucial. DFT was later extended to deal with ions and channels. The extension to ions and channels was arbitrary, although a sensible and natural first step. The extension did not include the components of conductance produced by distortion of the ionic atmosphere, etc. in the Fuoss and Onsager, and Justice treatments of bulk ionic conductance[7,252,254,331,332,492,736], nor did it identically satisfy sum rules and equivalent applications of Gauss' law. PNP-DFT—as the extension to channels was called by us[353]—has done very well in describing current flow through the ryanodine receptor, a channel in muscle, as shown by Gillespie and coworkers[348,349,351,352,357] where distortion of the ionic atmosphere by current flow is not likely to be a first order effect. PNP-DFT compares quite well with Monte Carlo simulations[356] at equilibrium where distortion of the ionic atmosphere by current flow does not occur. But PNP-DFT cannot be expected to deal with systems where distortion of the ionic atmosphere is important[7,252,254,331,332,492,736] because it does not allow current dependent effects on the shape and size of shielding or thus on the shape and size of the ionic atmosphere.

PNP-DFT cannot be considered a mathematically well defined or general approach for all these reasons, despite its evident success in dealing with a specific ion channel. We details that success later in the paper. PNP-DFT accounts for a wide variety of the properties of the ryanodine receptor so well that one must imagine that the ionic atmosphere screening ions in that channel is quite constant in shape and size.

**Variational Approach.** A variational approach designed to deal with strong interactions in complex fluids is likely to succeed where PNP-DFT fails, as well as where it succeeds.[348,349,351,352,357]

Ionic solutions in fact are a relatively simple complex fluid in some ways, because in the most important biological cases their microelements are hard spheres ($Na^+$, $K^+$, $Ca^{2+}$) or nearly hard spheres ($Cl^-$). Microelements of hard spheres are likely to produce much simpler behavior than the more complex fluids already successfully analyzed[11,151,244,249,464,465,552,573,574,582-584,765,766,806,932,939-941] by the theory of complex fluids.

Water in the variational approach is first analyzed as a conductive dielectric. Later treatments certainly need to be more realistic. Water can often be successfully described as a continuum in implicit solvent models of ionic solutions (also called 'the primitive model') and

proteins.[47,256,309,319,328,547,736,744,762,825,904]

We use[280,292,467,642] a theory of complex fluids based on the energy variational approach *EnVarA*. Liu has actually proved the existence and uniqueness theorems needed to make this approach mathematics.[581] He has also applied *EnVarA* to a variety of complex real systems[11,151,154,244,464,465,468,551,552,806]. This theory of complex fluids has dealt with systems with complex microelements: liquid crystals, polymeric fluids[88,89], colloids and suspensions[940,941] and electrorheological fluids [154,806]; magnetohydrodynamics systems [1]; systems with deformable electrolyte droplets that fission and fuse [766,940]; and suspensions of ellipsoids. The theory deals also interfacial properties of these complex mixtures, such as surface tension and the Marangoni effects of 'oil on water' and 'tears of wine'[323,897,940,941]. We try to create a field theory of ionic solutions that uses only a few fixed parameters to calculate most properties in flow and in traditional thermodynamic equilibrium, both in bulk and in spatially complex domains like pores in channel proteins.

The Energy Variational Principle is

$$\overbrace{\frac{\delta E}{\delta \vec{x}}}^{\text{Conservative Force}} = \overbrace{\tfrac{1}{2}\frac{\delta \Delta}{\delta \vec{u}}}^{\text{Dissipative Force}} \quad . \tag{2}$$

$$E\Big(\text{Primitive Phase}; t\Big) =$$

$$= \int_{\Omega} \left( \underbrace{\overbrace{\tfrac{1}{2}\rho\left|\vec{u}_{IP}\right|^2}^{\substack{\text{Hydrodynamic}\\ \text{Kinetic Energy}}} + \overbrace{w(\rho)}^{\substack{\text{Hydrodynamic}\\ \text{Potential Energy}\\ \text{Equation of State}}}}_{\text{Macroscopic (hydrodynamic)}} + \underbrace{\lambda \left[ \underbrace{\tfrac{1}{2}\varepsilon\left|\nabla\phi\right|^2}_{\text{Electrostatic}} + \underbrace{k_B T(c_n \log c_n + c_p \log c_p)}_{\text{Entropy}} + \underbrace{\psi\Big(\text{Spheres}\Big)}_{\text{Finite Size Effect}} \right]}_{\text{Microscopic (atomic)}} \right) d\vec{x} \tag{3}$$

The energy $E$ written in eq. (3) is used to describe finite sized ions in solution. The energy $E$ is a generalization of thermodynamic energy as used in variational theory in general. It is not the energy of the first law of thermodynamics. The electrostatic potential is $\phi$. The concentration of species $n$ is $c_n$. $k_B$ is the Boltzmann constant and $T$ the absolute temperature. $\psi(\text{Solid Spheres})$, $w(\rho)$, and $\frac{1}{2}\rho\,|\vec{u}_{IP}|^2$ are the contributions of the solid spheres, the hydrodynamic potential energy and the hydrodynamic kinetic energy to the energy function (as defined in [280,468]). The dissipation $\Delta$ is not hard to derive but is too complex to present in detail because of the finite size effects. It is described in full in[280,292,467]. A recent article[280] is meant to provide detail for the physical chemistry and biophysical communities.

The variational principle *EnVarA* combines the maximum dissipation principle and least action principle into a force balance law that expands the conservative conservation laws to include dissipation, using the generalized forces in the variational formulation of mechanics (p. 19 of reference[363]; also[27,87,345]). This procedure is a modern reworking of Rayleigh's dissipation principle—eq. 26 of reference[724]—motivated by Onsager's treatment of dissipation[680,681].

*EnVarA* optimizes both the action functional (integral) of classical mechanics[28,363,862] and the dissipation functional[540]. The variation of the action is taken with respect to the trajectory of particles. The variation of the dissipation is taken with respect to velocity. Both are written in Eulerian (laboratory) coordinates after the appropriate push forward or pull back of variables. The functionals of *EnVarA* can include entropy and dissipation as well as potential energy, and can be described in many forms on many scales from molecular dynamics calculations of atomic motion, to Monte Carlo MC simulations[244,249,939] to—more practically—continuum descriptions[574,940] of ions in water. We use a primitive model[47,256,309,547,548,701,762] of ions in an implicit solvent[25,213,259,360,663,751,917], adopting self-consistent treatments of electro-diffusion [51,283,284,286,484,801,889]—in which the charge on ions helps create their own electric field. In addition, we introduce the repulsion energy of solid spheres[102,103,106,273,630,664,748], using the variational calculus to extend the primitive model to spatially complex, nonequilibrium time dependent situations, creating a field theory of ionic solutions.

Energy functional integrals and dissipation functional integrals are written from specific models of the assumed physics of a multi-component system, as in references[154,574,766,806,940]. Components of the potential energy and dissipation functions are chosen so the variational procedure produces the drift diffusion equations of semiconductor physics[484,591,801,889]—sometimes called the Vlasov equations[90,114,364]—or the similar biophysical Poisson Nernst Planck equations—named PNP by reference[283]—and used by many biophysicists[36,118,124,196,272,276,284,286,288,374,445,446,471,472,535,602,666,798] and physical chemists.[286,762,51,660] The energy of the repulsion of solid spheres can be included in the energy functionals as Lennard-Jones spheres[574,806] giving (as their Euler-Lagrange equations) a generalization of PNP for solid ions. Boundary conditions tell how energy and matter flow into the system and from phase to phase and are described by a separate variational treatment of the 'interfacial' energy and dissipation. The resulting partial differential equations are analogous to the usual Euler Lagrange equations of variational calculus. They form the boundary value problems of our field theory of ionic solutions. ***They are derived by algebra and solved by mathematics—without additional physical approximations***—in spatially complex domains, that produce flow of nonideal mixtures of ions in solution.

*EnVarA* does not produce a single boundary value problem or field equation for ionic solutions. Rather, it produces different field equations for different models (of correlations produced by screening or finite size, for example), to be checked by experiment. In the biological and chemical context, *EnVarA* derives—it does NOT assume—systems of partial differential equations (i.e., field theories) of multiple interacting components and scales.

If a new component of energy (or dissipation) is added to a variational principle like *EnVarA*, the resulting partial differential equations that form the field theory of electrolytes—analogous to Euler Lagrange equations—change. The new field theory is derived by algebra and involves no further assumptions or parameters. The new field theory automatically includes all the interactions of the old and new components of the energy (and dissipation). This is an enormous advantage of variational principles and is probably the reason they are used so widely in physics. I am unaware of any other mathematical approach that forces field equations to be consistent with each other. The contrast between *EnVarA* and the usual approach to mixtures of ionic solutions, with their plethora of coupling coefficients[52,423,492,577,863,922], is striking. It is very difficult to determine those coupling coefficients, and even worse, the coupling coefficients are functions or functionals that depend on all the other parameters of the system, usually in an unknown way.

The variational principle can be applied to a primitive model of ionic solutions with a Lennard Jones treatment of excluded volume, and a selfconsistent computation of the electric field[280,467] as described in detail in[280,467]. A regularized repulsive interaction potential is introduced as

$$\Psi_{i,j}(|\vec{x}-\vec{y}|)=\frac{\varepsilon_{i,j}(a_i+a_j)^{12}}{|\vec{x}-\vec{y}|^{12}} \tag{4}$$

for the $i^{\text{th}}$ and $j^{\text{th}}$ ions located at $\vec{x}$ and $\vec{y}$ with the radii $a_i$, $a_j$, respectively, where $\varepsilon_{i,j}$ is an empirically chosen energy constant,. Then the contribution of repulsive potential $\Psi$ to the total (free) energy is

$$E_{i,j}^{repulsion}=\tfrac{1}{2}\iint \Psi_{i,j}(|\vec{x}-\vec{y}|)c_i(\vec{x})c_j(\vec{y})\,d\vec{x}d\vec{y} \tag{5}$$

where $c_i$, $c_j$ are the densities of $i$ th, $j$ th ions, respectively.

For the sake of simplicity in this derivation, we consider a two-ion system with the charge densities, $c_n$, $c_p$. All derivations and programs have been written for a multiple ion system, with ions of any charge[280,467]. Then, the total repulsive energy is defined by

$$E^{repulsion}=\sum_{i,j=n,p}E_{i,j}^{repulsion}=\sum_{i,j=n,p}\tfrac{1}{2}\iint\Psi_{i,j}(|\vec{x}-\vec{y}|)c_i(\vec{x})c_j(\vec{y})\,d\vec{x}d\vec{y}. \tag{6}$$

Now we take a variational derivative with respect to each ion, $(\delta E^{repulsion}/\delta c_i)=0$ to obtain the repulsive energy term and put it into the system of equations. This leads us to the following Nernst-Planck equations for the charge densities, $c_n$, $c_p$:

$$\frac{\partial c_n}{\partial t}=\nabla\cdot\left[D_n\left\{\nabla c_n+\frac{c_n}{k_BT}\left(z_ne\nabla\phi-\int\frac{12\varepsilon_{n,n}(a_n+a_n)^{12}(\vec{x}-\vec{y})}{|\vec{x}-\vec{y}|^{14}}c_n(\vec{y})\,d\vec{y}\right.\right.\right.$$
$$\left.\left.\left.-\int\frac{6\varepsilon_{n,p}(a_n+a_p)^{12}(\vec{x}-\vec{y})}{|\vec{x}-\vec{y}|^{14}}c_p(\vec{y})d\vec{y}\right)\right\}\right], \tag{7}$$

$$\frac{\partial c_p}{\partial t}=\nabla\cdot\left[D_p\left\{\nabla c_p+\frac{c_p}{k_BT}\left(z_pe\nabla\phi-\int\frac{12\varepsilon_{p,p}(a_p+a_p)^{12}(\vec{x}-\vec{y})}{|\vec{x}-\vec{y}|^{14}}c_p(\vec{y})\,d\vec{y}\right.\right.\right.$$
$$\left.\left.\left.-\int\frac{6\varepsilon_{n,p}(a_n+a_p)^{12}(\vec{x}-\vec{y})}{|\vec{x}-\vec{y}|^{14}}c_n(\vec{y})\,d\vec{y}\right)\right\}\right]. \tag{8}$$

The details of the derivation of the repulsive terms in the chemical potentials are presented in [280,467] We now have the coupled system including finite size effects. We call the system a modified PNP system. One advantage of the variational approach is the fact that the resulting system, the modified PNP, naturally satisfies the energy dissipation principle, the variational law eq. (2).

$$\frac{d}{dt}\int\left\{k_B T\sum_{i=n,p}c_i\log c_i+\frac{1}{2}\left(\rho_0+\sum_{i=n,p}z_i e c_i\right)\nabla\phi+\sum_{i,j=n,p}\frac{c_i}{2}\int\Psi_{i,j}c_j\,d\vec{y}\right\}d\vec{x}$$
$$=-\int\left\{\sum_{i=n,p}\frac{D_i c_i}{k_B T}\left|k_B T\frac{\nabla c_i}{c_i}+z_i e\nabla\phi-\sum_{j=n,p}\nabla\int\tilde{\Psi}_{i,j}c_j\,d\vec{y}\right|^2\right\}d\vec{x} \tag{9}$$

where $\tilde{\Psi}_{i,j}=12\Psi_{i,j}$ for $i=j$, and $\tilde{\Psi}_{i,j}=6\Psi_{i,j}$ for $i\neq j$.

As we have seen, these variational principles derive field equations that address (and I believe will probably some day solve) major problems in computational biology. The field theory *EnVarA* represents an ionic solution as a mixture of two fluids[583], a solvent water phase and an ionic phase. The ionic phase is a primitive model of ionic solutions[2,47,256,257,309,544,547,762,883]. It is a compressible plasma made of discrete[44] charged, solid (nearly hard) spheres. The ionic 'primitive phase' is itself a composite of two scales, a macroscopic compressible fluid and an atomic scale plasma of solid spheres in a frictional dielectric. Channel proteins are described in *EnVarA* by primitive ('reduced') models similar to those we have discussed at length. Similar models predicted complex and subtle nonequilibrium properties of the RyR ryanodine receptor before experiments were done in more than 100 solutions and in 7 mutations, some drastic, removing nearly all permanent charge from the 'active site' of the channel (see references in[348,357]).

I believe a variational method is required to deal with real ionic solutions because ionic solutions are dominated by interactions. Ionic solutions do not resemble the ideal simple fluids of traditional theory and the interactions between their components are not two-body, as assumed by the force fields of modern molecular dynamics. Indeed, **ions like $Na^+$ and $K^+$ have specific properties, and can be selected by biological systems, because they are non-ideal and have highly correlated behavior.** Screening[135] and finite size effects[47,309,547,548,700,701] produce the correlations more than anything else. Solvent effects enter (mostly) through the dielectric coefficient. Ionic solutions do not resemble a perfect gas[761] of non-interacting uncharged particles. Indeed, because of screening[135,403,610], the activity (which is a measure of the free energy) of an ionic solution is not an additive function as concentration is changed (Fig. 3.6 of reference[309]; Fig. 4.2.1 of reference[547]) and so does not easily fit some definitions of an extensive quantity (see p. 6 of the book of international standards for physical chemistry[173]).

Some correlations are included explicitly in our models as forces or energies that depend on the location of two particles. Other correlations are implicit and arise automatically as a mathematical consequence of optimizing the functionals *even if the models used in the functionals do not contain explicit interactions* of components. Kirchoff's current law (that implies perfect correlation in the flux of electrical charge[406,407]) arises this way as a consequence of Maxwell's equations[671,934] and does not need to be written separately.

Variational analysis is already an area of active research in modern mathematics. Our methods are also closely related to another exciting area of modern mathematical research, optimal control. *EnVarA* produces 'optimal' estimates of the correlations that arise from those interactions [508,725,827,845] (p. 42 of Gelfand and Fromin[345], see p. 11 of Biot[87]; and criticism[315] of the absence of mathematical analysis in Biot). All field equations arising from *EnVarA* optimize both the dissipation and the action integrals. Inadequate functionals can be corrected (to some extent) by adjusting effective parameters in the functionals. *EnVarA* produces optimal estimates of these parameters, because the mathematics of variational analysis is almost identical to the mathematics of optimal control. Both use variational methods that can act on the same functionals. *EnVarA* becomes optimal control when the functionals are combined in a more general way than just adding them, e.g., by using Lagrange multipliers or more sophisticated techniques. Inverse methods[118,304,493] could be used to provide estimators[393,837] of the parameters of *EnVarA* functionals with least variance or bias, or other desired characteristics. (This subject is discussed at length in reference[280].)

Effective parameters are needed to deal with ions in electrolytes because of the enormous range of scales involved (see Table 1). Effective parameters come along with reduced models. Effective parameters have in fact almost always used to describe complex interactions of ions in electrolyte solutions[219,252-254,257,492,498,700,701,736,863,3,47,256,257,309,319,328,492,495,530,548,660,700,735,762,310,311,402,519,594,688,689,914], e.g., the cross coupling Onsager coefficients[218,219,221,498,706] or Maxwell-Stefan coefficients[423,863].

*EnVarA* gives the hope that fewer parameters can be used to describe a system than in models[701] and equations of state[479,577,802] of ionic solutions, which involve many parameters. These parameters change with conditions and are really functions or even functionals of all the properties of the system. (It is important to understand that the parameters of classical models[701] and equations of state[425,514,710] almost always depend on the type and concentration of all ions, not just the pair of ions that are coupled.)

Of course, the variational approach can only reveal correlations arising from the physics and components that the functional actually includes. Correlations arising from other components or physics need other models and will lead to other differential equations. For example, ionic interactions that arise from changes in the structure of water would be an example of 'other physics', requiring another model, if they could not be described comfortably by a change in the diffusion coefficient of an ion or a change in the dielectric constant of water. Numerical predictions of *EnVarA* will be relatively insensitive to the choice of description (of pairwise interactions, for example) because the variational process in general produces the 'optimal' result[87,345,508,725,845] for each version of the model. (This is an important practical advantage of the variational approach: compare the success of the variational density functional theory of fluids[353,354,356,747] with the non-variational mean spherical approximation[2,47,252,254,256,257,309,544,547,736,823,883] that uses much the same physics.)

This variational approach can include energies of any type. It has in fact been used by Liu[244,249,939] to combine energies of reduced models and energies computed from simulations. It will be interesting to see how we can apply this approach to biological systems and how well it deals with the discrete ion effects[44] of the classical work on the capacitance of double layers, a glimpse of which can be found in the references[310,311,402,519,594,688,689,914].

**<u>Scaling in *EnVarA*</u>**. The variational approach deals with issues of scaling in a very different way from direct simulations. *EnVarA* has the great advantage of always being consistent. A model in *EnVarA* is the statement of energies and dissipation in eq. (2). Once that model is chosen, the rest is algebra. The resulting Euler Lagrange equations form a well posed boundary value problem, a field theory of (usually) partial differential equations and boundary conditions that account for all the behavior of the system described by the energy and dissipation. The field theory is much more general than the thermodynamic and statistical mechanical ideas of equilibrium and state. It includes flow and interactions of components automatically. If two of the components of the energy (and/or dissipation) are on different scales, *EnVarA* automatically produces Euler Lagrange equations that combine those scales selfconsistently. This is an enormous advantage compared to other multiscale methods.

*EnVarA* deals with interactions automatically but it does not deal with multiscale issues nearly as well. I will go through the issues one by one.

**<u>Scaling in Space in *EnVarA*</u>.** Spatial scaling and resolution are dealt with in *EnVarA* without error if the models of energy and dissipation include all scales at perfect resolution. Of course, that never happens! What typically happens is that part of the system is known well at one scale, part at another, and parts of the system are left out. Typically, one part of the system must be resolved on one scale and the other on another. Applying *EnVarA* to these situations is (reasonably) straightforward but the accuracy of the results can only be assessed after the fact by comparison with experiments. The basic approach is to write the energy and dissipation of each component of the model, of each scale, and combine them using Lagrange multiplier(s), or other penalty functions of optimal control. *EnVarA* guarantees that interactions will be dealt with correctly. *EnVarA* automatically deals with boundary conditions (once they are described with a model) and flow. These are important features not shared by many other methods.

But *EnVarA* cannot deal with phenomena that are not present in the models of the energy and dissipation and these can be important. For example, if *EnVarA* uses a Lennard-Jones description of spheres in a spatially uniform dielectric to describe an ionic solution, it cannot describe interactions that occur because of spatial variations in the dielectric properties of water. In general, *EnVarA* (particularly when implemented numerically) may not be able to resolve steep phenomena and gradual phenomena well enough to estimate their interactions correctly. *EnVarA* will double count phenomena that are described in more than one component of a model. For example, if an equation of state is used to deal with the finite volume of ions (on the macroscopic scale) and Lennard Jones potentials are used to deal with the finite volume of ions (on the atomic scale), double counting can be expected. The Lagrange multipliers (or penalty functions of optimal control) and variational process minimize the effect of the double counting (by choosing optimal parameters that minimize the functionals) but the residual effects may be significant. We are in unknown territory here. We know how to investigate but we do not know the results of the investigation.

**<u>Scaling in time in *EnVarA*</u>**. Time dependence in *EnVarA* is produced by the dissipation function and so depends on the accuracy of the model of dissipation. It is obvious that the linear frictional model used in *EnVarA* (and in Rayleigh and Onsager's dissipation principles) is inadequate. Friction is not proportional to velocity in general. The consequences of the oversimplified model of dissipation are not known.

One important characteristic of *EnVarA* arises from its time dependence and is both a curse and a blessing. The blessing is that *EnVarA* computes time dependence at all. The curse is that it must compute time dependence starting at time zero. Steady states only arise from transient computations. This property of the Euler Lagrange equations makes computation less efficient. One must approach the steady state. One cannot just arrive there.

**Scaling of Parameters in *EnVarA*.** Parameters arise in *EnVarA* from the models of energy and dissipation and in general appear as parameters in the Euler Lagrange equations that specify the resulting field problem. Parameters are handled as well or as badly as they are in other partial differential equations. Analytically, parameters of any scale are handled 'perfectly', but numerical issues of stiffness and dynamic range can easily arise and be limiting. Each case must be studied as a separate numerical system because each case can have quite different qualitative behavior. The numerical schemes must be adapted to the qualitative behavior.

The very generality of the *EnVarA* approach causes considerable difficulty. The behavior of the system with all its interactions is often unknown in initial calculations. If reduced models with effective parameters are used (as they should be in early survey calculations), it is hard to know what 'region of phase space'— i.e., what qualitative range of behaviors—one is seeing. Dealing with an *EnVarA* calculation is much like a survey experiment in biology. You have to determine what is going on and you have to learn to simplify the calculation or experiment by choosing parameter ranges or setups in which the interesting phenomena dominate.

For example, computations of current flow through channels using *EnVarA* always produce charging phenomena at short times (because such must be present in any calculation that includes the electric field consistently), flow through the channel at intermediate times, and accumulation of ions outside the channel as the flow continues into long times. The charging phenomena and accumulation are peripheral to one's initial main interest in the channel itself, but the numerical procedures must deal with them correctly and efficiently. Experimental scientists can take years to learn to isolate the phenomena of interest. Numerical analysts using *EnVarA* face similar prospects. In this regard, *EnVarA* adds complexity as well as power to the present state of the art. One would of course like to add the power without the complexity but I do not know how to do that.

The power of *EnVarA* is again a blessing and a curse. It is a blessing because it forces the theorist to deal with phenomena well known in the laboratory (i.e., the difficulty of actually keeping solutions well stirred at constant temperature) but often not advertised in experimental papers. The curse is the difficulty of computation and the efforts needed to isolate important special cases.

Despite these difficulties—that are described here in vivid detail so we do not mislead the reader into thinking *EnVarA* is a magical solution for all problems—computations with *EnVarA* are possible for real systems. Many have been done in physical systems[1,88,89,154,323,766,806,897,940,941] and a substantial number have been done with some success in ionic solutions.[280,468]

**Scaling of the protein.** The above discussion does not deal with the multiscale issues of describing the protein, whether channel or enzyme. I do not know how to do that in a general way even for channels, where covalent bond changes and orbital delocalization are not involved, let alone for enzymes where covalent bond changes are what the system is all about. (See reference[285] for a discussion of 'Channels as Enzymes' and reference[270] for a discussion of channels as transistors.) Reduced models have been built in many ways, using quantum mechanics (references in[896]), reduced models with water detail (references in[111,113,750]), and

reduced models with implicit models of water[19,103,106,107,193,197,273,274,602,665,913]. I apologize for the many references I have unknowingly omitted.

There seems to be no *a priori* way to choose between the different reduced models of channel proteins. I would use the fits to experimental data as the test for such models, although others prefer a more structural approach, arguing (understandably enough) that considerable structural detail is needed to deal with water and side chains of proteins. Each perspective emphasizes what the investigator can best do.

What seems clear is that each system of proteins will need separate treatment, with generalizations emerging only after enough special cases are studied. We do not know at this stage if we are dealing with biological transport systems that are fundamentally alike (like audio, signal, and power amplifiers), or systems that are fundamentally different (like analog and digital integrated circuits).

A variational approach is likely to have an important role in building these models of transporters and channels. After all, channels and transporters were originally defined by the interactions of the ions that flow through them. A mathematics that deals automatically and consistently with the interactions of flows will obviously be useful. This mathematics can be applied to physical models of reduced complexity, such as those discussed at length here. It can also be used in principle with mixtures of physical models and atomic scale simulations (as Liu has done in physical systems[252,254,736]). And it can also be used to combine chemical kinetic models with physical models and atomic scale simulations. These chemical kinetic models are widely used to describe covalent bond changes, or even changes in high energy hydrogen bonds where orbital delocalization is important, although they should not be used to describe low energy hydrogen bonds where orbital delocalization is much less important[224]. One must be careful in the use of chemical kinetic models, however, since the law of mass action is difficult to apply for nonideal solutions. **The rate constants of chemical kinetic models must never be assumed to be constants independent of experimental conditions**, lest the kinetic model contradict established physical laws. These issues are discussed in the Appendix.

**Outlooks: unsolved problems in physical chemistry.** The Editor of Advances in Chemical Physics, Stuart Rice, suggested that a discussion of unsolved problems might be worthwhile. It is good he did not ask me to look to the future, because the future is so often dominated by unforeseeable chaotic and stochastic events, and thus predictions are always wrong. In particular, science is often dominated by what is possible. What is possible is often dominated by technology and funding, and both are wildly unpredictable. Who in 1955 could have predicted Moore's law, let alone its continuation for decades? Indeed, Moore did not[638,639], thinking in 1965 that exponential growth was about to end then. Since then there have been some [45 years/(1.5 years per cycle)] = 15 cycles of doubling of density and speed, an increase of capability by 33,000×, an increase unprecedented in human history. No scientist predicts the future well, including a brilliant scientist and humanist like Moore who analyzed the then past history of a technology he helped to create. Aging scientists like me are particularly hampered by their necessarily foreshortened view.

Thus, I was glad to be asked to deal with unsolved problems (of the past) rather than unknowable advances of the future. Indeed, solving unsolved problems in a mature science like physical chemistry may be important for its future, as Stuart hinted. Technological advances are crucial for an infant science like computational biology, and an adolescent science like molecular biology. They may be somewhat less important for an adult science like physical chemistry.

Scientists understandably can easily overlook—thereby denying—the unsolved problems of past generations. Scientists, like all people, have enormously strong mechanisms of denial necessary for their collective survival. Survival would be threatened by depression, I fear, if we were continually conscious of all we cannot do.

**Unsolved problems in physical chemistry of solutions.** From my outsider's point of view, the unsolved problems in physical chemistry start with some of the oldest.[23,390,594,735]

The unfortunate fact is that Werner Kunz' remark previously cited (p.10 of ref[531]) is an understatement. Indeed, it is " … easier to fly to the moon than to describe the free energy of even the simplest salt solutions beyond a concentration of 0.1M …" But it is even harder to describe the free energy of mixtures of ions in biological systems. The descriptions are inadequate at any concentration. Physical chemistry started by measuring the colligative properties of mixtures of ions (as I understand it). Mixtures are not understood much better today than they were many years ago.

The first unsolved problem then is to use a variational method to (try to) compute the properties of mixtures of electrolytes, starting with the simplest colligative properties, moving to equilibrium properties in general and then to nonequilibrium properties of diffusion and conductance in mixed solutions. The scientific questions are (1) how successful is the *EnVarA* theory of primitive solutions? (2) What has to be added to it to deal with the complexity of concentrated and mixed solutions of electrolytes?

In practical terms, some who know how to write and integrate variational models and Euler Lagrange equations will need to learn to deal with the libraries of measurements made by physical chemists in the last century. Those libraries include

(1) colligative properties summarized in existing equations of state,[23,479,577,802]

(2) dielectric properties of homogeneous and inhomogeneous systems measured in great detail by Barthel[46,47] and Macdonald[44,45,593,594] (among many others no doubt) in the last 50 years,

(3) activities of ions in innumerable conditions,[2,47,76,256,257,309,310,319,328,329,390,453,486,489,492,494,518,528,530,547,548,580,594,660,700-702,728,735,764,904]

(4) conductive properties of ions reported in the literature of polarography[10,38,178,553] and electrochemistry in general [38,93,94,183,762,790]

and

(5) measurements of many other types I regrettably do not know enough even to cite.

One must also deal with the regrettable but understandable fact that large schools of science have developed using treatments of ionic solutions that more or less ignore the excess chemical potential of real ionic solutions and thus treat the solutions as if they are ideal. Many such treatments use the Poisson Boltzmann or even the linearized Poisson Boltzmann equation,[25,33,116,213,239,259,319,358-361,394,395,448-450,469,510,580,637,663,722,751,763,800,803-805,826,850,868,915-918,950] or simplified extensions[503,504,846] of them, that do not even seek to fit the data that Kunz discussed[47,256,257,309,530,531,547,762]. There is little that can be said politely here: ignoring the excess chemical potential of ionic solutions is ignoring the properties of real ionic solutions.[44,47,256,257,309,328,329,390,453,486,487,490,492,494,502,518,528,547,548,594,700-702,728,735,762,904,943]

Thus an important unsolved problem, and future goal for physical chemistry, should be the use of variational methods to deal with the nonideal properties of ionic solutions.

Two other unsolved problems need to be mentioned. The RISM (reference-interaction-site-model) approach to ionic solutions and spatially complex systems like proteins—championed by the Hirata group[421,475,693,935]—deals with a number of the issues raised here. Its integration with *EnVarA* and the advantages it promises for ions, proteins, and channels remain to be explored.

Wetting and dewetting of small capillaries by water and ionic solutions is a problem that has received a great deal of attention and we have recently suggested[749], along with many others,[24,53-58,176,220,392,455,509,781,839,840,875-878,908,924,951] that the spontaneous openings and closings of channels might be produced this way. (Note the suggestion is made about spontaneous gating, not all kinds of gating.) Wetting and dewetting includes interactions of ions, water and membranes and it seems likely that an *EnVarA* approach might be useful. The enormous literature of wetting and dewetting might be well ordered by a variational approach, particularly one that included phase field methods to describe membrane phenomena. The reader should note that the extensive listing of references in reference[749], inadvertently omitted many papers of importance, including those of workers known to me in other contexts, to whom I again offer apologies.

**Unsolved problems in applied mathematics.** The most immediate problem facing the variational approach is that of dealing with three spatial dimensions on the scales implied by Table 1. Until *EnVarA* can be computed quickly in three dimensions, the variational field theory of ionic solutions will be difficult to compare with established results of Monte Carlo simulations or experiments.

Numerical issues abound. How should electrostatics be computed efficiently? Can these be reconciled with the particle mesh methods of computational electronics or the Ewald sum methods of molecular dynamics? Can user-friendly packages of software be written that will allow experimentalists easy access to the results of an *EnVarA* treatment, whether in biophysics and biology, or physical chemistry?

The variational approach obviously must reach beyond the primitive model of solutions to include richer descriptions of water as a time (or frequency) dependent dielectric including its particulate properties. Indeed, the inclusion of energetics of particles is an important frontier for variational methods. Chun Liu and collaborators[244,249,765,939] have already included simulations in an *EnVarA* analysis. Can this initial work be extended to include the traditional simulations of molecular dynamics? Can low energy[224] or even high energy hydrogen bonds be included in a variational treatment? Indeed, how would one include classical chemical reactions or even quantum chemical energies? Does one deal only with the energy of such reactions? Or does one include the free energy? If so, how does one avoid double counting of entropic effects, in the entropy and in finite size effects? How does one deal with dissipation in a traditional chemical model, let alone one with a quantum chemical basis?

**Unsolved problems in molecular biology and biophysics.** Interactions of ions, water flow, cell volume, and tissue deformation are some of the classical questions of physiology.[110,214,404,440,478,687,692,709] Indeed, such physiological questions were an important motivation for early work in irreversible thermodynamics[498,499] that linked biophysics and physical chemistry.[218,306,500,595,680-682,706,952]. (Aharon Katchalsky told me of this motivation when I first met him at Harvard in the spring of 1962 and later at an NSF summer workshop in Boulder in the summer of 1966.) Sadly, this work was less useful than originally hoped because different forces and fields were combined without a variational principle, as I learned to my dismay[276] from the Gaussian-Markoff assumption in deGroot and Mazur[219] and my later attempt with Schuss[294] to provide a proper stochastic basis for biological models of flux. I finally realized that even in the most sophisticated stochastic models we could build[42,797] (also see[294,651-653,796-799,828,830,831]) forces were combined in a way that was not consistent or unique and could not deal with the correlations introducing by time varying electric fields or finite diameter of ions, for that matter. For example, trajectories produced stochastically varying concentrations of charged species but electrical forces were treated as independent of time. Each combination of forces and flows was treated in an ad hoc way. Mean field results required coefficients that were unknown functions or functionals and so different combinations of forces and fields could not be told apart.

Variational methods like *EnVarA* seek similar goals to the classical work of irreversible thermodynamics, but they are always selfconsistent and combine the energies and dissipations of different process with many fewer unknown coefficients. Thus, variational methods can automatically produce (as outputs) the correlations that arise from the electric field and diameter of ions. Indeed, once the energies and dissipations are chosen, the field equations combining forces and flows of different types—are unique.

A variational method may thus be able to deal with the classical physiological problems that involve coupling of very different kinds of forces and energies—e.g., of water flow and ion flow, of blood flow, including plasma, red blood cells, and other cell types, including deformation of blood vessel walls. Can *EnVarA* be extended to deal with water flow in cells and tissues, including blood flow, urine production and so on?[642]

On a smaller scale, interactions of membrane deformation, water flow, and ionic movement produce biological processes of great importance, from vesicle formation, to endocytosis, to the fusion of viruses to cells. Indeed, such interactions are involved in cellular locomotion in general.

In fact, variational methods have already been used to deal with closely related problems in physical systems, arising in the theory of complex fluids, like oil droplets in water, or liquid crystals. Phase field methods have been introduced by Chun Liu and collaborators[245-248,583,765-768,940] in an extension of the *EnVarA* approach to deal with different phases, including those of immiscible fluids like oil and water. Chun Liu, Rolf Ryham, and I are working with Fred Cohen a biophysicist familiar with these systems to see how the phase field approach can be extended to deal with membrane bound systems like cells and organelles.

An important unanswered question arises: can phase field methods be extended to deal with viral and vesicle fusion, or cell motion in general? Can phase field methods allow computational biology to link the physics of water flow, ionic movement, membrane deformation with the most biological of processes, flow, movement and deformation of cells? Can the classical behavior of single cells like ameba and paramecia, observed by millions of biologists in microscopes since van Leeuwenhoek invented the microscope (around 1600), be computed with a physical model with few arbitrary parameters, using phase field *EnVarA* inspired methods?

On a more molecular scale, the reduced models and *EnVarA* need to confront the literature of binding of ions to proteins, for example, zinc binding proteins,[65,506,628,723,911] thrombin,[228,229,685,686] and calmodulin[170,201,625,793] along with many others I know too little to cite.

In each of these cases, it should be possible to make reduced models involving some specific representation of crowded charges, and the energetics of conformation change (if need be), and combine them with *EnVarA* to make a specific field theory of that binding system. Solving the Euler Lagrange equations of that field theory for the appropriate boundary conditions will allow the reduced model to be checked against experiment in the same spirit that the model of the calcium, sodium, and ryanodine receptor channel was checked.

**Unsolved problems in channels.** There are a number of classical unsolved problems in channel and transport biophysics that can be approached by physical chemists. First, the one dimensional treatment of *EnVarA* must be developed and calibrated. No matter how successful mathematicians and numerical analysts are in speeding up three dimensional models, one dimensional models will remain very much faster. One dimensional models will be needed to deal with inverse problems of parameter and model estimation, which is a fancy way to say one dimensional models will be needed if we wish to understand and manipulate actual biological systems and experiments. One dimensional models will remain the natural obvious way to describe the input output relations of ion channels and transporters. Just as they have remained the obvious way to describe amplifiers and integrated circuits, even when full circuit diagrams are available.

Specifically, one must learn to calibrate one dimensional *EnVarA* models of channels to be sure that the steady buildup of ions (just outside the channel) is a good representation of the particular channel system being studied. The model is likely to be different for different channels and transporters. Once a model is constructed for the EEEE type calcium channel, for example, the results of the *EnVarA* calculation need to be checked against experimental results in the steady state (where the results are least ambiguously interpreted). The same procedure of constructing a one dimensional model, calibrating it, checking it against steady state data, and refining must be repeated for each type of channel. For example, the process needs to be repeated for the sodium channels and the ryanodine receptor discussed at length in this paper. The process will be more efficient as it is repeated but it will not give identical results. The sodium channel, the calcium channel, and the ryanodine receptor have quite different functions. They will mix ionic diffusion inside and outside the channel in different ways. A three dimensional calculation might capture this uniquely, without much individual tailoring or focused attention, if enough structural detail can be included. But one dimensional models must be tailored. There are many different reduced models of transistors taught and used by engineers, each specialized for a particular purpose.

*EnVarA* will provide time dependent results beyond the steady state results. It will predict transients, including gating currents (for an open channel of one conformation). The current voltage and current vs. time predictions of *EnVarA* need to be compared with 'the real thing'. *EnVarA* will show currents on the full time scale from atomic motions, to the fastest components of gating currents, to gating currents of the classical slow type, to changes in ionic 'conductances' with time, to accumulation of ions inside channels, to accumulation of ions outside channels. Each time dependence will vary with the conditions the channel is in.

Each type of channel will need a different model even in three dimensions because each channel does different things (compare *L*-type calcium channels and RyR calcium channels, for example). The physical chemist will despair of the complexity. The biologist will rejoice. The engineer will understand, because the diversity of change in currents with time and conditions is what channels and transporters are all about. They are devices built to make and control those changes in current, flux and concentration. Of course, there are many of them, just as there are many types of transistors, integrated circuits, amplifiers, and so on.

Thus, there is no shortage of other channels on which a similar analysis will be revealing. One may find surprises concerning the origin of gating, gating currents, inactivation, slow inactivation, even of the rapid opening and closing of channels. Some of these phenomena may look very differently when viewed through the selfconsistent gaze of a variational treatment. Some may not. Only specific investigation will tell.

**Unsolved problems in transporters.** One of the great unsolved problems in biology is the mechanism of coupling of fluxes in transporters. Flux coupling is by its very definition a study of interactions. Much of the early work on nonequilibrium and irreversible thermodynamics was motivated by understanding these interactions.[498,499] So far no mathematics or physical chemistry has been up to the task[218,306,500,595,680-682,706,952], in my view, because none of it was automatically selfconsistent. It was never clear when additional coefficients or parameters were justified. A variational approach always produces consistent Euler Lagrange equations. A variational approach always deals consistently with the electric field and with boundary conditions. Specific chemical and physical models of transporters will then produce specific predictions of the coupling of macroscopic measurable fluxes. 'Guess and check' will soon enough converge to the proper chemical and physical model, if the properties of the model can be calculated and compared systematically to experiments.

**Unsolved problems in protein biology.** Another great mystery in biology is the proper representation of the conformational changes that abound in the protein kingdom. Most workers hope to compute these directly with all atom simulations or to observe them in some way or other through structural methods. The range of scales in Table 1 shows the difficulty in such computations. The special needs of structural measurements make it quite unlikely that many conformational changes can be studied in full atomic detail as they are used in biology.

In my view, what is needed are specific reduced models of the energetics of computational change, formulated with *EnVarA*, combining atomic scale particle representations with the important continuum fields. These can be converted to specific predictions through their Euler Lagrange equations. 'Guess and check' will hopefully lead us to useful representations.

Channel gating is another closely related mystery. The sudden opening and closing of channels is clearly a sudden change in the conformation of forces in the channel, as is the sudden opening and closing of a semiconductor diode. It may be a sudden change in the conformation (i.e., location) of atoms of the protein as well. In some gating systems, one kind of conformation change may dominate, in others, another. In any case, a specific model of the energetics is needed, in which everything is coupled to everything else. That model will start with structures of the machine, shown to us by structural biology. These structures are hardly known today, and then only in a few special channels.[78,79,129,130,336,627,898] The parts list is a little clearer. Perhaps a combination of inspired guesswork, led by insights from structural biology, and using the careful mathematics of a variational approach will work here as it has in understanding selectivity in some calcium and sodium channels. But most channels gate specifically in response to different stimuli. So one must expect an enormous diversity of gating mechanisms, albeit sharing some common structural themes, and of course always common physics and chemistry.

In general there are no shortage of unsolved problems in physical chemistry and biology that involve interactions and can be attacked afresh with variational methods. Of course, variational methods may not be the only or even the best way to deal with these problems. New technologies may provide enormous shortcuts that make some theory unnecessary. But when interactions dominate, as they seem to do in the crowded conditions near active sites, electrodes, and in channels, a theory that deals with interactions first, and always selfconsistently, and always with boundary conditions, and usually with flow, seems necessary. *EnVarA* is a first step in that direction. Others will follow, no doubt, but probably along the same path. The journey of a thousand kilometers will begin with a single step, or stumble. It seems unlikely that the journey can succeed if the steps are angstroms long, of femtosecond duration, as in atomic scale simulations. Life is too short, the journeys are too long for that.

**Conclusion.** It seems that issues of ions in channels remain inextricably bound to issues of ions in solution, as they have been for more than a century. The biological systems are much less general than physical systems, because they are built to function in a specific way in living creatures and plants, according to reasonably robust input output relations. These input output rules provide useful reduced models of the biological system and can be linked to reduced models of the atomic scale behavior of ions in channels, at least in some favorable cases.

Interestingly, analysis of the biological system forces physical chemistry to confront situations that have not been understood. Ions in channels flow in gradients of concentration, electrical, and chemical potential. Ions in channels are at enormous concentrations. Ions in channels interact with each other.

Theory and simulations must deal with these issues if ions in channels are to be understood. Indeed, I suspect these issues must be dealt with in many other cases. I suspect that many of the defining characteristics of devices that use ions occur in crowded regions that are nothing like ideal solutions.

A variational approach attacks all these issues all at once, but it is in its infancy, only a few years past conception, and so its success it not yet known. In my view, the issues of ions crowded in nonideal systems are the essential ones that must be solved as physical chemistry deals with electrochemical devices, whether biological or technological. Specifically,

(1) Theories and simulations of ionic solutions must deal with boundary conditions including spatially nonuniform boundary conditions that produce flow.

(2) Theories and simulations of ionic solutions must deal with highly concentrated solutions because these occur in the nanovalves that control the properties of biological systems.

(3) Theories and simulations of ionic solutions must deal with many interacting components because that is what occurs in biological systems. Theories and simulations must be selfconsistent.

(4) Theories and simulations of ionic solutions should allow easy introduction of new components and physics as these are discovered to be important.

A field theory of ionic solutions has most of these properties. A variational treatment has all of them, in principle, subject to the limitations discussed previously. Biologists, physical chemists and mathematicians together need to discover what else is needed to make a variational theory of ionic solutions as successful and useful as the variational theories of fluid dynamics.

**A full circle.** It seems we have come full circle. We started by using physical chemistry to deal with biological problems in the particular context of ion channels. We held channels up for close inspection, and saw the importance of interactions. We looked through them to see the importance of interactions in general in ionic solutions.

Motivated by the ambiguities of classical treatments—and biological necessity—we adopted a mathematical variational approach that deals naturally and automatically with interactions, by algebra alone. Given models of the energies of components, it produces the partial differential equations that describe the entire system of interacting components.

We suggest that the variational procedure will be useful in physical chemistry in general. We hope so, because interacting systems in physical chemistry and in biology need to be analyzed by powerful—hopefully indisputable—mathematics if they are to be effectively controlled and used for technological and medical purposes.

The exponential development of semiconductor technology was catalyzed by the successful mathematics of semiconductor physics and computational electronics. A similar development of biotechnology will occur if it can find a mathematics appropriate for computational biology. That mathematics will be a successful mathematics of physical chemistry. Interactions dominate.

## Appendix
## Models of Chemical Kinetics and the Law of Mass Action

The law of mass action is taught early and often in the education of chemists and biologists. The law is taught as a commandment without derivation or discussion, as a glance at the textbook literature will show. Commandments taught early in one's education have a particular continuing impact on thought processes, as seen in the history of human behavior, religious, political, and social. They tend to be forever unquestioned. Commandments have their uses, but in the scientific tradition it is important that they be questioned just like everything else.

The law of mass action says that the flux of a species over a potential barrier into a solution of zero concentration (i.e., into an absorbing boundary) is proportional to the number density of that species. This 'law' is certainly a reasonable initial working hypothesis. It in fact can be proven to be true—as a matter of mathematics, not science—for systems of stochastic trajectories of uncharged particles satisfying Langevin equations with high friction[294,513].

$$m\frac{d^2}{dt^2} = \text{force on particle } \mathbf{X} = \bar{\beta}\,\frac{d\mathbf{X}(t)}{dt} + ze\Phi'(x) - \sqrt{2\bar{\beta}k_BT}\;\dot{w} \qquad (10)$$

Here $\bar{\beta} = \text{friction} = m\left(k_BT/D\right)$; $m = \text{mass}$; $D = \text{Diffusion coefficient}$; $ze\Phi'(x)$ is the electrical force produced by the electric potential field $\Phi(x)$; $\sqrt{2\bar{\beta}k_BT}\;\dot{w}$ is the Gaussian white noise process that makes the Langevin equation a stochastic differential equation, with weighting chosen to satisfy the fluctuation dissipation theorem. Details are in[294,513] and explanations in [340,794,795].

The derivation of eq. (10) must use the properties of doubly conditioned trajectories if it is to deal with different concentrations in different locations. It must use the version of the Langevin equation with a second derivative. The version used by Einstein with one derivative does not allow two boundary conditions. Two boundary conditions are needed to account for the diffusion process addressed by Fick's law.

The theory[234,413] needs revision to avoid (misleading if not artifactual) boundary layers near electrodes[828,830] that fortunately have only small effects. The theory can easily be revised to describe the four electrode method widely used in experiments [38,660,790] because it avoids boundary layers altogether.

A careful derivation of the law of mass action leads to beautifully simple expressions. The solutions of the Langevin equation eq. (10) theory can be rewritten in an appealingly simple way[271,288] when concentrations are specified on either side of the channel, in the high friction limit, but without further approximation. $J_k$ is the flux of species *k*. The unidirectional flux is defined precisely in operational terms in appendix[148] and in mathematics in[294]. Roughly speaking it is the flux of a tracer into a region with negligible tracer concentration (in the language of tracer experiments) or the flux into an absorbing boundary (in the language of stochastic processes).

$$J_k = \underbrace{C_k(L)}_{\substack{\text{Source}\\ \text{Concentration}}} \overbrace{\underbrace{\left(\frac{D_k}{l}\right)}_{\substack{\text{Diffusion}\\ \text{Velocity}}} \underbrace{\text{Prob}\{R|L\}}_{\substack{\text{Conditional}\\ \text{Probability}}}}^{\text{Unidirectional Efflux}} - \overbrace{C_k(R)\left(\frac{D_k}{\underset{\substack{\uparrow\\ \text{Channel}\\ \text{Length}}}{l}}\right)\text{Prob}\{L|R\}}^{\text{Unidirectional Infflux}} \tag{11}$$

or

$$J_k = \overbrace{l \cdot k_f C_k(L_{eft})}^{\substack{\text{Unidirectional Efflux}\\ J_{out}}} - \overbrace{l \cdot k_b C_k(R_{ight})}^{\substack{\text{Unidirectional Infflux}\\ J_{in}}} \tag{12}$$

Here

$$\begin{aligned} k_f &\equiv \frac{J_{out}}{C_k(L_{eft})} = k\{R_{ight}|L_{eft}\} = \frac{D_k}{l^2}\text{Prob}\{R_{ight}|L_{eft}\} = \frac{D_k}{l^2}\frac{\exp(z_k F V_{trans}/RT)}{\frac{1}{l}\int_0^l \exp(z_k F\phi(\zeta)/RT)d\zeta}; \\ k_b &\equiv \frac{J_{in}}{C_k(R_{ight})} = k\{L_{eft}|R_{ight}\} = \frac{D_k}{l^2}\text{Prob}\{L_{eft}|R_{ight}\}\frac{D_k}{l^2}\frac{1}{\frac{1}{l}\int_0^l \exp(z_k F\phi(\zeta)/RT)d\zeta}. \end{aligned} \tag{13}$$

$R$ is the gas constant, $F$ is Faraday's constant, $T$ is the absolute temperature, $V_{trans}$ is the electrical potential across the channel, left minus right. Note the typo in eq. 14 of ref[202,], corrected here.

These equations can be written exactly as a chemical reaction in the usual mass action form, without further approximation,

$$L_{eft} \underset{k_b}{\overset{k_f}{\rightleftarrows}} R_{ight} \tag{14}$$

where

$$J_k = \overbrace{l \cdot k_f \cdot C_k(L_{eft})}^{\substack{\text{Unidirectional Efflux}\\ J_{out}}} - \overbrace{l \cdot k_b \cdot C_k(R_{ight})}^{\substack{\text{Unidirectional Infflux}\\ J_{in}}} \tag{15}$$

This looks like a beautiful and clear result. A diffusion model can be written (nearly) exactly as a chemical reaction. But it is highly misleading.

**Rate constants vary.** The difficulty is in the properties of the rate constant. The rate constant is nearly always treated as a constant independent of concentration, for example, for admirable reasons. Experimentalists, or young scientists learning the law for the first time, fear that allowing the rate constant to vary will introduce a 'fudge factor' that "lets them fit anything". They wish to avoid such arbitrary behavior and so in the name of good science, they make the rate constant constant, as the name implies.

What is rarely realized, however, is that making a rate constant constant directly contradicts physical facts of great importance. Consider the situation of interest in this paper, concentrated solutions of ions, flowing through channels, from one mixed solution to another.

The rate constant in such a system obviously must depend on the concentrations of each species. **Any property of the solutions on either side of the channel depends on the concentration of each and every species of ions.** This is the fundamental property of nonideal solutions described in many textbooks of physical chemistry for many years.[38,47,75,93,183,256,257,262,309,328,390,530,547,548,660,700,701,735,790,854] Only in ideal solutions of uncharged particles at infinite dilution are properties independent of concentration. Only in ideal solutions does the free energy of one ion depend only on the concentration of that type of ion. In real solutions and mixtures, everything interacts with everything else and all rate constants are variables depending on all concentrations. In particular, the conditional probabilities that appear in the law of mass action depend on the concentration of every ion. If those concentrations are changed, the rate constant must vary.

Indeed, even in dilute (say 1 mM) solutions of $Na^+Cl^-$, rate constants are variable. The properties of such solutions are described decently by the Debye-Hückel model of shielding and screening. In that model, ions are not ideal. Their electrochemical potential has a crucial term that varies as the square root of ionic strength. Indeed, in any system of mobile charge, screening of this sort is a crucial, if not dominant determinant of physical behavior.[135] Thus, whenever mobile charges are present, rate constants will vary with ionic strength, and with all the variables that determine ionic strength. The rate constant will not be constant.

As is discussed several times in the text, most ionic solutions do not follow the Debye-Hückel theory and have much more complex behavior. Their behavior deviates from the law of mass action in a profound way.

**Rate Constants are not Constant in Crowded Conditions.** Experimental conditions can be found, of course, in which the rate constant is constant, and those are just the conditions established in experiments designed to test the law of mass action or to use it to describe classical enzyme kinetics. But those conditions are remarkably far from the conditions in which the kinetic models are used, at least in ion channels, and probably in enzymes, and other applications in physical chemistry I am not familiar with. In ion channels, ions flow from mixed solutions, with nonideal properties, through regions of enormous concentration in which everything interacts with everything else, under the influence of large densities of charge and enormous electric fields, in systems so crowded that everything competes for the same tiny volume. Conditions of this sort are present not just to make our theories and simulations difficult. These special conditions are present in channels so a tiny valve can control macroscopic flows. One can expect crowded conditions whenever ions in small structures are used to control large flows.

Crowded conditions of this sort characterize any valve. Any valve uses small forces in small regions to control large flows in big regions. The nanovalves of life are no exception. Extreme conditions of crowding are present in ion channels because they are the conditions that allow a few atoms to control macroscopic flows of current. Extreme conditions allow robust and sensitive control of macroscopic biology by a few atoms of a protein.

Ion channels are an extreme system. They are as small as they can be, given the particulate nature of matter. Ion channels are atomic valves that allow a handful of atoms to control macroscopic flows of current, and thus macroscopic properties of cells, tissues, animals and life. They do this by working at the extremes of forces as well as sizes. They have enormous densities of ions crowded into tiny spaces with huge electric and chemical fields and forces of excluded volume. I believe ion channels will prove to be extraordinarily strong and often rigid proteins (although I hasten to say this is an idea unproven and even untested as of now, as far as I know).

Traditional chemical theories are designed for the opposite extreme, for the case of dilute noninteracting solutions that are hardly ionic. Traditional chemical theories fail altogether when used with rate constants that are constants to describe systems that are wildly nonideal.

**Mathematics must deal with interactions**. In my view, many proteins, like channels, must be analyzed with a mathematics that deals naturally with the real properties of ions, that allows everything to interact with everything else. The mathematics should deal with interactions in a natural way. Interactions should be at the core of the mathematics. They should not require *ad hominem* (or worse *ad hominiculum*) arguments that are different for each type of interaction. The mathematics should not start with ideal fluids. It cannot use the law of mass action with constant constants. Of course, not all interactions occur everywhere. Interactions that are not important in a particular system can be ignored, as PNP-DFT ignores some interactions and yet succeeds magnificently with the ryanodine receptor. Of course, it is much safer for the mathematics to include insignificant interactions than it is to ignore them a priori, if the numerical and computational complexities can be handled.

I suspect that most enzymes will use crowded ions to control flows of substrates to products, as channels use crowded ions to control flows from 'substrates' (i.e., ions outside the cell) to 'products' (i.e., ions inside the cell).

The analogy between channels and enzymes[285] has deep evolutionary origins, I suspect, since life before membranes must have used electrostatics to 'confine' its crucial molecules.

It is clear that life existed for millions or billions of years before cells were invented. Pre-cellular life was probably an RNA universe. That RNA cell free universe was devoured and encompassed by today's cellular based organisms. Today, cells use their membranes to confine the 'expensive' macromolecules that allow life to reproduce. These macromolecules of nucleic acids—RNA and DNA—and proteins are for that very real reason the essential components of life.

The crucial macromolecules of life must be confined close together if they are to function. Membranes of cells and organelles provide that confinement today. The question is what provided that confinement in life before cells existed?

I propose that the electric charge of nucleic acids and their surrounding electric field was the main confining agent before membranes took on that role. The density of mobile charge within a Debye length of RNA is ~10 molar.

I propose that the confinement motif of pre-cellular life was then used inside proteins in their active sites. The same motif would be repeated in binding proteins, enzymes, and channels inside and on the boundaries of cells, and so on and so forth, in my fanciful view of life's evolution.

In this view, enzymes, like channels and proteins, use confined ions to force everything to interact with everything else. Those interactions are central to the functioning of some channels, as we have seen. I suspect those interactions of crowded charges will prove to be central to the function of binding proteins, and enzymes as well.

If everything interacts with everything else in a way important for function, the mathematics used to describe everything must deal naturally with interactions. In that case, a variational approach like *EnVarA* becomes the natural mathematics of physiological function, as it is the natural mathematics of interaction. The mathematics should deal with interactions. It should not start with noninteracting particles of perfect fluids. It should not start with perfect fluids and perturb them because interactions dominate. They are not perturbations. The mathematics cannot use the law of mass action with constant constants.

On a larger scale, we know that most biological systems (of organelles, cells, and tissues, even organs) involve water flows, mechanical forces, membrane and cellular movements,[110] along with the ionic flows discussed in this paper. I suspect all these flows interact with each other. If they do, I know they must be analyzed with a mathematics built for interactions, like a variational approach.[642] In that case, a variational approach like *EnVarA* becomes the natural mathematics of organ function, as it is the natural mathematics of channel function, and perhaps enzyme function as well. A variational approach is needed when interactions dominate.

In electrochemistry it is clear that ions near electrodes determine many of the characteristic properties of electrochemical systems. These crowded environments are crucial to the function of electrochemical systems and to many other properties of ionic systems used in chemical engineering, I suspect. The crowded environment guarantees that everything talks to everything else. In that case, a variational approach like *EnVarA* is the natural mathematics of electrochemical function, as it is the natural mathematics of many biological functions. Wherever in physical science or engineering ions are concentrated, ions interact and a variational approach is needed, in my view. A variational approach unites physical and biological science whenever ions are concentrated and often that is where they are most important.

## Acknowledgement

I am most grateful to my many colleagues and collaborators who have made possible my journey from biophysical chemistry to physiology to physical chemistry, by way of molecular biology and channels. They have been the joy of my life, along with my extraordinary editor Ardyth Eisenberg, who has made all this worthwhile.

Financial support from the NIH, and many other sources through the years, is gratefully acknowledged. NSF funding allowed me to move, through a critical point, from muscle physiology to molecular biology.

# **Figures**

Fig. 1

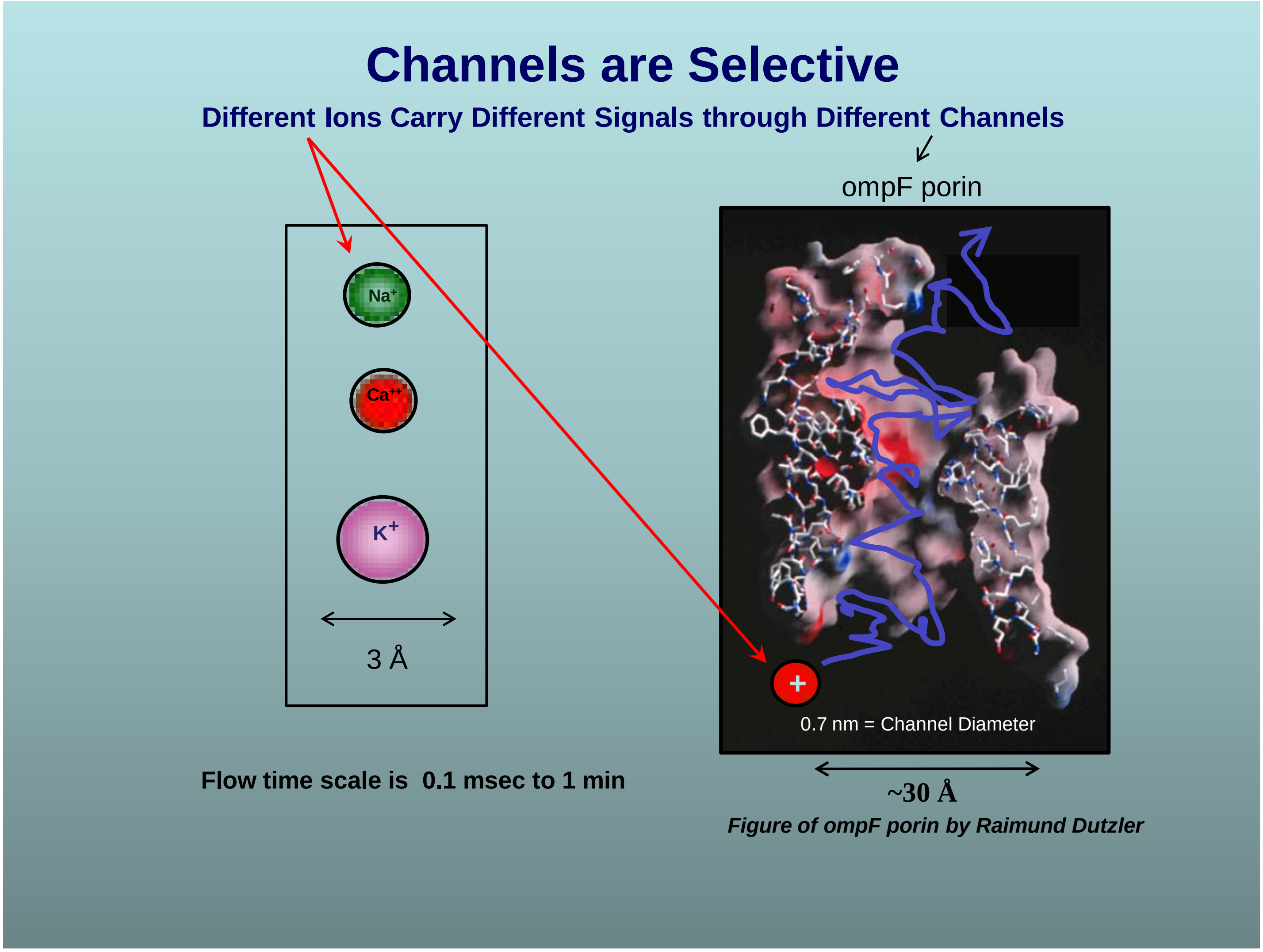

Fig. 1 A ‘longitudinal’ section of a channel structure drawn to emphasize the selectivity properties of ion channels. The structure of the channel OmpF is known from crystallography (see text for references).

Fig. 2

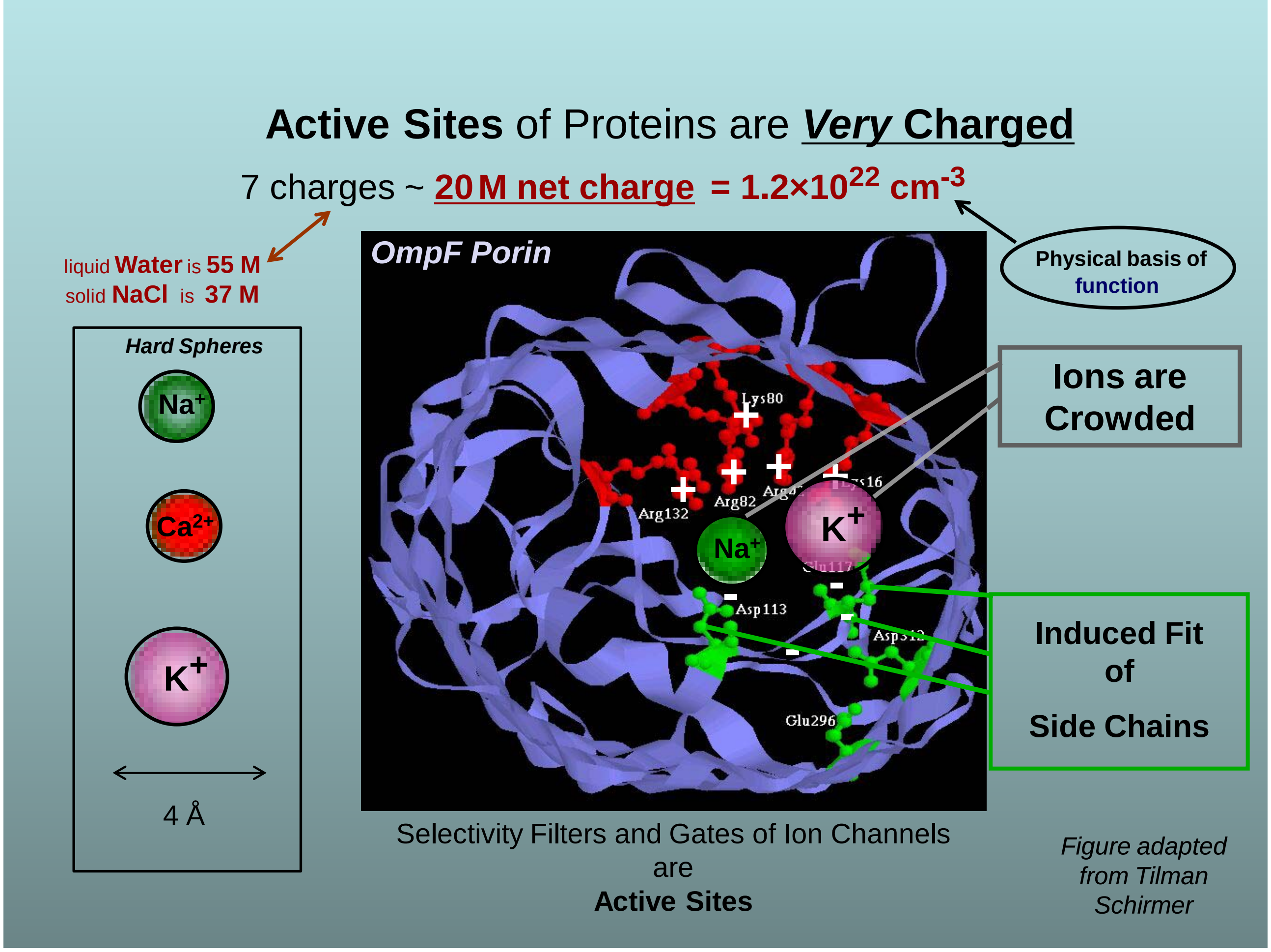

Fig. 2. A 'cross section' of a channel structure drawn to emphasize the crowding of ions in a channel. The structure of the channel OmpF is known from crystallography._ENREF_200[482,587,699,771,772,786,787].

Fig. 3

# Channel Structure Does Not Change

## once the channel is open

Typical Raw Single Channel Records

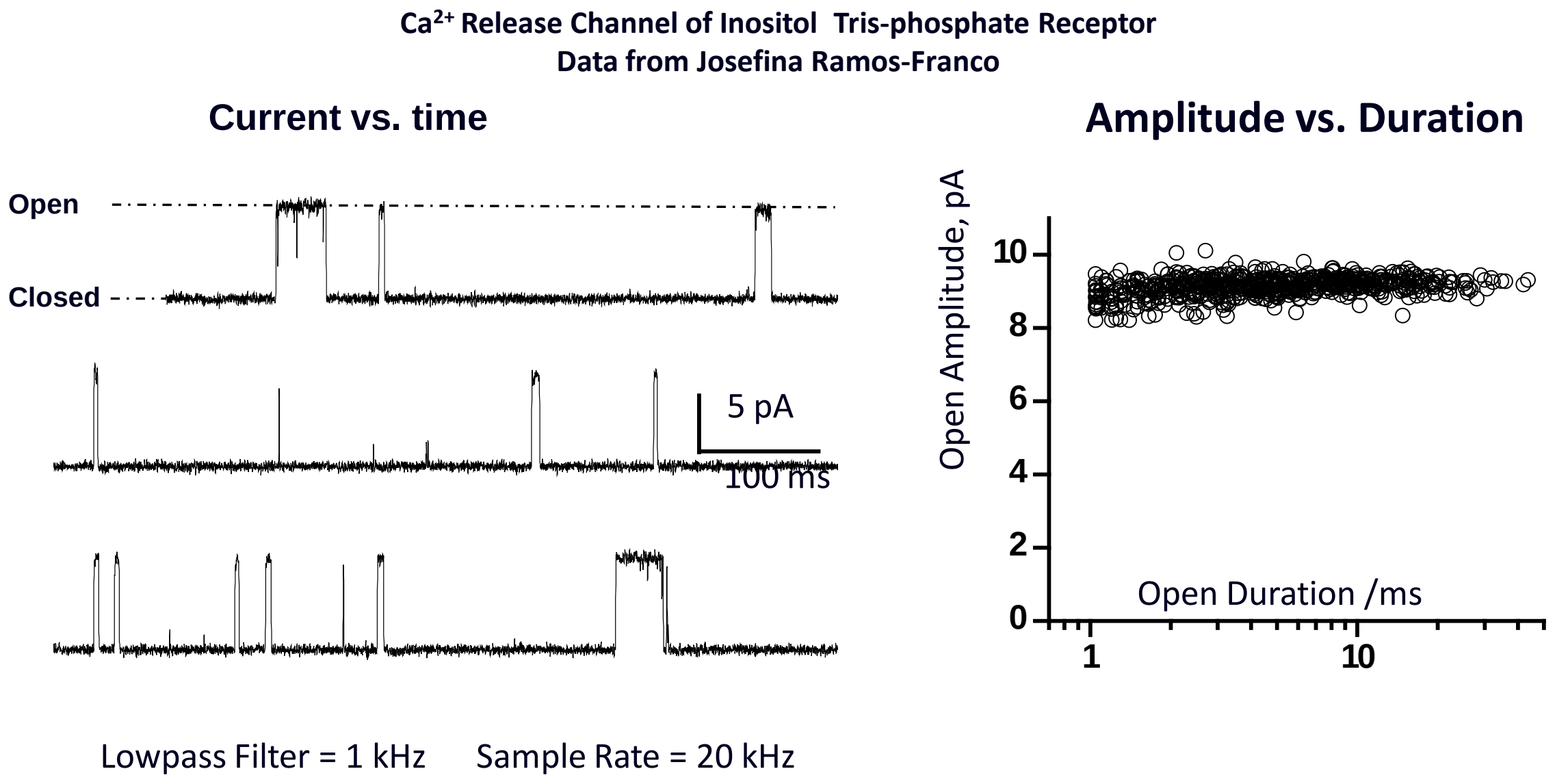

Fig. 3. Typical ‘raw’ recording of current through a single channel. The left hand panel shows a plot of current vs. time recorded by Dr. Ramos-Franco from a single channel calcium release channel, of the insoitol tris-phosphate receptor with a sampling rate of 20 k samples per second, through a 1 kHz low pass filter. I am grateful for her permission to show these results.

Under the conditions shown, the channel is closed most of the time, opening suddenly from a level of nearly zero, to a level of approximately 9 pA. The openings occur at stochastic intervals, and have stochastic durations. Successive records are not identical, but are reproducibly distributed around a mean value.

The right hand panel shows that the amplitude of the open (single) channel current is independent of duration. This is a general property of single channel recordings and is nearly their ‘operational definition’. The text argues that the amplitude can be independent of duration only if the ‘structure’ of the channel does not change significantly: if the structure changed by even 0.1 Å, the current would change because the charges of the protein are so close to the ions in the channel. Indeed, the probably are mixed with them in an ionic and “electric stew” (references in text). The word ‘structure’ means the average location of atoms, averaged over the duration of a few sampling intervals, here say $2\times50 = 100\,\mu$sec.

Fig. 4

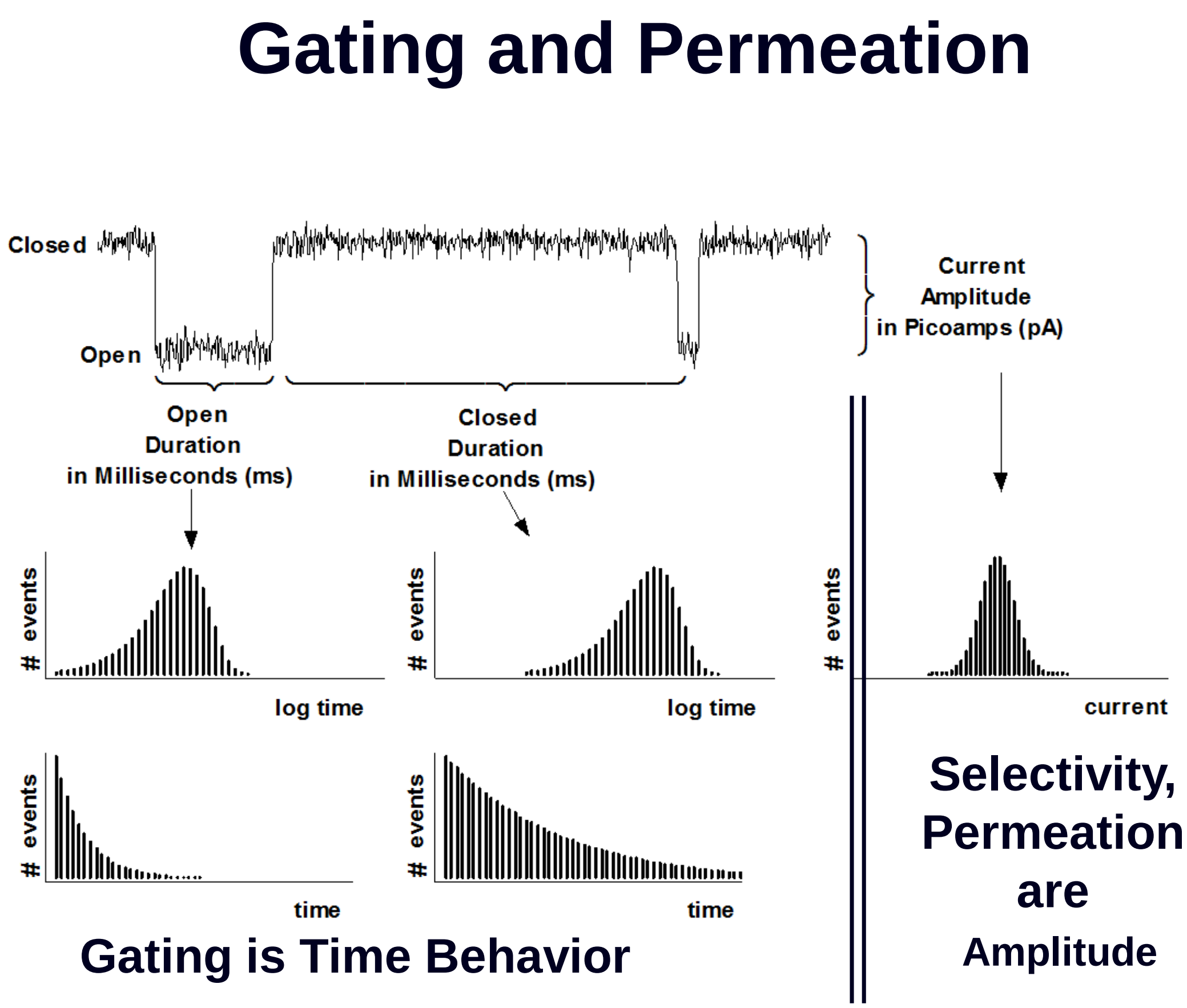

Fig. 4. Gating, Permeation, and Selectivity. The figure shows a typical single channel record (upper left, amplitudes are typically 10 pA, durations typically 10 msec). The histograms show the distribution of amplitudes (on the right hand side) and durations (on the left hand side) in log an linear plots. Selectivity properties of channels measure the current flow or binding in the open channel. They are properties of the amplitude of the open single channel current. Gating properties are produced by different mechanisms, with different structures, pharmacology, voltage, and time dependence. Most of this review is about selectivity. Not enough is yet known to make physical models of gating.

Fig. 5

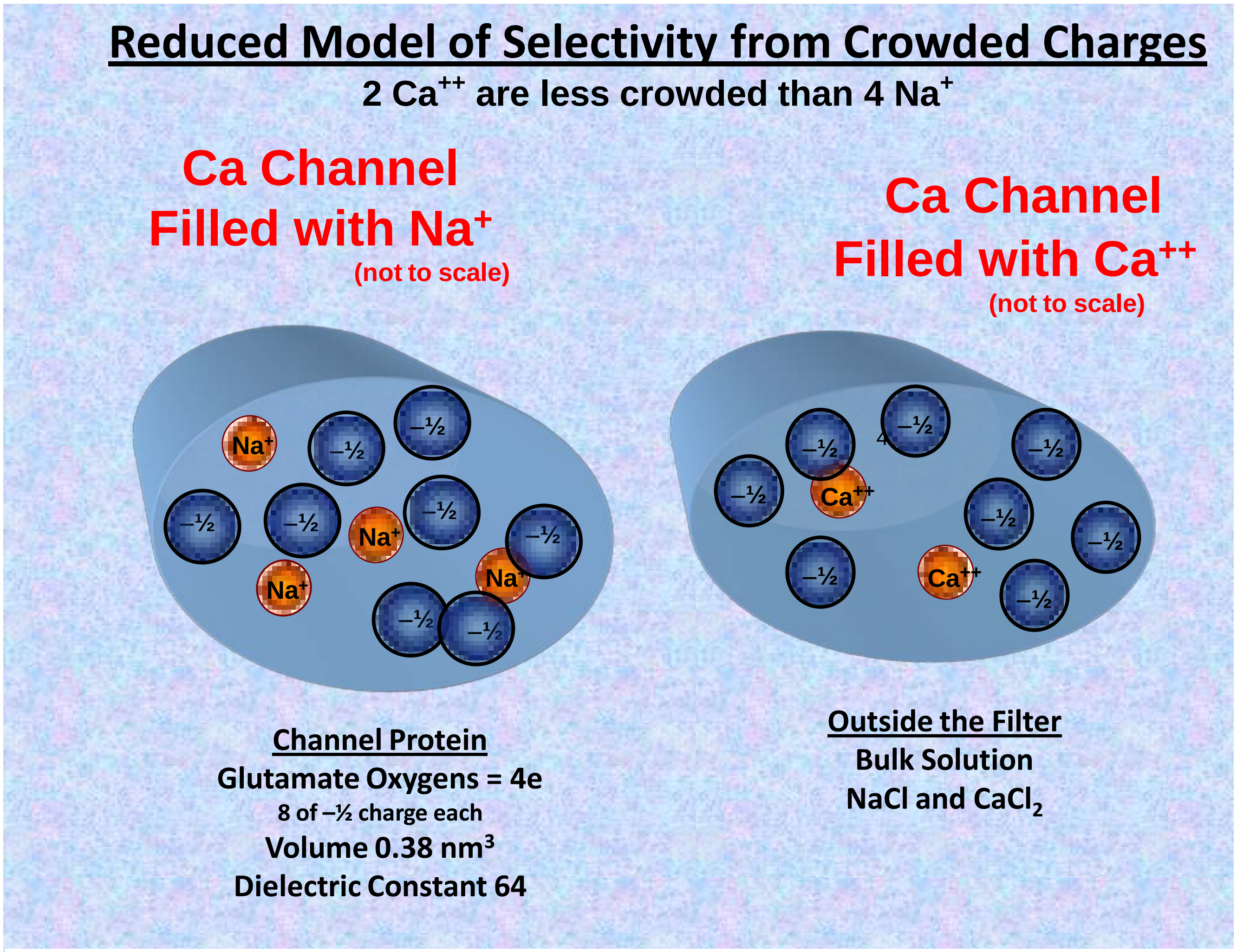

Fig. 5. Reduced model of the crowded charge model of selectivity. The model is shown for the L type calcium channel. The channel protein is represented as a right circular cylinder (the oval shape in the figure is for artistic effect) containing 8 half charged oxygens that represent the side chains of the glutamate amino acids (‘residues’) known to be responsible for the selectivity of this channel. The ‘side chains’ are treated as mobile ions, except they are not allowed to leave the channel. In a Monte Carlo simulation the side chains are distributed according to a Boltzmann distribution in the set of locations that provides lowest free energy for the system. This set of locations changes significantly even dramatically when the ionic concentrations in the baths are changed.

Fig. 6

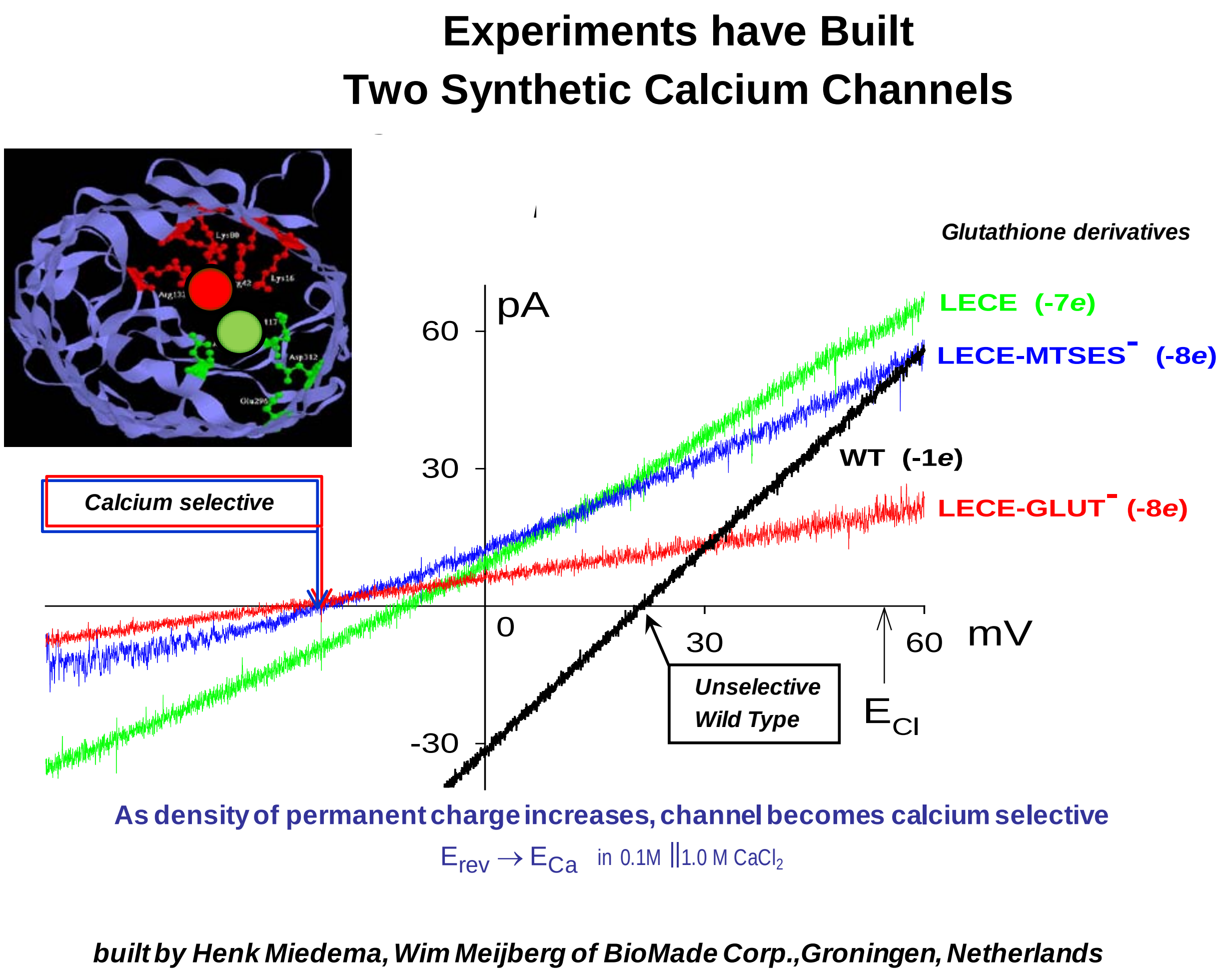

Fig. 6. Current voltage relations recorded from several mutants of OmpF porin. The details of the mutants are described in papers referred to in the text and they are important. The wild type of porin is unselective and so has a 'reversal potential' which is channel language for the gradient of chemical potential of permeant ions of +25 mV under these conditions. The mutants with large densities of glutamates and small volumes (because of the glutathione derivatives) are calcium selective and have reversal potentials of approximately -25mV. See the original papers [629-631,906] for details. The graph is redrawn from data in those papers.

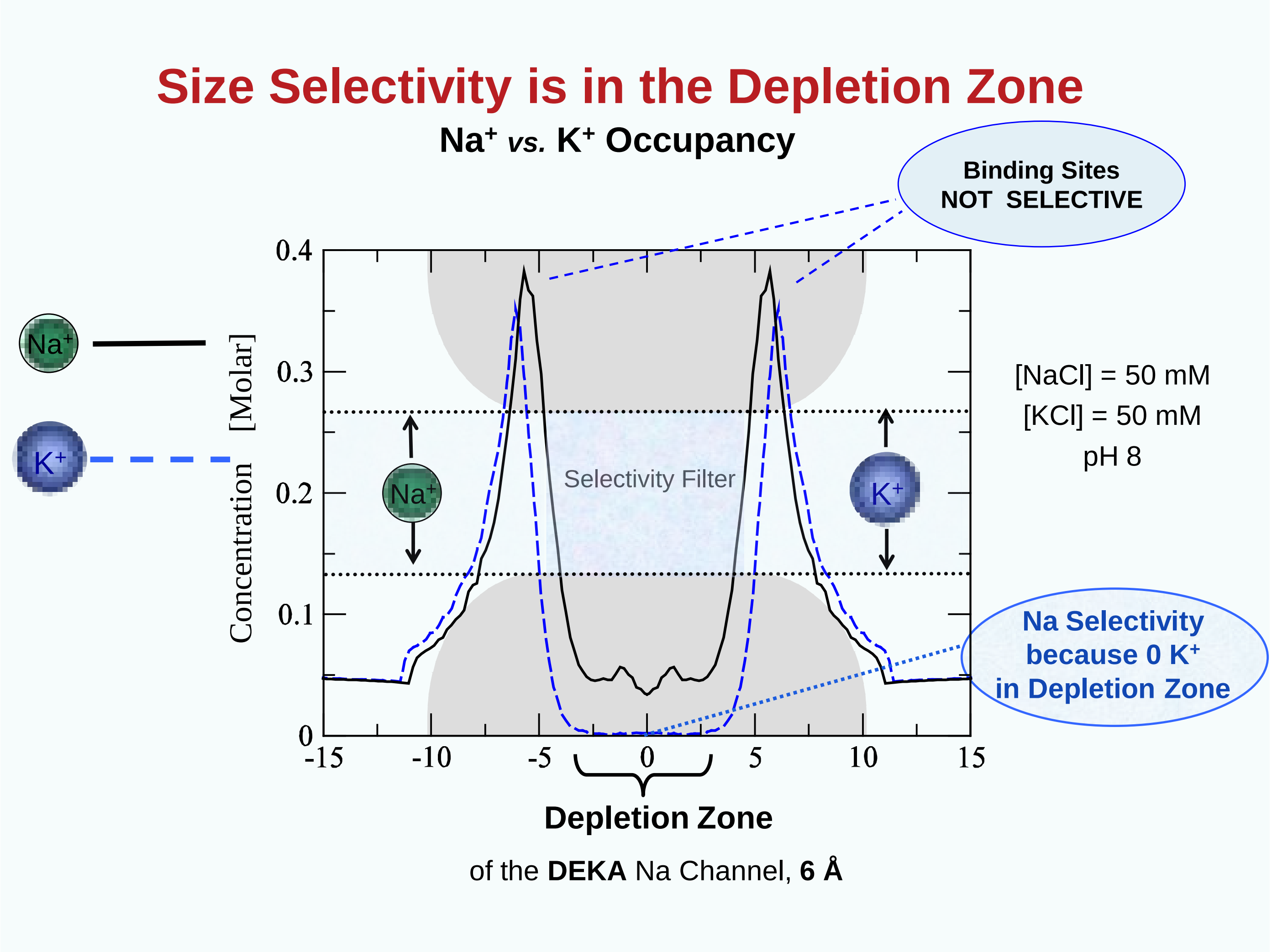

Fig. 7

Fig. 7. The number density ('concentration') of ions in the selectivity filter of a DEKA (glutamate aspartate lysine alanine) sodium channel. Note the binding sites (i.e., regions of high number density) are not selective. Remember that the binding sites are the consequence of the forces in the model. No arbitrary free energies of binding are in the model. These are outputs of the simulation. Selectivity in this channel between $Na^+$ and $K^+$ arises in the depletion zone.

Fig. 8

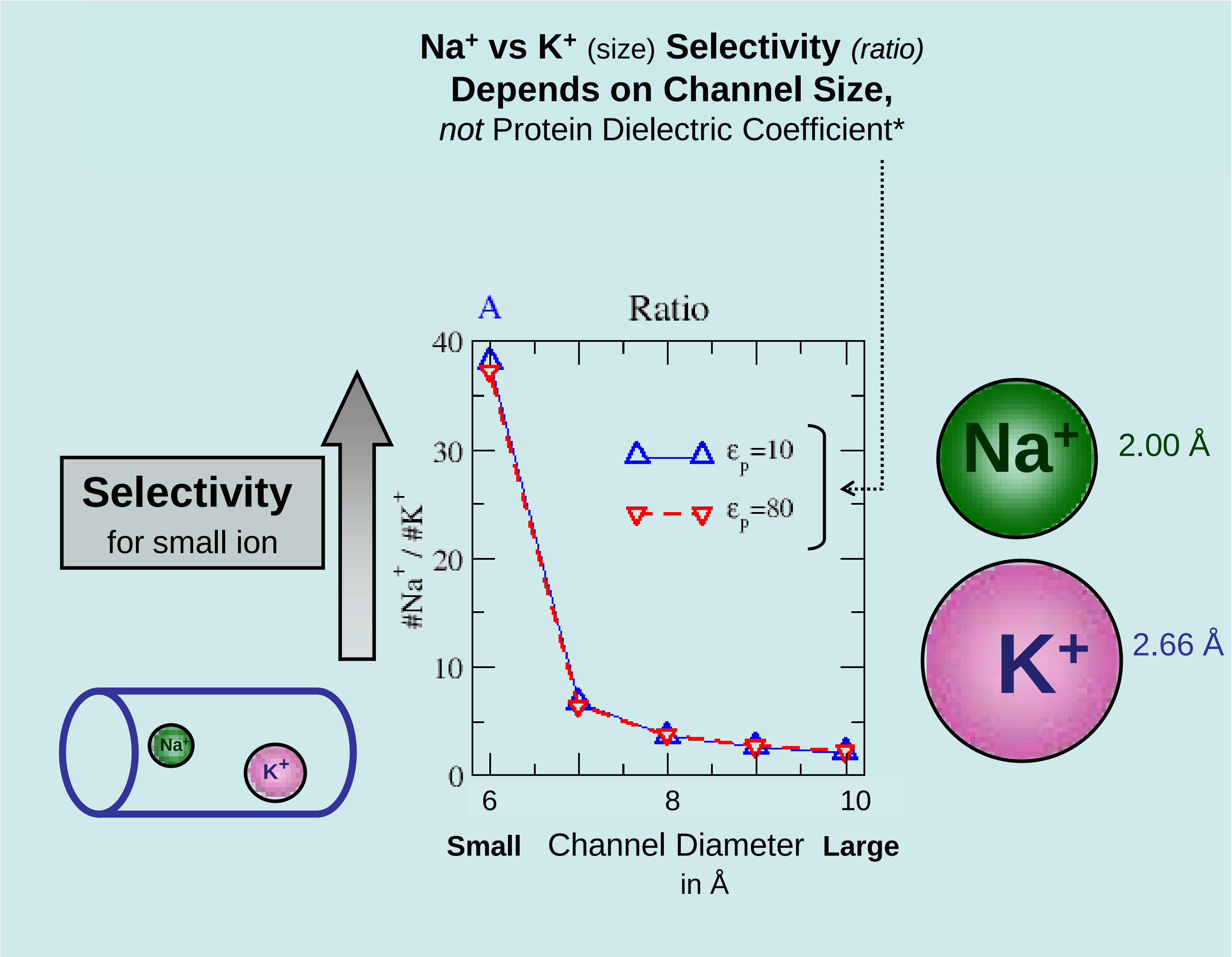

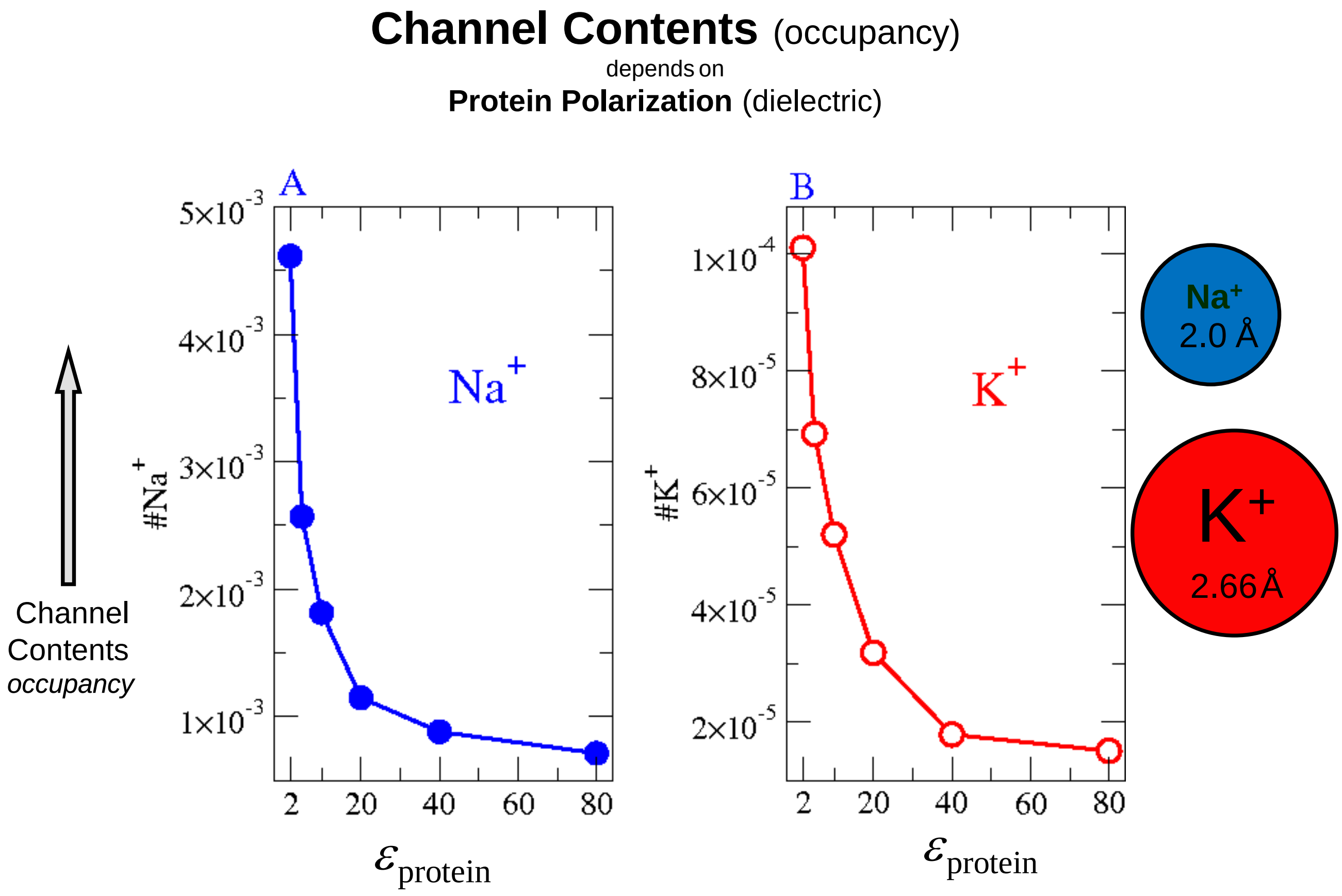

Fig. 9

Fig. 10

**Current Voltage Relations of Ryanodine Receptor Fit with Gillespie's Reduced DEDDE Model**

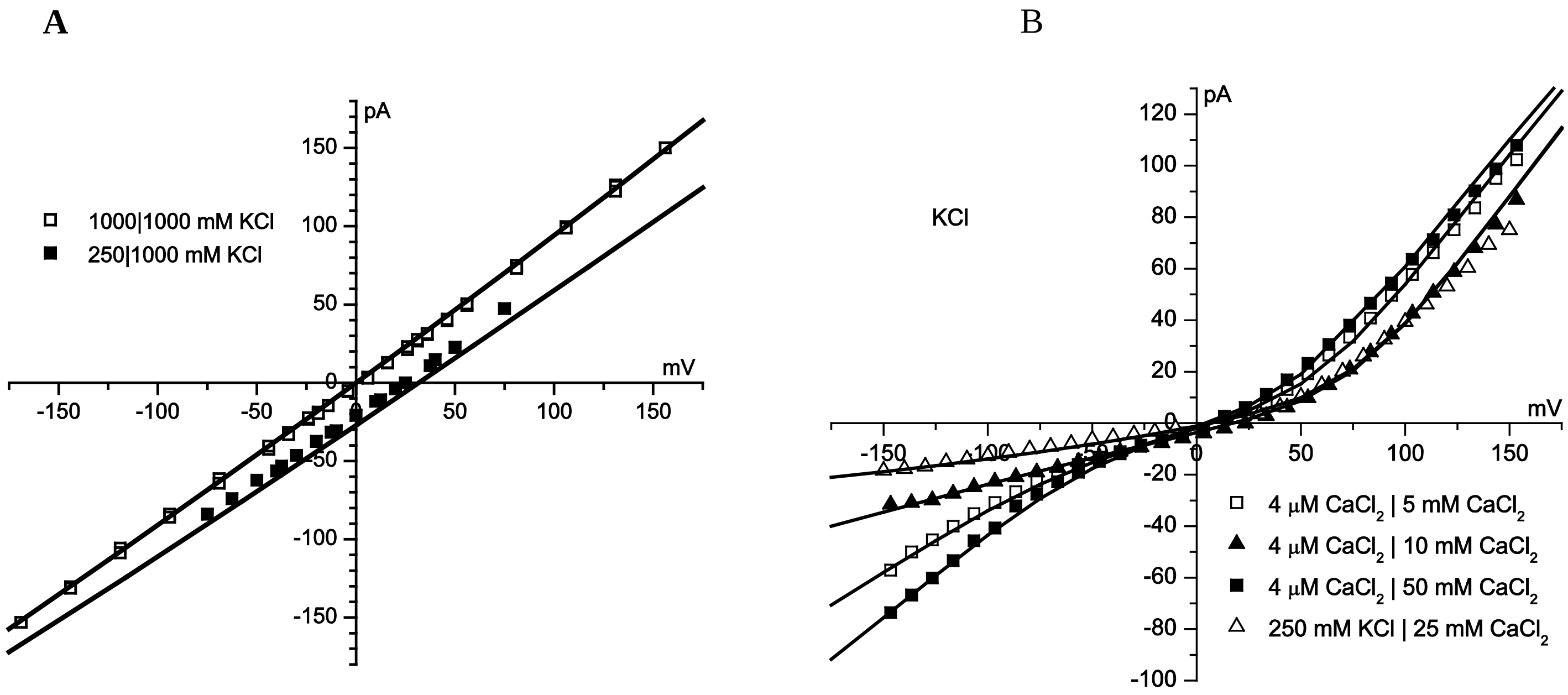

Fig. 10. Current Voltage Relations and fit with Gillespie's DEDDE Model. Details discussed are in text. Figure is redrawn from Fig. S1-A and Fig. S-9-A of supplementary material of[339,348]. Data was originally published in[143] and/or [141]. I thank Dirk Gillespie for providing the data and reading my discussion of his RyR results.

Fig. 11

**Current Voltage Relations of RyR Mutants Fit with Gillespie's Reduced DEDDE Model**

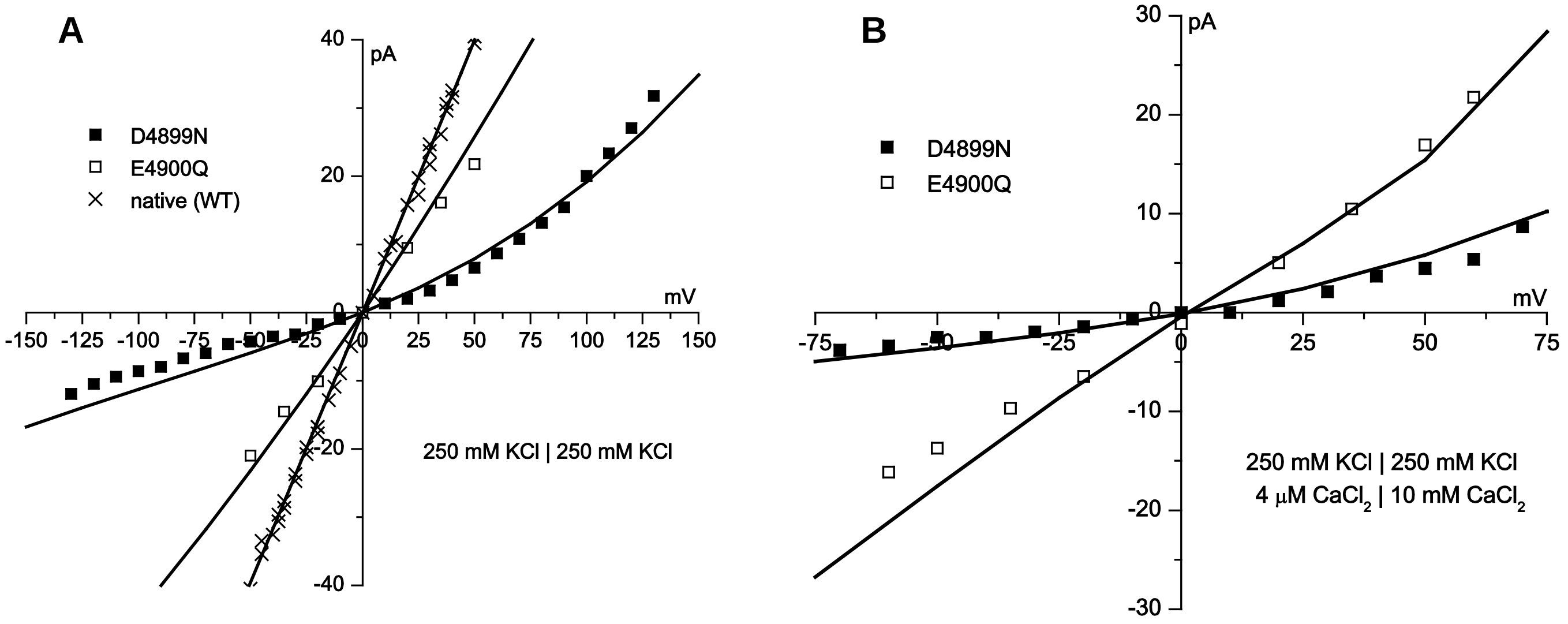

Fig. 11. Current Voltage Relations of Mutants fit with Gillespie's DEDDE Model. Details are discussed in text. The mutants are drastic as described in[913,929] and involve large changes in the density of permanent charge of the order of 13M.[357] Nonetheless, the same model with the same parameters fits the data remarkably well in different solutions. Evidently, not even the diameter of the channel changes significantly when these drastic mutations occur. Or, more precisely, whatever structural changes occur with these drastic mutations do not disturb the energetics of the channel as discussed at length in [348]. Figure is redrawn from Fig. S-1A and Fig. S9-A of supplementary material of [348]. Wild type data was originally published in [143], and/or [141], and mutation data in [357]. I thank Dirk Gillespie for providing the data and reading my discussion of his RyR results.